\documentclass[reprint,aps,rmp,amsmath,amssymb,floatfix]{revtex4-1}

\pdfoutput=1

\usepackage{graphicx}
\usepackage[pdftex,breaklinks=true,bookmarksopen=true,bookmarksopenlevel=3,bookmarksnumbered=true,colorlinks=true,urlcolor= magenta,citecolor=blue,linkcolor=blue]{hyperref}
\usepackage{dcolumn}

\begin{document}

\title{Femtosecond x rays from laser-plasma accelerators}

\author{S. Corde}
\author{K. Ta Phuoc}
\author{G. Lambert}
\author{R. Fitour}
\author{V. Malka}
\author{A. Rousse}
\affiliation{Laboratoire d'Optique Appliqu\'ee, ENSTA ParisTech - CNRS UMR7639 - \'Ecole Polytechnique, Chemin de la Huni\`ere, 91761 Palaiseau, France}
\author{A. Beck}
\author{E. Lefebvre}
\affiliation{CEA, DAM, DIF, 91297 Arpajon, France}

\begin{abstract}
Relativistic interaction of short-pulse lasers with underdense plasmas has recently led to the emergence of a novel generation of femtosecond x-ray sources. Based on radiation from electrons accelerated in plasma, these sources have the common properties to be compact and to deliver collimated, incoherent and femtosecond radiation. In this article we review, within a unified formalism, the betatron radiation of trapped and accelerated electrons in the so-called bubble regime, the synchrotron radiation of laser-accelerated electrons in usual meter-scale undulators, the nonlinear Thomson scattering from relativistic electrons oscillating in an intense laser field, and the Thomson backscattered radiation of a laser beam by laser-accelerated electrons. The underlying physics is presented using ideal models, the relevant parameters are defined, and analytical expressions providing the features of the sources are given. Numerical simulations and a summary of recent experimental results on the different mechanisms are also presented. Each section ends with the foreseen development of each scheme. Finally, one of the most promising applications of laser-plasma accelerators is discussed: the realization of a compact free-electron laser in the x-ray range of the spectrum. In the conclusion, the relevant parameters characterizing each sources are summarized. Considering typical laser-plasma interaction parameters obtained with currently available lasers, examples of the source features are given. The sources are then compared to each other in order to define their field of applications.
\end{abstract}

\maketitle

\tableofcontents

\section{Introduction}
\label{chap1}

X-ray radiation has been, ever since its discovery over a century ago, one of the most effective tools to explore the properties of matter for a broad range of scientific research. Successive generations of radiation sources have been developed providing radiation with always higher brightness, shorter wavelength and shorter pulse duration \cite{Synchrotron}. Despite remarkable progress on x-ray generation methods, there is still a need for light sources delivering femtosecond pulses of bright high-energy x-ray and gamma-ray radiation, emitted from source size of the order of a micron \cite{Science2002Service, RPP2006Pfeifer}. Indeed, the intense activity on the production of such radiation is motivated by countless applications in fundamental science, industry or medicine.\cite{PROC1992Ultrafast,WorldScientific1997Zewail,RMP1999Bloembergen,RMP2001Rousse}. For example, in the studies of structural dynamics of matter, the ultimate time scale of the vibrational period of atoms is a few tens of femtoseconds. Fundamental processes such as dissociation, isomerization, phonons, and charge transfer evolve at this time scale. High-energy radiation is used to radiograph dense objects that are opaque for low-energy x rays, while micron source size allows one to obtain high-resolution images and makes possible phase contrast imaging to see what is invisible with absorption radiography.
Several techniques are being developed to produce femtosecond x rays. In the accelerator community, large-scale free-electron laser facilities can now deliver the brightest x-ray beams ever, with unprecedented novel possibilities \cite{Nature2000Neutze, NatPhys2006Chapman, Science2007Brock, Science2007Fritz, Science2007Gaffney, NatPhot2008Barty, NatPhot2008Marchesini}. The slicing technique, combining a conventional accelerator with a femtosecond laser to isolate short electron slices, allows synchrotrons to produce radiation pulses with duration of the order of 100 fs \cite{Science2000Schoenlein}. High-energy radiation can be delivered by radioactive sources, x-ray tube, and Compton scattering sources based on a conventional accelerator. However, even if widely used these high-energy radiation sources have limitations in terms of storage, pulse duration, spectrum tunability, energy range, and source size.

In parallel, alternative and complementary methods based on laser-produced plasmas  have been developed to produce ultrashort compact radiation sources covering a wide spectral range from the extreme ultraviolet (XUV) to the gamma rays. While several laser-based source schemes were proposed in the early 1970s, this field of research has seen rapid development when lasers have been able to produce intense femtosecond pulses \cite{OC1985Strickland,Science1994Perry}. At laser intensities on the order of $10^{14}$ W/cm$^2$, XUV radiation, in the few tens of electronvolts (eV) energy range, can be produced using the mechanism of high-order harmonics generation from gas targets \cite{PRL1993Corkum, RPP1997Protopapas, RMP2000Brabec, RMP2009Krausz} or by XUV laser amplification in a laser-produced plasma \cite{RPP2002Daido}. These sources can deliver, in most recent configurations, up to a microjoule of radiation within a beam of a few milliradians divergence. At laser intensities on the order of $10^{16}$ W/cm$^2$, x-ray sources from laser solid target interaction can produce a short pulse of $K_\alpha$ line emission, emitted within $4\pi$ steradians \cite{Science1991Murnane, PhysFluids1993Kieffer, PRE1994Rousse}. Discovered more than a decade ago, these sources have been widely developed and have led to the first structural dynamics experiments at the femtosecond time scale \cite{Nature1997Rischel, Science1999Siders, Nature1999RosePetruck, PRL2001Sokolowski, PRL2001Cavalleri, Nature2001Rousse, Nature2003Sokolowski}. With recent developments, laser systems can deliver focused intensities above $10^{18}$ W/cm$^2$ and the laser-plasma interaction has entered the relativistic regime \cite{JPD2003Umstadter,RMP2006Mourou}. At this light intensity, relativistic effects become significant.  Electrons can be accelerated within the laser field or in the wakefield of the laser up to relativistic energies \cite{PRL1979Tajima, Nature1984Joshi, Nature1994Everett, Nature1995Modena, Science1996Umstadter, IEEE1996Esarey, Science2002Malka, Nature2007Patel, RMP2009Esarey}. In particular, laser wakefield acceleration has led to the production of high-quality femtosecond relativistic electron bunches \cite{Nature2004Mangles, Nature2004Geddes, Nature2004Faure, Nature2006Faure} created and accelerated up to the gigaelectronvolt level \cite{NatPhys2006Leemans, NatPhot2008Hafz, PRL2009Kneip} within only a few millimeters or centimeters plasma. 
Using these relativistic electrons, several novel x-ray source schemes have been proposed over the past decades to produce collimated and femtosecond radiation in a spectrum ranging from the soft x rays to gamma rays. Most of these schemes are based on the wiggling of relativistic electrons accelerated in a laser wakefield. In this article, the physics of these sources is reviewed, and the opportunities offered by these relativistic electrons to generate ultrashort x-ray radiation \cite{MST2001Catravas, APL2003Fritzler, IEEE2005Leemans, PRSTAB2007Hartemann, APB2007Gruner, NatPhys2008Nakajima, NatPhys2008Malka,  Phil2006Jaroszynski} are highlighted. These sources can deliver x rays or gamma rays as short as a few femtoseconds, as they inherit the temporal profile of the laser-plasma electron bunch, whose few-femtosecond duration was recently experimentally demonstrated \cite{NatPhys2011Lundh}.

%Compared to conventional synchrotron facilities, the laser-plasma approach has the main advantage of reducing the size of these facilities from hundreds of meters to the university scale laboratory and its cost from multi-hundred million dollars to multi million dollars. It can grant university scale laboratories access to X-ray sources with peak spectral brightness comparable to those of conventional facilities. In addition, these sources have the intrinsic properties of being femtosecond and synchronized with exciting laser pulses for pump/probe experiments. For conventional synchrotron facilities, external femtosecond synchronization is difficult and the X-ray pulse duration is in the picosecond timescale for most facilities, even if it can be reduced to the femtosecond timescale with the most recent technological developments. However, the 10 Hz repetition rate of current laser systems is still very low compared to synchrotron facilities which work at MHz repetition rate allowing data accumulation.

\bigskip
The aim of this article is to review the novel x-ray sources based on relativistic laser and underdense plasma interaction and to highlight their similitude by using a common formalism for their description. The paper is organized as follows. In Sec. \ref{chap2}, the general formalism of radiation from an accelerated relativistic electron is presented, which provides a framework for the description of the sources discussed throughout the paper. From this formalism, the relevant parameters describing the properties of the radiation, such as its spectrum, divergence, number of emitted photons and duration, can be extracted. As the x-ray sources presented here are based on electrons accelerated by laser wakefields [\textit{i.e.}, by the laser wakefield accelerator (LWFA)], a description of the most efficient laser-based electron accelerator to date is given in Sec. \ref{chap3}.

In Secs. \ref{chap4}, \ref{chap5} and \ref{chap6}, different methods for the production of incoherent x rays from relativistic electrons are reviewed; the objective is to define the relevant regimes to accelerate and wiggle electrons in such a way that they emit x rays. In Sec. \ref{chap4}, betatron radiation is described. In that case, a plasma cavity created in the wake of an intense laser pulse acts as both an electron accelerator and a wiggler (referred to as a plasma undulator). The scheme presented in Sec. \ref{chap5} relies on the use of laser wakefield accelerated electrons, transported and wiggled in a meter-scale periodic arrangement of permanent magnets (conventional undulator). This method is the closest to synchrotron technology. In Sec. \ref{chap6}, the nonlinear Thomson scattering and the Thomson backscattering sources are reviewed. In nonlinear Thomson scattering, electrons are directly accelerated and wiggled in an intense laser field. For the Thomson backscattering case, the plasma is used to accelerate electrons which are then wiggled in a counterpropagating electromagnetic wave (EM undulator).

In Sec. \ref{chap7}, an introduction to the topic of the free-electron laser is given and the conditions for realizing such an ultrahigh-brightness and coherent source of radiation with wavelength down to the angstrom (hard x rays) from laser-accelerated electrons are discussed for different types of undulators (plasma, conventional, and EM).

Other radiation sources based on the laser-plasma interaction have been developed and could provide photons in the keV range. For example, high-order harmonics from gas \cite{PRL1993Corkum, RPP1997Protopapas, RMP2000Brabec, RMP2009Krausz} or solid targets \cite{PRA2000Tarasevitch, NatPhys2006Dromey, NatPhys2007Thaury, RMP2009Teubner}, sources based on the flying mirror concept \cite{RMP2006Mourou, PRL2003Bulanov, PRL2007Kando, PRL2009Esirkepov}, or the $K_\alpha$ source \cite{Science1991Murnane, PhysFluids1993Kieffer, PRE1994Rousse} can produce femtosecond XUV or x-ray radiation. However, they are not based on the same physical principle of acceleration and wiggling of relativistic electrons, and will therefore not be reviewed here.

\section{General Formalism: Radiation from Relativistic Electrons}
\label{chap2}

In this section, the radiation from relativistic electrons is introduced. A qualitative understanding of the phenomenon is highlighted and analytical results for the relevant general parameters determining the radiation features are given. The formalism described here is general and will be common to all the sources presented throughout the article. We follow the approach of \textcite{Jackson} and provide the basic results necessary for the remainder of the paper. The interested reader is referred to \textcite{Wiedemann} for a complete and detailed description of synchrotron radiation from bending magnets, undulator, and wiggler insertion devices (e.g. for the angular distribution of individual undulator harmonics, for polarization or spatial and temporal coherence properties).

\subsection{Radiation features}
\label{chap2secA}

Relativistic electrons can produce bright x-ray beams if their motion is appropriately driven. For all the laser-based x-ray  sources discussed in this article, the radiation mechanism is the emission from accelerated relativistic electrons. The features of this relativistically moving charge radiation are directly linked to the electron trajectories. Obtained from the Li\'enard-Wiechert field, the general expression that gives the radiation emitted by an electron, in the direction of observation $\vec{n}$, as a function of its position, velocity, and acceleration along the trajectory is written \cite{Jackson}
\begin{equation}
\label{chap2eq1}
\begin{split}
&\frac{d^2I}{d\omega d\Omega} = \frac{e^2}{16 \pi ^3 \epsilon_0 c} \\
&\quad\times \left |  \int_{-\infty}^{+\infty} e^{i \omega [t-\vec{n}. \vec{r}(t)/c]}
\frac{\vec{n} \times \left [ (\vec{n}-\vec{\beta}) \times
\dot{\vec{\beta}} \right ]}{(1-\vec{\beta}.\vec{n})^2} dt \right
|^2.
\end{split}
\end{equation}
This equation represents the energy radiated within a spectral band $d\omega$ centered on the frequency $\omega$ and a solid angle $d\Omega$ centered on the direction of observation $\vec{n}$. Here $\vec{r}(t)$ is the electron position at time $t$, $\vec{\beta}$ is the velocity of the electron normalized to the speed of light $c$, and $\dot{\vec{\beta}}=d\vec{\beta}/dt$ is the usual acceleration divided by $c$. We stress that this expression assumes an observer placed at a distance far from the electron so that the unit vector $\vec{n}$ is constant along the trajectory. The expression (\ref{chap2eq1}) for the radiated energy shows an important number of generic features:

\begin{enumerate}

\item
When $\dot{\vec{\beta}}=0$, no radiation is emitted by the electron. This means that the acceleration is responsible for the emission of electromagnetic waves from charged particles.

\item
According to the term $(1-\vec{\beta}.\vec{n})^{-2}$, the radiated energy is maximum when $\vec{\beta}\,.\,\vec{n}\rightarrow 1$. This condition is satisfied when $\beta\simeq 1$ and $\vec{\beta}\:\|\:\vec{n}$. Thus, a relativistic electron ($\beta\simeq1$) will radiate orders of magnitude higher than a nonrelativistic electron, and its radiation will be directed along the direction of its velocity. This is simply the consequence of the Lorentz transformation: for an electron emitting an isotropic radiation in its rest frame, the Lorentz transformation implies that the radiation is highly collimated in the small cone of typical opening angle of $\Delta\theta=1/\gamma$ around the electron velocity vector, when observed in the laboratory frame (see Fig. \ref{chap2fig1}). In the following, we consider ultrarelativistic electrons, $\gamma\gg1$, and all angles which will be defined are supposed to be small so that $\tan\theta\simeq\sin\theta\simeq\theta$.

\item
The term $(\vec{n}-\vec{\beta}) \times \dot{\vec{\beta}}$, together with the relation $\dot{\vec{\beta}}_\| \propto \vec{F}_\|/\gamma^3$ and  $\dot{\vec{\beta}}_\bot \propto \vec{F}_\bot/\gamma$ between applied force and acceleration (respectively for a force longitudinal or transverse with respect to the velocity $\vec{\beta}$), indicate that applying a transverse force $\vec{F}  \, \bot \, \vec{\beta}$ is more efficient than a longitudinal force. The term also shows that the radiated energy increases with the square of the acceleration $\dot{\vec{\beta}}$. More precisely, $P\propto F_{\|}^2$ and $P \propto \gamma^2F_{\bot}^2$, where $P$ is the radiated power. Thus, it is much more efficient to use a transverse force in order to obtain high radiated energy.

\item
The phase term $e^{i \omega [t-\vec{n}. \vec{r}(t)/c]}$ can be locally approximated by $e^{i \omega(1-\beta)t}$. The integration over time will give a nonzero result only when the integrand, excluding the exponential, varies approximately at the same frequency as the phase term which oscillates at $\omega_{\varphi}=\omega(1-\beta)$. Given that the velocity $\vec{\beta}$ of the electron varies at the frequency $\omega_{e^{-}}$, the condition $\omega_{\varphi}\sim\omega_{e^{-}}$ is required to have a nonzero result. The electron will radiate at the higher frequency $\omega=\omega_{e^{-}}/(1-\beta)\simeq2\gamma^2\omega_{e^{-}}$. Thus, the usual Doppler upshift is directly extracted from this general formula. This indicates the possibility to produce x-ray beams ($\omega_X\sim10^{18}\:\text{s}^{-1}$) by wiggling a relativistic directional electron beam at a frequency far below the x-ray range: $\omega_{e^{-}}\simeq\omega_X/(2\gamma^2)$.
\end{enumerate}

This analysis underlines the directions for the production of x rays from relativistic electrons: the goal for x-ray generation from relativistic electrons is to force a relativistic electron beam to oscillate transversally. This transverse motion will be responsible for the radiation. This is the principle of synchrotron facilities, where a periodic static magnetic field, created by a succession of magnets, is used to induce a transverse motion to the electrons. The laser-based sources presented here rely on this principle. In the next sections, the properties of moving charge radiation in two different regimes are reviewed. The different laser-based x-ray sources which will be discussed can work in both regimes depending on the interaction parameters.

\subsection{Two regimes of radiation: undulator and wiggler}
\label{chap2secB}

We consider ultrarelativistic electrons with a velocity along the direction $\vec{e}_z$ executing transverse oscillations in the $\vec{e}_x$ direction. Two regimes can be distinguished.

\begin{figure}
\includegraphics[width=8.5cm]{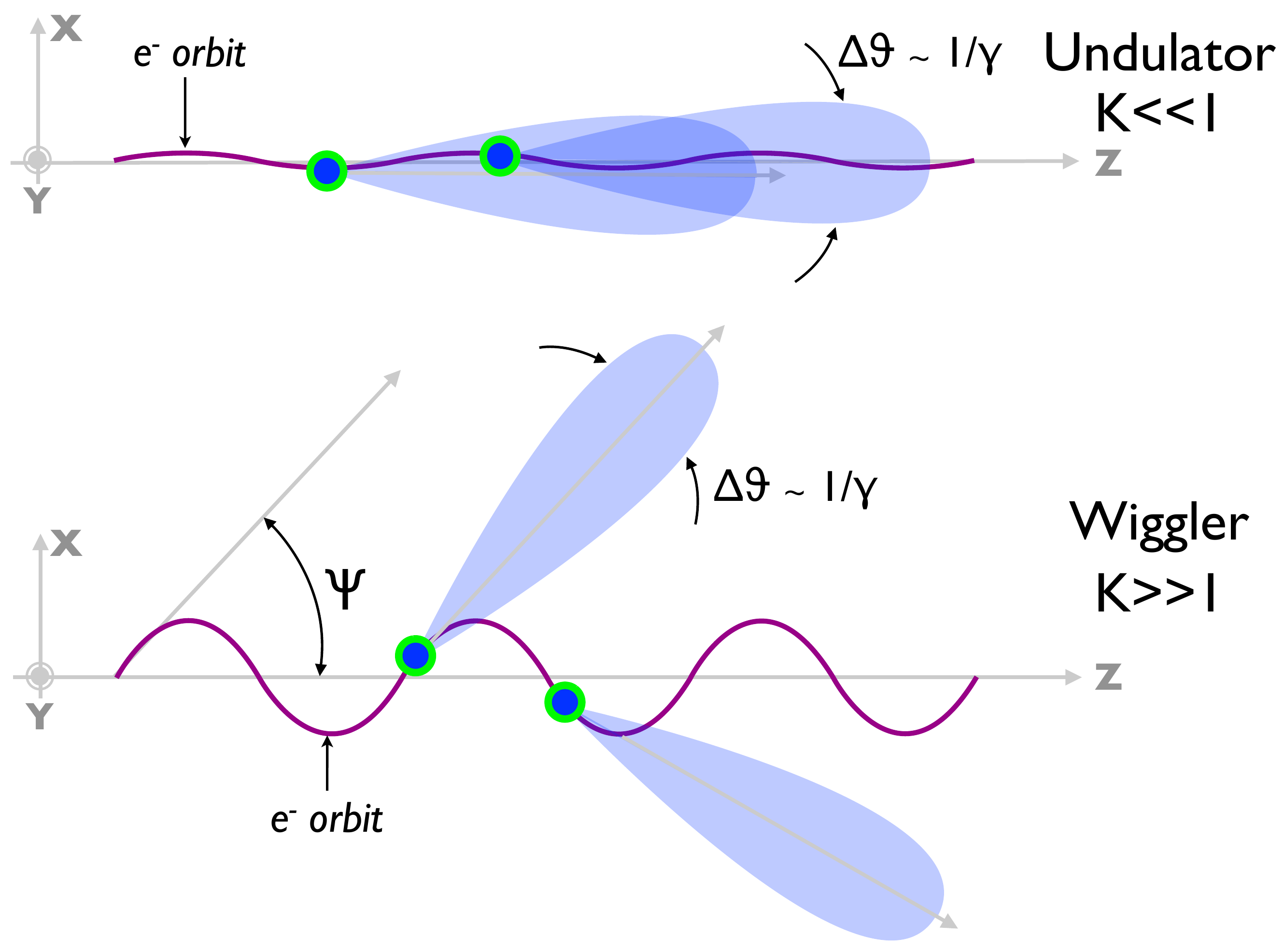}
\caption{Illustration of the undulator and wiggler limits, at the top and the bottom, respectively. The lobes represent the direction of the instantaneously emitted radiation. $\psi$ is the maximum angle between the electron velocity and the propagation axis $\vec{e}_z$ and $\Delta\theta$ is the opening angle of the radiation cone. When $\psi\ll\Delta\theta$ (undulator), the electron always radiates in the same direction along the trajectory, whereas when $\psi\gg\Delta\theta$ (wiggler), the electron radiates toward different directions in each portion of the trajectory.}
\label{chap2fig1}
\end{figure}

The undulator regime corresponds to the situation where an electron radiates in the same direction at all times along its motion, as shown in Fig. \ref{chap2fig1}. This occurs when the maximal angle of the trajectory $\psi$ is smaller than the opening angle of the radiation cone $\Delta\theta=1/\gamma$.
The wiggler regime differs from the undulator by the fact that the different sections of the trajectory radiate in different directions. Thus, emissions from the different sections are spatially decoupled. This occurs when $\psi\gg1/\gamma$.
The fundamental dimensionless parameter separating these two regimes is $K=\gamma\psi$. 
The radiation produced in these two regimes have different qualitative and quantitative properties in terms of spectrum, divergence, and radiated energy and number of emitted photons.
 
\subsection{Qualitative analysis of the radiation spectrum}
\label{chap2secC}

The shape of the radiation spectrum can be determined using simple qualitative arguments. 
In most of the cases discussed throughout the article the electron trajectory can be approximated by a simple transverse sinusoidal oscillation of period $\lambda_u$ at a constant velocity $\beta$ and constant $\gamma$. The orbit can be written as
\begin{equation}
\label{chap2eq2}
x(z)=x_0\sin(k_uz)=\frac{\psi}{k_u}\sin(k_uz)=\frac{K}{\gamma k_u}\sin(k_uz),
\end{equation}
where $k_u=2\pi/\lambda_u$ is the wave-vector norm, $x_0$ is the transverse amplitude of motion, and $\psi$ is the maximum angle between the electron velocity and the longitudinal direction $\vec{e}_z$. Since the electron energy is constant, an increase of the transverse velocity leads to a decrease of the longitudinal velocity. This can be explicitly derived from the assumed trajectory (\ref{chap2eq2}),
\begin{align}
\beta_z&\simeq \beta\left(1-\frac{K^2}{2\gamma^2}\cos^2(k_uz)\right),\\
\overline{\beta_z}&\simeq \beta\big(1-\frac{K^2}{4\gamma^2}\big)\simeq1-\frac{1}{2\gamma^2}\big(1+\frac{K^2}{2}\big).
\end{align}

\begin{figure}
\includegraphics[width=8.5cm]{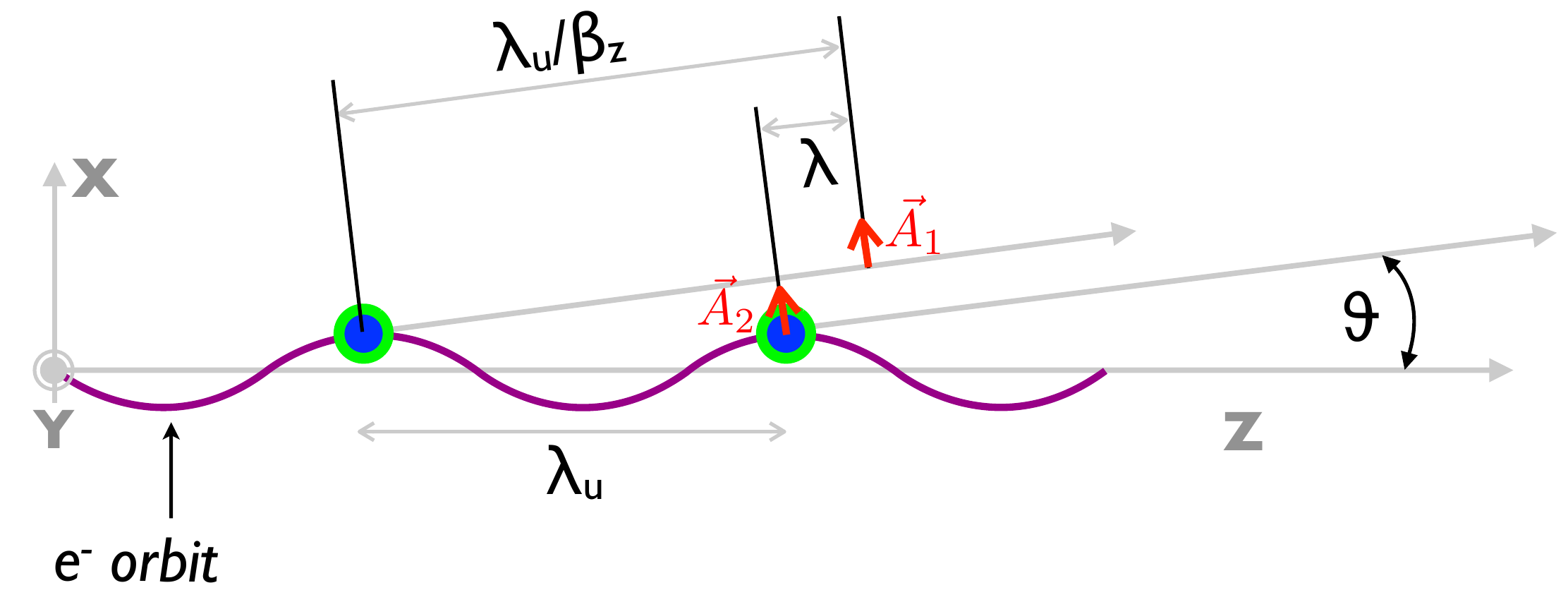}
\caption{Schematic for the calculation of the spatial period $\lambda$ of the radiation emitted toward the direction of observation $\vec{n}$, forming an angle $\theta$ with the $\vec{e}_z$ direction. At the two positions marked by a blue point, the electron radiates the same field amplitude. The distance between these two amplitudes  at a given time corresponds to $\lambda$.}
\label{chap2fig2}
\end{figure}

With the trajectory of the electron periodic, the emitted radiation is also periodic since each time the electron is in the same acceleration state, the radiated amplitude is identical. The period of the radiated field can be calculated to obtain the fundamental frequency of the radiation spectrum. The radiation emitted in the direction $\vec{n}$, forming an angle $\theta$ with the $\vec{e}_z$ direction, is considered, as represented in Fig. \ref{chap2fig2}. The field amplitude $\vec{A}_1$ radiated in the direction $\vec{n}$ by the electron at $z=0$ and $t=0$ propagates at the speed of light $c$.  At $z=\lambda_u$ and $t=\lambda_u/(\overline{\beta_z}c)$, the electron radiates an amplitude $\vec{A}_2=\vec{A}_1$.
The distance separating both amplitudes ($\vec{A}_1$ and $\vec{A}_2$) corresponds to the spatial period $\lambda$ of the radiated field and is given by
\begin{equation}
\label{chap2eq3}
\lambda=\frac{\lambda_u}{\overline{\beta_z}}-\lambda_u\cos\theta\simeq\frac{\lambda_u}{2\gamma^2}(1+\frac{K^2}{2}+\gamma^2\theta^2).
\end{equation}
The radiation spectrum consists necessarily in the fundamental frequency $\omega=2\pi c/\lambda$ and its harmonics. To know if harmonics of the fundamental are effectively present in the spectrum, it is instructive to look at the electron motion in the average rest frame, moving at the velocity $\overline{\beta_z}$ in the $\vec{e}_z$ direction with respect to the laboratory frame.

If $K\ll1$, the longitudinal velocity reduction is negligible: $\overline{\beta_z}\simeq\beta$ and $\gamma_z=1/\sqrt{1-\overline{\beta_z}^2}\simeq\gamma$. The motion contains only the fundamental component. Indeed, the motion reduces to a simple dipole in the average rest frame. The spectrum consists in a single peak at the fundamental frequency $\omega$ which depends on the angle of observation $\theta$. As $K\rightarrow1$, radiation also appears at harmonics.

\begin{figure}
\includegraphics[width=8.5cm]{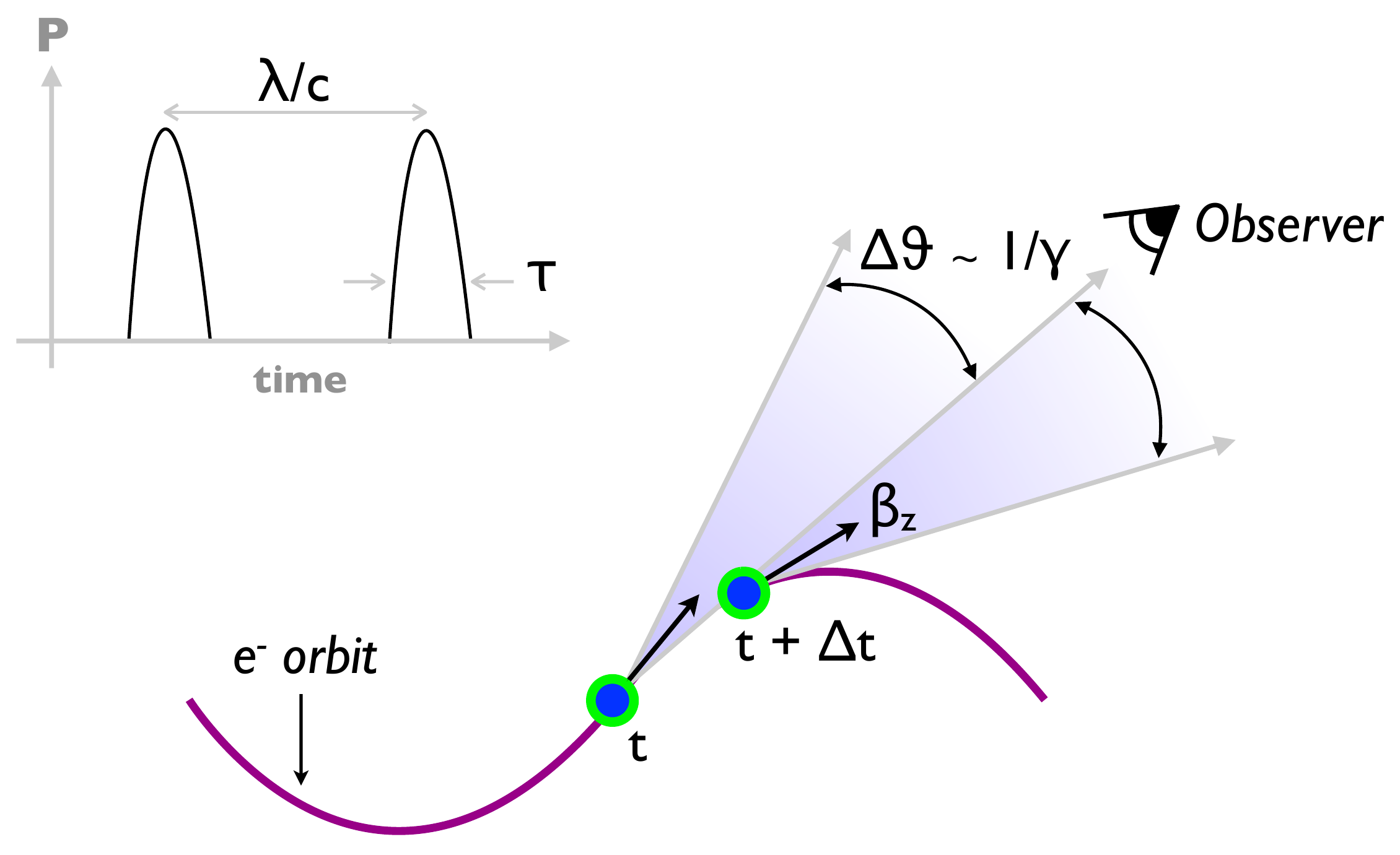}
\caption{In the wiggler limit, the radiation cone points toward the observer during a time $\Delta t$, which corresponds to a duration $\tau$ for the emitted radiation. This is repeated at each period: the observer receives bursts of radiation separated by a time $\lambda/c$. The inset gives the temporal profile of the radiation power seen by the observer.}
\label{chap2fig3}
\end{figure}

If $K\gg1$, the longitudinal velocity reduction is significant: $\gamma_z=\gamma/\sqrt{1+\frac{1}{2}K^2}$. In the average rest frame, the trajectory is a figure-eight motion. It can contain many harmonics of the fundamental. In the laboratory frame, this can be explained by the fact that an observer receives short bursts of light of duration $\tau$. Indeed, the instantaneous radiation is contained within a cone of opening angle $\Delta\theta=1/\gamma$ centered on $\vec{\beta}$ and points toward an observer positioned in the direction $\vec{n}$ during a time $\Delta t$ (see Fig. \ref{chap2fig3}), corresponding to the variation of $\vec{\beta}$ by an angle $\Delta\theta=1/\gamma$. Locally, a portion of the trajectory can be approximated by a portion of a circle of radius $\rho$, such that the direction of the velocity $\vec{\beta}$ changes by an angle $\Delta\theta$ when the electron travels a distance $d_e=2\pi\rho\times(\Delta\theta/2\pi)=\rho/\gamma$, which corresponds to a time $\Delta t = t_e = d_e/(\beta c)$. During the time $\Delta t$, the radiation has covered a distance $d_\gamma=2\rho\sin(1/2\gamma)$ corresponding to a propagation time of $t_\gamma=d_\gamma/c$. The radiation burst duration $\tau$ as observed by an observer reads
\begin{align}
\tau & =  t_e - t_\gamma \simeq \frac{13\rho}{24\gamma^3c}.
\end{align}
The temporal profile (see the inset of Fig. \ref{chap2fig3}) of the radiation emitted in the wiggler regime has been qualitatively obtained. The Fourier transform of this typical profile gives a precise representation of the radiation spectrum.
With the time profile a succession of bursts of duration $\tau$, the spectrum will contain harmonics up to the critical frequency
\begin{equation}
\omega_c\sim1/\tau\sim\gamma^3\frac{c}{\rho}.
\end{equation}
Note that the spectrum that arises from a complete calculation of the radiation emitted by a relativistic charged particle in instantaneously circular motion \cite{Jackson} is in agreement with the above estimation. Such calculation yields the synchrotron spectrum, which is written in terms of radiated energy per unit frequency and per unit time,
\begin{align}
\nonumber
\frac{dP}{d\omega}&=\frac{P_\gamma}{\omega_c}\:S(\omega/\omega_c),\\
\nonumber
S(x)&=\frac{9\sqrt{3}}{8\pi}x\int_x^{\infty}K_{5/3}(\xi)d\xi,\\
\nonumber
P_\gamma&=\frac{e^2c\gamma^4}{6\pi\epsilon_0\rho^2}=\frac{2e^2\omega_c^2}{27\pi\epsilon_0 c\gamma^2},\\
\label{chap2eq4}
\omega_c&=\frac{3}{2}\gamma^3\frac{c}{\rho},
\end{align}
where we introduced the exact definition of the critical frequency $\omega_c$ of the synchrotron spectrum which will be used throughout the review, $P_\gamma=\int dP/d\omega$ is the radiated power, and $K_{5/3}$ is the modified Bessel function of the second kind.

The expression for the radius of curvature $\rho$ can be obtained for an arbitrary trajectory. Its value can be calculated at each point of the trajectory, corresponding to a particular direction of observation. In the particular case of a sinusoidal trajectory, the radius of curvature reads $\rho(z)=\rho_0[1+\psi^2\cos^2(k_uz)]^{3/2}/|\sin(k_uz)|$ and it is minimum when the transverse position is extremum ($x=\pm x_0$) and its value at this point reads
\begin{equation}
\rho_0=\frac{\lambda_u}{2\pi\psi}=\gamma\frac{\lambda_u}{2\pi K},
\end{equation}
leading to the following expression for the critical frequency:
\begin{align}
\omega_c & = \frac{3}{2}K\gamma^22\pi c/\lambda_u\\
& = \frac{3}{4} K \omega{_{\{K \ll 1,\theta=0\}}}.
\end{align}
Note that for a nonplanar and nonsinusoidal trajectory, the spectrum extends up to a critical frequency determined by the minimal radius of curvature of the trajectory. 

The parameter $K$ can be considered as the number of decoupled sections of the trajectory. With each section radiating toward a different direction, the radiation is spatially decoupled and leads to bursts of duration $\tau$ in each direction and to a broad spectrum with harmonics of the fundamental up to $\omega_c$.

\bigskip
An infinite periodical motion has been considered so far, leading to harmonics that are spectrally infinitely thin. For a finite number of oscillation periods $N$, Fourier transform properties imply that the harmonic of number $n$ and of wavelength $\lambda_n=\lambda/n$ has a width given by
\begin{equation}
\frac{\Delta\lambda_n}{\lambda_n}=\frac{1}{nN}.
\end{equation}
This corresponds to the harmonic bandwidth in a given direction characterized by the angle $\theta$. Since the wavelength $\lambda_n$ depends on the angle $\theta$ [see Eq. (\ref{chap2eq3})], the integration of the radiation over a small aperture of finite dimension broadens each harmonic. When integrating over the total angular distribution, harmonics overlap and the radiation spectrum becomes continuous, but keeps the same extension (up to $\omega_c$). In the undulator limit, in which only the fundamental wavelength is present, the bandwidth is highly degraded when integrating over the total angular distribution.

\subsection{Duration and divergence of the radiation}
\label{chap2secD}

The pulse duration and the divergence of the radiation emitted by a single electron can be deduced from the previous analysis. If $N$ is the number of periods of the trajectory, the radiation consists in $N$ periods of length $\lambda$ and the total duration of the pulse is $\tau_r{_{|_{N_e=1}}}=N\lambda/c$. 

For an undulator, $K\ll1$, the direction of the vector $\vec{\beta}$ varies along the trajectory by an angle $\psi$ negligible compared to $\Delta\theta=1/\gamma$. Hence, the radiation from all sections of the trajectory overlap and the typical opening angle of the radiation is simply $\theta_{r}=1/\gamma$~\footnote{For $K<1$, the root mean square angle of the angular distribution of the radiated energy at the fundamental frequency is $\langle\theta^2\rangle^{1/2}=1/\gamma_z$, which simplifies to $1/\gamma$ for $K\ll1$~\cite{Jackson}.}. For a wiggler, $K\gg1$, the divergence increases in the direction of the transverse oscillation $\vec{e}_x$, but remains identical in the orthogonal direction $\vec{e}_y$. In the direction of the motion $\vec{e}_x$, the typical opening angle of the radiation is $\theta_{Xr}=\psi=K/\gamma$~\footnote{Here the typical opening angle is defined as the maximum deflection angle of the electron trajectory $\psi$, such that the full width of the angular distribution of the radiated energy in the direction $\vec{e}_x$ is $2K/\gamma$.}, which is greater than $\Delta\theta=1/\gamma$, while the typical opening angle of the radiation in the direction $\vec{e}_y$ is $\theta_{Yr}=1/\gamma$. For the general case of a transverse motion occurring in the $\vec{e}_x$ and $\vec{e}_y$ directions, the angular profile can take various shapes depending on the exact three-dimensional (3D) trajectory.

\subsection{Analytical formulas for the total radiated energy and the number of emitted photons}
\label{chap2secE}

Analytical calculations provide simple expressions for the radiated energy and the number of emitted photons per period. Using the expression of the radiated power by an electron $P(t)=(e^2/6\pi\epsilon_0c)\gamma^2[(d\hat{\vec{p}}/dt)^2-(d\gamma/dt)^2]$ ($\hat{\vec{p}}$ is the momentum normalized to $mc$), the averaged radiated power $\overline{P}_\gamma$ and the total radiated energy per period $I_\gamma$ can be derived for both the undulator and the wiggler case for an arbitrary trajectory. In the case of a planar sinusoidal trajectory, the result is
\begin{align}
\label{eq_chap2_power}
\overline{P}_\gamma &=  \frac{\pi e^2c}{3\epsilon_0}\:\frac{\gamma^2K^2}{\lambda_u^2},\\
I_\gamma &= \frac{\pi e^2}{3\epsilon_0}\:\frac{\gamma^2K^2}{\lambda_u}.
\end{align}

To obtain an estimation of the number of emitted photons $N_\gamma $, a mean energy of photons must be determined. For $K\ll1$, the spectrum is quasimonochromatic in the forward direction and the mean energy of photons, after integrating over the angular distribution, is equal to $\hbar\omega_{\theta=0}/2$. The number of emitted photons reads
\begin{equation}
N_\gamma   =  \frac{2\pi}{3}\alpha K^2,
\end{equation}
where $\alpha=e^2/(4\pi\epsilon_0\hbar c)$ is the fine structure constant.\\
For $K\gg1$, the spectrum is synchrotronlike with the critical frequency $\omega_c$. Using the fact that for a synchrotron spectrum $\langle\hbar\omega\rangle=(8/15\sqrt{3})\hbar\omega_c$, the following estimation is derived:
\begin{equation}
N_\gamma  = \frac{5\sqrt{3}\pi}{6}\alpha K.
\end{equation}

\subsection{Radiation from an ideal electron bunch}
\label{chap2secF}

The radiation from a single electron has been discussed so far. We now take into account the fact that there are $N_e$ electrons contained in the bunch, assuming that they all have exactly the same energy and the same initial momentum (zero emittance). The radiation from several electrons is obtained by summing the contribution of each electron before taking the squared norm,
\begin{equation}
\begin{split}
&\frac{d^2I}{d\omega d\Omega} = \frac{e^2}{16 \pi ^3 \epsilon_0 c} \\
&\quad\times \left | \sum_{j=1}^{N_{e}}\int_{-\infty}^{+\infty} e^{i \omega (t-\vec{n}. \vec{r}_j(t)/c)}
\frac{\vec{n} \times \left [ (\vec{n}-\vec{\beta}_j) \times
\dot{\vec{\beta}}_j \right ]}{(1-\vec{\beta}_j.\vec{n})^2} dt \right
|^2.
\end{split}
\end{equation}
This formula expresses the coherent addition of the radiation field of each electron. It can be considerably simplified by considering that all electrons follow similar trajectories linked to each other by a spatiotemporal translation $( t_j,\vec{R}_j)$ of a reference trajectory $\vec{r}(t)$:
\begin{align}
\begin{split}
\vec{r}_j(t)   &= \vec{R}_j+\vec{r}(t-t_j),\\
\frac{d^2I}{d\omega d\Omega}  &= \left | \sum_{j=1}^{N_{e}} e^{i\omega (t_j-\vec{n}.\vec{R}_j/c)} \right |^2
\frac{e^2}{16 \pi ^3 \epsilon_0 c}\\
&\quad\times \left | \int_{-\infty}^{+\infty} e^{i \omega (t-\vec{n}. \vec{r}(t)/c)}
\frac{\vec{n} \times \left [ (\vec{n}-\vec{\beta}) \times
\dot{\vec{\beta}} \right ]}{(1-\vec{\beta}.\vec{n})^2} dt \right | ^2.
\end{split}
\end{align}
The radiated energy per unit frequency and unit solid angle from an electron bunch is equal to the radiated energy from a single electron following the trajectory $\vec{r}(t)$ multiplied by the coherence factor
\begin{equation}
c(\omega)=\left | \sum_{j=1}^{N_{e}} e^{i\omega (t_j-\vec{n}.\vec{R}_j/c)} \right |^2.
\end{equation}
The value of $c(\omega)$ depends on the electron distribution in the $(t_j,\vec{R}_j)$ space. For a uniform distribution, $c(\omega)=0$, whereas for a random distribution, $c(\omega)=N_e$ on average. If the distribution is microbunched at the wavelength $\lambda_b=2\pi c/\omega_b$, the summation is coherent for the frequency $\omega_b$ and its harmonics, $c(n\omega_b)=N_e^2$ for $n\in\mathbb{N}^*$.

In large accelerators or in laser-plasma accelerators, electrons are randomly distributed inside the bunch at the x-ray wavelength scale, and the radiation is incoherently summed,
\begin{align}
c(\omega)&=N_e,\\
\frac{d^2I}{d\omega d\Omega}{_{\big|{N_e}}}&=N_e\frac{d^2I}{d\omega d\Omega}_{\big|{N_e=1}}.
\end{align}
The spectrum shape and the radiation divergence $\theta_r$ remain unchanged for an electron bunch. The temporal profile of the radiation is given by the convolution between the electron bunch temporal profile and the radiation profile from a single electron. Since the typical electron bunch length is in the micron range, whereas the radiation length from a single electron $l_r{_{|_{N_e=1}}}=N\lambda$ is in the nanometer range for x rays, the duration of the radiation from an electron bunch is in most cases approximately equal to the bunch duration, $\tau_r{_{|_{N_e}}}=\tau_b$.

\bigskip
In addition, the electron experiences the radiation from other electrons in the case of a bunch, which can modify its motion and its energy. This interaction of the bunch with its own radiation can lead to a microbunching of the electron distribution within the bunch at the fundamental wavelength of the radiation and its harmonics. This is the free-electron laser (FEL) process which produces coherent radiation. Here the coherence factor $c(n\omega_b)$ is $N_e^2$ instead of $N_e$, indicating that the FEL radiates orders of magnitude higher than conventional synchrotrons. However, the FEL effect requires stringent conditions on the electron beam quality and an important number $N$ of oscillations. At the present status of laser-plasma accelerators, realizing a FEL represents a technological challenge. In the following sections, the interaction between the electron bunch and its radiation will not be taken into account because in these schemes the conditions required for the FEL are not fulfilled: the electron distribution remains random along the propagation and the radiation is incoherent. In Sec. \ref{chap7}, the underlying physics of the free-electron laser is presented in more detail and the possible realization of such high-brightness coherent radiation using laser-plasma accelerators is discussed. 

\subsection{Radiation reaction}
\label{chap2secRR}

The radiation reaction (RR) corresponds to the effect of the electromagnetic field scattered by an electron on itself, the so-called self-interaction  \cite{Jackson, Hartemann, LandauLifshitz}. RR effects can modify the electron trajectory and its energy and it is therefore important to define the range of parameters for which these effects come into play. The main steps, followed by Dirac to derive the relativistically covariant form of the self-force associated with RR, are to solve for the self-quadripotential $A_\mu^s$ scattered by the electron in terms of Green's functions, and then to calculate the associated electromagnetic force on the electron, $F_\mu^s=-e(\partial_\mu A_\nu^s-\partial_\nu A_\mu^s)u^\nu$, where $u^\nu=dx^\nu/d\tau$ is the electron quadrivelocity with $\tau$ the electron proper time. This leads to the Dirac-Lorentz equation of motion for a pointlike electron,
\begin{align}
\frac{dp_\mu}{d\tau} &= -eF_{\mu \nu} u^\nu + \tau_0 \left [ \frac{d^2 p_\mu}{d\tau^2} - \frac{p_\mu}{m^2c^2} \left ( \frac{dp_\nu}{d\tau} \frac{dp^\nu}{d\tau} \right) \right ],
\label{chap2secRReq1}
\end{align}
where the first term is the Lorentz force with $F_{\mu \nu}$ the external electromagnetic field tensor, the last two terms that define the RR self-force correspond, respectively, to the Schott term and the radiation damping term, $p_\mu=mu_\mu$ is the electron quadrimomentum and $\tau_0=2r_e/(3c)=e^2/(6\pi\epsilon_0mc^3)=6.26\times10^{-24}\:\text{s}$, with $r_e$ the classical electron radius. The Schott term $\tau_0d_\tau^2p_\mu=-d_\tau G_\mu$ accounts for the change of energy momentum $G_\mu$ of the external field, while the radiation damping term $-\tau_0 p_\mu (d_\tau p_\nu d_\tau p^\nu) /(m^2c^2)=-d_\tau H_\mu$ accounts for the change of energy-momentum $H_\mu$ of the scattered electromagnetic wave \cite{Hartemann}.

Several difficulties appear in the point electron model described by the Dirac-Lorentz equation (\ref{chap2secRReq1}). First, it admits unphysical runaway solutions with exponentially increasing acceleration, which can be eliminated by requiring the Dirac-Rohrlich asymptotic condition $\lim_{\tau\rightarrow\pm\infty}d_\tau p_\mu=0$. Second, physical solutions presents acausal preacceleration, \textit{i.e.}, that electron momentum changes before an external force is suddenly applied, on a time scale $\tau_0$. Equation (\ref{chap2secRReq1}) can be approximated by evaluating the RR self-force with the solution of the zeroth-order equation $d_\tau p_\mu = -eF_{\mu \nu} u^\nu$ \cite{LandauLifshitz}. This yields the Landau-Lifshitz equation which neither admits runaway solutions nor presents acausal preacceleration behavior.

For the case of an electron undulating according to Eq. (\ref{chap2eq2}), it is important to define the range of parameters for which radiation reaction comes into play and has to be included in the description of the electron motion and its radiation. For relativistic electrons, the dominant term in the radiation reaction comes from the energy momentum transferred to the scattered electromagnetic wave [while for rest electrons, the radiation reaction describes the direct exchange of energy momentum between the external field and the scattered wave \cite{Hartemann}]. The rate of energy loss $\nu_\gamma$ for the electron can be estimated from $mc^2d\gamma/dt=-\overline{P}_\gamma$, with $\overline{P}_\gamma$ the average power radiated by the electron given by Eq. (\ref{eq_chap2_power}). It leads to $\gamma(t)=\gamma_0/(1+\nu_\gamma t)$ with
\begin{align}
\nu_\gamma = \frac{\tau_0}{2} \gamma_0 K^2 \left ( \frac{2\pi c}{\lambda_u} \right)^2,
\end{align}
and $\gamma_0$ is initial gamma factor of the electron \cite{PRL1997Telnov, PRL1998Huang, NIMA2000Esarey, PoP2005Koga, PRE2006Michel}. Therefore, the radiation reaction can be neglected when the interaction duration or equivalently the number of oscillations satisfy respectively, $\tau\ll\nu_\gamma^{-1}$ and $N\ll N_\text{RR}=\lambda_u/(2\pi^2c\tau_0\gamma_0 K^2)$. With conventional undulators (see Sec. \ref{chap5}), RR will always be negligible; for $\sim10$ GeV electrons, $\lambda_u\sim1$ cm and $K\sim1$, the limiting number of period $N_\text{RR}$ is on the order of $1.3 \times10^7$. For current and short-term laser-based betatron experiments (see Sec. \ref{chap4}), RR is negligible but in the long term or for electron beam driven plasma accelerators [e.g., parameters of the Facility for Advanced Accelerator Experimental Tests (FACET) \cite{NJP2010Hogan}], $\sim10$ GeV electrons, $\lambda_u\sim1$ cm, and $K\sim100$ lead to $N_\text{RR}\sim 1.3 \times10^3$. For Thomson backscattering (see Sec. \ref{chap6}), GeV electrons colliding with a laser pulse of strength parameter $K\sim10$ and wavelength $\lambda=0.8$ $\mu$m ($\lambda_u=\lambda/2$) leads to $N_\text{RR}\sim60$. The radiation reaction can be neglected in Thomson backscattering for sub-GeV electron beams and laser pulses of strength parameter on the order of unity, for which $N_\text{RR}\gtrsim 6\times10^3$.

In the realm of quantum electrodynamics (QED), the radiation reaction corresponds to the recoil experienced by an electron due to consecutive incoherent photon emissions~\cite{PRL2010Piazza}. Quantum effects become important when the electron energy loss associated with the emission of a photon is on the order of the electron energy. Signatures of quantum effects can be observed before entering this quantum regime. Indeed, quantum fluctuations, which imply that different electrons emit a different number of photons (that carry different energies) and hence lose a different amount of energy, can lead to an observable increase of the electron beam energy spread \cite{NIMA2000Esarey}. Experimentally, a study of the quantum regime is accessible in the framework of Compton scattering (Sec. \ref{chap6}). QED effects, such as nonlinear Compton scattering~\cite{PRL1996Bula} and the production of electron-positron pairs from light~\cite{PRL1997Burke}, were observed in the SLAC E-144 experiment, where 46.6 GeV electron beams collided with relativistic laser pulses with intensities of $10^{18}$ W.cm$^{-2}$.

\subsection{Real electron bunch: longitudinal and transverse emittance}
\label{chap2secG}

In the previous section, ideal bunches with electrons at the same energy and same momentum have been considered in order to simply discuss the effect of summation of the radiation from each electron. However, in realistic bunches, electrons have slightly different energies and  momenta. More precisely, inside the bunch, electrons that are at the same location can have different energies and momenta. This implies, for example, that the bunch cannot be focused and compressed on an infinitely small point. This limitation is fundamental and inherent to the bunch, it does not depend on practical realization. The parameter which accounts for that is called the emittance and is related to the volume occupied by the electrons in the 6D phase space $(x, y, z, p_x, p_y, p_z)$ at a given time \cite{Humphries}. The 6D phase volume is constant in time if only smooth external forces are applied and if collisions are neglected (the phase volume conservation is a consequence of the collisionless Boltzmann equation). The emittance reflects the quality of the electron beam because it quantitatively indicates if electrons have the same coordinates, direction, and energy.

For a relativistic electron beam traveling in the $\vec{e}_z$ direction, an emittance is defined for each dimension: the $z$ one is called longitudinal and two others are transverse ($x$ and $y$). For a uniform distribution with sharp boundary, the \textit{normalized} emittance $\epsilon_{aN}$ is defined as the area occupied by electrons in the $(a, p_a/mc)$ space divided by $\pi$ $(a=x,y,z)$. But because realistic electron beams have diffuse boundaries, the normalized root-mean-square (rms) emittance is used and defined as
\begin{equation}
\epsilon_{aN}=\sqrt{\langle \Delta a^2\rangle\,\langle \Delta p_a^2\rangle -\langle \Delta a \Delta p_a\rangle^2}/mc,
\end{equation}
with $a=x,y,z$ and where $\Delta a=a-\langle a\rangle$, $\Delta p_a=p_a-\langle p_a\rangle$.
For transverse dimensions, it is convenient to use the \textit{unnormalized} emittance $\epsilon_a$, which is related to the area occupied by electrons in the $(a, a^{\prime})$ trace space $(a=x,y)$, where $a^{\prime}\simeq p_a/p_z$ is the transverse angle with respect to the propagation axis $z$,
\begin{equation}
\epsilon_{a}=\sqrt{\langle \Delta a^2\rangle\,\langle \Delta a^{\prime}{^2}\rangle -\langle \Delta a \Delta a^{\prime}\rangle^2}, \qquad \qquad a=x,y.
\end{equation}
It is generally expressed in $\pi.\text{mm}.\text{mrad}$. For cylindrically symmetric beams, the emittance $\epsilon_r$ (defined in the same way by putting $a=r$) can be used. The normalized emittance is related to the unnormalized one by $\epsilon_{N}=\gamma\beta\epsilon$ and has the advantage of being conserved during acceleration (in systems which preserve the emittance). In a focal plane, where there is no correlation between position and angle, the emittance is simply the product of the rms transverse size $\sigma_a$ by the rms angular dispersion $\sigma_{a^\prime}$:
\begin{equation}
\epsilon_a=\sigma_a \sigma_{a^\prime}, \qquad \qquad a=x,y,r.
\end{equation} 

For incoherent radiation of electrons oscillating in undulator or wiggler devices, the emittance and energy spread have the following effects. The electron beam is focused in the device, with a transverse size $\sigma$ and a divergence $\theta$ satisfying $\epsilon=\sigma\theta$. Because of the angular spread, the radiation angular distributions from each single electron are slightly shifted from one another, leading to a redshifted broadening of harmonic bandwidths [an electron with direction $\theta$ with respect to the propagation axis contributes on axis with higher wavelength $\lambda=\lambda_{\theta=0}(1+\gamma_z^2\theta^2)$, see Eq. (\ref{chap2eq3})]. The energy spread effect is straightforward: electrons with different energies radiate at slightly different wavelengths, leading to a broadening of harmonic bandwidths. These effects result in a modified bandwidth given by $(\Delta\lambda_n/\lambda_n)^2=(1/(nN))^2+(2\Delta\gamma/\gamma)^2+(\gamma_z^2{\epsilon}^2/{\sigma^2})^2$. Hence, the bandwidth of an harmonic at a given direction comes from three different effects: the finite number of periods, the energy spread, and the angular spread (which depends on the emittance).

The transverse emittance is essential for transport considerations and applications such as the free-electron laser. Concerning the FEL application, required conditions on the transverse emittance and the energy spread will be given in Sec. \ref{chap7}. A smaller transverse emittance permits one to transport or focus the electron beam on a smaller focal spot size.

\section{Electron Acceleration in Plasma}
\label{chap3}

The possibility to accelerate electrons in laser-produced plasmas was originally proposed by \textcite{PRL1979Tajima}. They suggested to use the intense electric field of a relativistic plasma wave, created in the wake of an intense laser pulse, to accelerate electrons to relativistic energies.  The main advantage of plasmas relies on their ability to sustain an accelerating gradient much larger (on the order of 100 GeV/m) than a conventional radio frequency accelerating module (on the order of 10 MeV/m). This means that electrons could be accelerated up to 1 GeV in millimeter- or centimeter-scale plasmas \cite{NatPhys2006Leemans, NatPhot2008Hafz, PRL2009Kneip,PRL2009Froula,PRL2010Clayton} while a few tens of meters would be necessary to reach the same energy in conventional accelerators. 

This acceleration method has experienced a remarkable development over the past decades, mainly thanks to the advent of high-intensity lasers and to a better understanding of the physical mechanisms driving the acceleration. Different plasma accelerator schemes have been developed over the years, \footnote{For an overview of the historical development of the field, see \textcite{PRL1979Tajima,Nature1984Joshi,PRL1992Kitagawa,PRL1993Clayton,Nature1994Everett,PRL1995Nakajima,PRL1995Amiranoff,Nature1995Modena,Science1996Umstadter,IEEE1996Esarey,PRL1997Wagner,PRL1997Moore,PRL1998Gordon,PRL1998Amiranoff,PRL1999Gahn,PRL2001Santala,PRL2002Leemans,Science2002Malka,PRL2005Mangles,RPP2003Pukhov,PPCF2004Bingham,Nature2007Patel,PoP2007Joshi,RMP2009Esarey,PoP2003Najmudin,PoP1997Ting,PRL1995Coverdale}.} leading to electron bunches with ever-increasing quality. The most efficient to date is the so-called bubble, blowout or cavitated wakefield regime \cite{APB2002Pukhov,PPCF2004Pukhov,PRL2006Lu,PoP2006Lu, PRSTAB2007Lu}. In that regime, depending on the chosen parameters, electron bunches can now be produced with tunable energy in the hundreds of MeV range \cite{Nature2006Faure}, low divergence (mrad), a relatively high charge ($\sim 100$ pC), and a bunch duration of less than 10 fs \cite{Nature2004Faure, Nature2004Mangles, Nature2004Geddes, PRL2004Tsung, PoP2006Tsung, PRL2006Mangles, PRL2007Thomas, PRL2007Glinec, PoP2008Davoine, PRL2006VanTilborg, NatPhys2011Lundh}. Because the x-ray sources which will be reviewed are based on laser-plasma accelerators, this section is dedicated to a short description of wakefield acceleration in the cavitated regime. In particular two important physical mechanisms are introduced: the ponderomotive force and the plasma wave. After a brief description of the characteristics of the acceleration mechanism, recent experimental progress is presented. We refer the interested reader to the recent article of \textcite{RMP2009Esarey} for a complete review of laser-plasma electron accelerators.

\subsection{Ponderomotive force and plasma waves}
\label{chap3secA}

The ponderomotive force is a force associated with the intensity gradients in the laser pulse, that pushes both electrons and ions out of the high-intensity regions. Ions, being much heavier than electrons, still remain for short interaction times whereas electrons are cast away. This leads, in an underdense plasma, to the formation of a relativistic plasma wave whose fields can accelerate electrons. Here a short description of the ponderomotive force and of the excitation of a plasma wave~\cite{Kruer} is given.

%The motion of an electron fluid element can be decomposed into two different components: the leading order motion corresponding to the fast oscillation of the electrons in the laser pulse and the second order motion corresponding to a slow drift of the electrons from the laser high-intensity regions to the laser low-intensity regions. A fully relativistic treatment gives the following decomposition:
Because of the mass of the plasma ions, they can be considered motionless for short interaction times. Considering a fluid description for the plasma electrons, the equation of motion for a fluid element submitted to the electromagnetic force reads~\footnote{Equation (\ref{chap3eq1}) assumes that $\vec{\nabla}\times(\hat{\vec{p}}-\vec{a})$ is initially zero, which is the case in practice because both $\hat{\vec{p}}$ and $\vec{a}$ are zero before the passage of the laser pulse in the plasma.}
\begin{align}
\label{chap3eq1}
\frac{\partial\hat{\vec{p}}}{\partial t} &= \frac{\partial\vec{a}}{\partial t} + c\vec{\nabla}(\phi-\gamma),
\end{align}
where $\hat{\vec{p}}=\vec{p}/mc$ is the normalized momentum of an electron fluid element, $\gamma$ is the relativistic factor of an electron fluid element, and $\phi=eV/mc^2$ and $\vec{a}=e\vec{A}/mc$ are, respectively, the normalized scalar potential and the normalized vector potential of the electromagnetic fields. $\vec{a}$ describes the high-frequency laser pulse, $c\vec{\nabla}\phi$ is the Coulomb force associated with the charge distribution, and $-c\vec{\nabla}\gamma$ is the relativistic ponderomotive force which expels electrons away from the laser pulse. In the absence of Coulomb and ponderomotive forces, the equation simplifies to $\hat{\vec{p}}=\vec{a}$, corresponding to the fast electron oscillation in the laser pulse. Depending on the amplitude of the laser pulse normalized vector potential $a_0$, plasma electrons oscillate with relativistic velocities $|\vec{v}|\simeq c$ ($a_0>1$) or with velocities much smaller than $c$ ($a_0\ll1$). An inhomogenous laser intensity distribution leads to an inhomogenous $\gamma$ distribution and to a ponderomotive force that pushes plasma electrons from the high $\gamma$ region (corresponding to high $a_0$) to the low $\gamma$ region (low $a_0$). This slow drift motion of the plasma electrons leads to a charge density distribution $\rho$ responsible for a Coulomb force $c\vec{\nabla}\phi$ (by virtue of the Poisson equation $\triangle\phi=\rho/\epsilon_0$).

On the other hand, a small charge density perturbation in a plasma oscillates at a characteristic frequency, the plasma frequency $\omega_p=\sqrt{n_ee^2/m\epsilon_0}$, where $n_e$ is the electron density of the plasma. For example, if we translate an electron slice (continuously, without crossing between different electrons) from its initial position, the slice will oscillate around its initial position at the plasma frequency because of the restoring Coulomb force from the plasma ions. These plasma oscillations, called plasma waves, are excited by the ponderomotive force of the laser pulse, which creates the initial charge density perturbation. The phase velocity of the plasma wave $v_\phi$, excited in the wake of the laser pulse, is approximately equal to the laser group velocity.

For a low-intensity laser pulse, with a normalized vector potential amplitude $a_0 \ll1$, the excited plasma wave is linear and has a sinusoidal shape at the plasma frequency, $\phi(\vec{r},t)=\phi_0(\vec{r})\sin(\omega_p t)$. For a higher intensity, $a_0>1$, the plasma wave becomes highly nonlinear and can involve transverse currents and a quasistatic magnetic field~\cite{PRL1996Gorbunov}.

Relativistic plasma waves can be produced in various regimes depending on the laser and plasma parameters. For appropriately chosen parameters, electrons can be trapped at a proper phase of the plasma wave and experience its field over a distance sufficiently long to be accelerated up to relativistic energies.

\bigskip
In practical units, the plasma wavelength $\lambda_p=2\pi c/\omega_p$ and the laser strength parameter $a_0$ are given by
\begin{align}
& \lambda_p[\mu\text{m}] = 3.34\times10^{10} / \sqrt{n_e[\text{cm}^{-3}]},\\
& a_0 =0.855 \sqrt{I [10^{18}\:\text{W/cm}^2] \lambda_L^{2} [\mu\text{m}]},
\end{align}
where $I$ is the laser intensity and $\lambda_L$ is the laser wavelength.

\subsection{The cavitated wakefield or bubble regime}
\label{chap3secB}

To date, the most efficient mechanism to accelerate electrons in a plasma wave is called the bubble, blowout or cavitated wakefield regime. In this regime the wake consists of an ion cavity having a spherical shape \cite{APB2002Pukhov,PRL2006Lu,PoP2006Lu,PRSTAB2007Lu}.

\begin{figure}
\includegraphics[width=8.5cm]{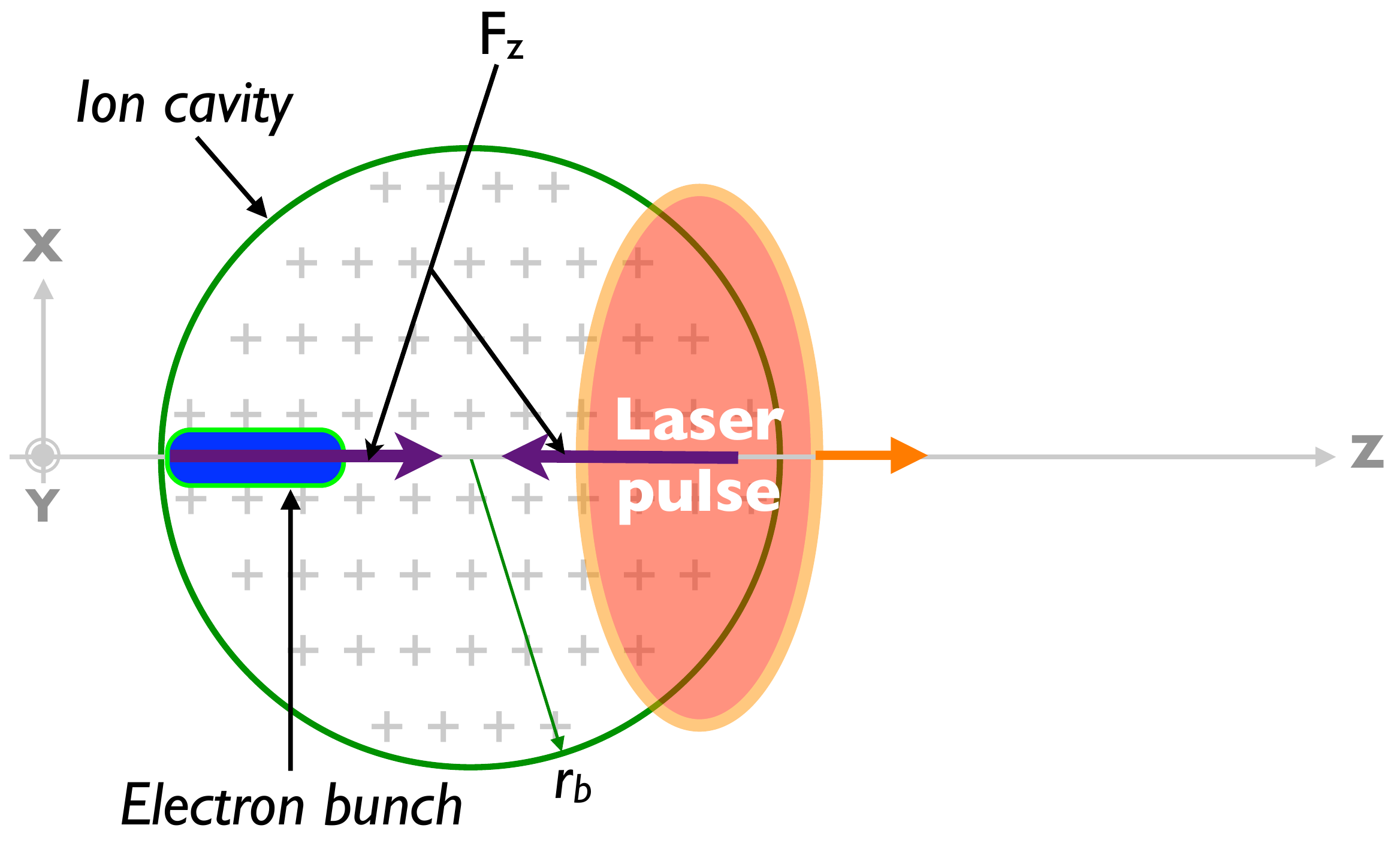}
\caption{Principle of the electron acceleration in the bubble regime. The laser pulse expels all plasma electrons out of the focal spot, leaving in its wake an ion cavity of radius $r_b$. The longitudinal force $F_z$ accelerates the self-injected electron bunch in the first half of the bubble, while the bunch decelerates in the second half of the bubble.}
\label{chap3fig1}
\end{figure}
This regime is reached when the waist $w_0$ of the focused laser pulse becomes matched to the plasma ($k_p w_0=2\sqrt{a_0},\:\text{with}\;k_p=\omega_p/c$) and if the pulse duration is of the order of half a plasma wavelength ($c\tau\sim\lambda_p/2$). In addition, the laser intensity must be sufficiently high ($a_0>2$) to expel most of the electrons out of the focal spot. If these conditions are met, an ion cavity is formed in the wake of the laser pulse, as represented in Fig. \ref{chap3fig1}. Electrons can be trapped at the back of the cavity and accelerated by the high electric field (the space-charge force) until they reach the middle of the cavity where they start to decelerate. The distance over which electrons must propagate before they reach that point is called the dephasing length $L_d$ and is much larger than the bubble radius $r_b$, as electrons travel almost at the same speed as the wave phase velocity. If the process is turned off at that time, by setting the acceleration length close to the dephasing length, electrons exit the plasma with the maximum energy gain. The maximum electric field and the radius of the cavity are, respectively \cite{PRSTAB2007Lu},
\begin{align}
E_m &= m\omega_pc\sqrt{a_0}/e,\\
r_b &= w_0=(2/k_p)\sqrt{a_0}.
\end{align}
For the typical experimental condition accessible with present lasers, $a_0 =4$ with a 30 fs laser, the maximum electric field is $E_m \sim 600$ GeV/m and the radius of the cavity is $r_b \sim 7\:\mu$m for an electron density of $n_e=1\times10^{19}\:\text{cm}^{-3}$. 

\subsection{Experimental production of relativistic electron bunches}
\label{chap3secC}

Several techniques have been explored to produce plasma waves appropriate to accelerate electrons. A typical experiment for laser-plasma acceleration consists in focusing a laser pulse into a gas jet, the interaction parameters being essentially the laser intensity, focal spot size and duration, the propagation length, and the plasma density. Depending on the choice of these parameters the features of the produced electron bunch can be very different.
Prior to 2004, experiments used relatively high  plasma densities ($>10^{19}$ cm$^3$) and the produced electron bunches were characterized by broadband spectra, extending up to about 100 MeV. These spectra were either nearly Maxwellian in the direct laser acceleration (DLA) regime \cite{PRL1999Gahn, PoP1999Pukhov, PoP2000Tsakiris, RPP2003Pukhov, PRL2005Mangles, PRL2008Kneip} or non-Maxwellian in the self-modulated laser wakefield accelerator (SM-LWFA) and forced laser wakefield (FLWF) regimes \cite{Nature1995Modena, Science1996Umstadter, PRL2001Santala, PRL1997Wagner, PoP2003Najmudin, PoP1997Ting, PRL1995Coverdale, Science2002Malka}.
In 2004, major advances were made by setting the laser pulse duration close to the plasma period, increasing the interaction length and matching the dephasing to the propagation length. Three groups reported simultaneously on the production of monoenergetic electrons in the 100 MeV range \cite{Nature2004Mangles,Nature2004Geddes,Nature2004Faure}, collimated within a few milliradians and with charge on the order of 100 pC. However, despite these remarkable progresses, the high nonlinearity of the mechanism resulted in electron bunches which were neither stable nor tunable in energy; in addition electron energies remained below the GeV level.

To overcome these limitations, improved schemes have recently been developed. An external and controlled electron injection into the wakefield was proposed \cite{PRL1996Umstadter, PRL1997Esarey, PRE2004Fubiani} and recently demonstrated \cite{Nature2006Faure, PPCF2007Faure,PoP2009Malka}. It consists of colliding the main laser pulse generating the plasma wave with a second laser pulse, creating a beat wave whose ponderomotive force can preaccelerate and locally inject background electrons in the wakefield \cite{PoP2008Davoine}. The energy of the electron bunch can be tuned by varying the collision position and therefore the acceleration length. Experiments demonstrated that stable electron bunches can be produced with energy continuously tunable from a few tens of MeV to above 200 MeV \cite{Nature2006Faure, PPCF2007Faure,PoP2009Malka} and an energy spread on the order of 1\%  \cite{PRL2009Rechatin1}. The few femtosecond duration and the few kA current of these electron bunches were experimentally demonstrated by~\textcite{NatPhys2011Lundh}. A cold optical injection providing narrow energy spread for GeV electrons has been proposed \cite{PRL2009Davoine}.\\
Several other possibilities to control electron injection have been investigated: the use of a plasma density downramp \cite{PRE1998Bulanov, PRL2001Suk, PRSTAB2002Hemker, PoP2008Brantov, PRL2008Geddes, PoP2010Faure}, ionization-induced injection \cite{PRL2010McGuffey, PRL2010Pak, PRL2010Clayton, PRL2011Pollock} or magnetically controlled injection \cite{PRL2011Viera}.\\
Laser-plasma accelerators have also recently reached the GeV level, using either external laser guiding \cite{NatPhys2006Leemans, PoP2007Nakamura} or higher laser power \cite{NatPhot2008Hafz, PRL2009Kneip,PRL2009Froula,PRL2010Clayton}. In the first case, the experiment relied on channeling a few tens of TW-class laser pulses in a gas-filled capillary discharge waveguide in order to increase the propagation distance, and so the accelerator length, to the centimeter scale \cite{NatPhys2006Leemans,PoP2007Nakamura,PRL2008Rowlands}. In the experiment of \textcite{NatPhys2006Leemans}, electrons have been accelerated up to a GeV.

In order to further improve the quality of electrons from laser-plasma accelerators, several routes are proposed from the use of a PW-class laser to multistaged acceleration schemes \cite{PoP2005Gordienko, PoP2005Lifschitz, PRSTAB2006Malka, PRSTAB2007Lu, NatPhys2010Martins}. These foreseen developments are of major importance for the production of a free-electron laser based on a laser-plasma accelerator.

\section{Plasma Accelerator and Plasma Undulator: Betatron Radiation}
\label{chap4}

\begin{figure}
\includegraphics[width=8.5cm]{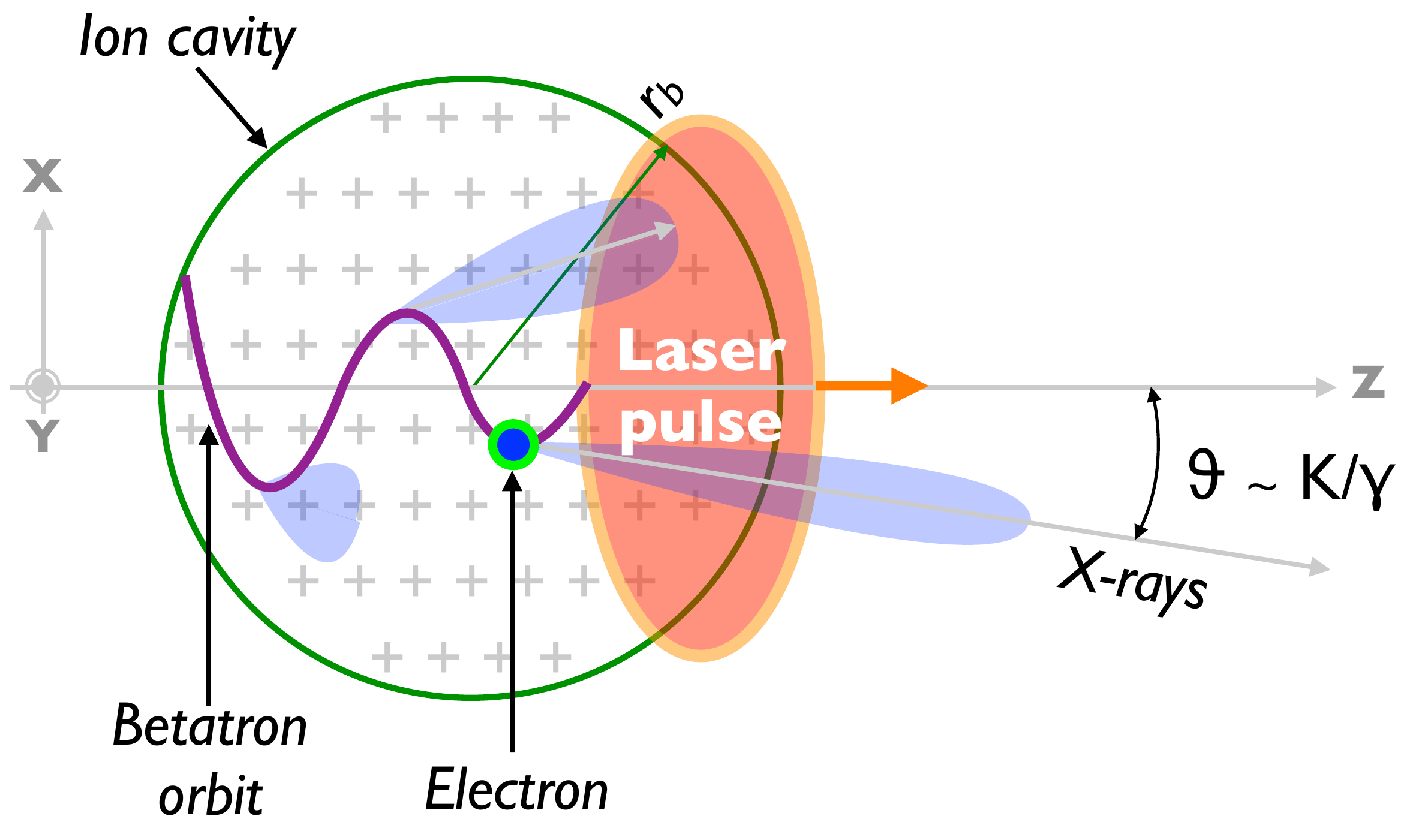}
\caption{Schematic of the betatron mechanism. When an electron is injected in the ion cavity, it is submitted not only to the accelerating force, but also to a restoring transverse force, resulting in its wiggling around the propagation axis. Because of this motion, the electron radiates x rays whose typical divergence is $\theta=K/\gamma$.}
\label{chap4fig1}
\end{figure}

In the bubble acceleration regime, the plasma cavity can act as a wiggler in addition to being an accelerator, reproducing on a millimeter scale the principle of a synchrotron to produce x rays \cite{PRL2004Kiselev, PRL2004Rousse}. In this section, we will show a laser-produced ion cavity drives the electron orbits in such a way that a short pulse of collimated x-ray radiation is emitted.
Figure \ref{chap4fig1} represents the principle of the mechanism. As discussed, the bubble regime is reached when an intense femtosecond laser pulse, propagating in an underdense plasma, evacuates plasma electrons from the high-intensity regions and leaves an ion cavity in its wake. In addition to the longitudinal force, responsible for the acceleration discussed above, the spherical shape of the ion cavity results in a transverse electric field producing a restoring force directed toward the laser pulse propagation axis. Therefore, electrons trapped and accelerated in the cavity are also transversally wiggled. The conditions for an efficient production of accelerated charged particle radiation, discussed in Sec. \ref{chap2}, are therefore met. A collimated beam of x-ray radiation is emitted by the electron bunch. This radiation, which can be directly compared to a synchrotron emission in the wiggler regime, is called betatron radiation.

The betatron radiation from laser-produced plasmas was simultaneously proposed and demonstrated in 2004 \cite{PRL2004Kiselev, PRL2004Rousse}. This represented a major step forward in the field of plasma x-ray sources since it was the first method allowing one to produce bright collimated x-ray (keV) beams from laser-plasma interactions. Since then, this radiation has been measured and widely characterized in interaction regimes from multiterawatt \cite{PoP2005TaPhuoc, PRE2006Shah, PRL2006TaPhuoc, PoP2007TaPhuoc, PoP2008TaPhuoc1, PRE2008Albert} to petawatt lasers \cite{PRL2008Kneip}. According to the theory, femtosecond pulses of x rays up to a few tens of keV could be produced within tens of milliradian beams.
In the following sections, the properties of the betatron mechanism are reviewed. Following the approach of Sec. \ref{chap2}, the electron orbit is first calculated using an ideal model and then the features of the emitted radiation are derived. Simulations based on this model and a particle in cell (PIC) simulation are presented. Finally, after a summary of the experimental results, we conclude with the short term developments foreseen.

\subsection{Electron orbit in an ion cavity}
\label{chap4secA}

An idealized model of a wakefield in the bubble regime based on the phenomenological description developed by \textcite{PRL2006Lu, PoP2006Lu, PRSTAB2007Lu} is assumed. The phase velocity of the wake is expressed as $v_\phi=v_g-v_{\text{etch}}\simeq c(1-3\omega_p^2/2\omega_L^2)$, where $v_g =c\sqrt{1-\omega_p^2/\omega_L^2}$ is the laser group velocity in the plasma, $v_{\text{etch}}\simeq c\omega_p^2/\omega_L^2$ is the etching velocity due to local pump depletion \cite{PRSTAB2007Lu, PRL1994Decker, PRE1995Decker, PoP1996Decker}, $\omega_p=\sqrt{n_e e^2/m\epsilon_0}$ is the plasma frequency, $n_e$ the electron density of the plasma, and $\omega_L$ the central frequency of the laser field. The ion cavity is assumed to be a sphere of radius $r_b = (2/k_p)\sqrt{a_0}$ ($k_p=\omega_p/c$) and a cylindrical coordinate system $(r,\theta,z)$, where $r$ is the distance from the laser pulse propagation axis is used. The comoving variable is defined as $\zeta=z-v_\phi t$ so that ($\zeta=0,r=0)$, $(\zeta= r_b,r=0)$, and  $(\zeta=-r_b,r=0)$, respectively, correspond to the center, the front, and the back of the cavity. The electromagnetic fields of the wake are an axial electric field $E_z$, a radial electric field $E_r$, and an azimuthal magnetic field $B_\theta$. They are given by $E_z/E_0=k_p\zeta/2$, $E_r/E_0=k_pr/4$, and $B_\theta / E_0=-k_pr/(4c)$, where $E_0=m\omega_p c/e$ is the cold nonrelativistic wavebreaking field \cite{PPCF2004Pukhov, PRL2006Lu, PoP2006Lu, PoP2004Kostyukov, PoP2007Xie}. The equation of motion of a test electron in the cavity is then:
\begin{equation}
\label{chap4eq1}
\frac{d\vec{p}}{dt}=-e(\vec{E}+\vec{v}\times\vec{B})=\vec{F}_\parallel+\vec{F}_\perp
\simeq-\frac{m \omega_p^2}{2}
(\zeta\vec{e}_z+r\vec{e}_r),
\end{equation}
where $\vec{p}$ is the momentum of the electron. The last expression assumes $p_{\bot}\ll p_z$. The electron is initially injected at the back of the cavity with space-time coordinates $(t_i=0, x_i, y_i,z_i)$ such that $x_i^2+y_i^2+z_i^2=r_b^2$ and $\zeta_i<0$, and with a energy-momentum quadrivector ($\gamma_imc^2,\vec{p}_{i}$) such that $v_z>v_\phi$.

The term $\vec{F}_\parallel$ is responsible for the electron acceleration in the longitudinal direction $\vec{e}_z$. As the electron becomes relativistic, its velocity becomes greater than $v_\phi$; the term $-\zeta=v_\phi t - z$ decreases and the accelerating force is reduced. The length for the electron to reach the middle ($\zeta=0$) of the cavity and to become decelerated ($\zeta>0$) corresponds to the dephasing length $L_d$.
The term $\vec{F}_\perp$ is a linear restoring force that drives the transverse oscillations of the electron across the cavity axis at the betatron frequency $ \omega_{\beta} \simeq \omega_p /\sqrt{2\gamma}$.

To derive the analytical expressions of the electron orbit, we start the integration of the equation of motion from an initial state where $\gamma_{zi}=1/\sqrt{1-(v_{zi}/c)^2}\gg\gamma_\phi$ in order to ensure that the slippage between the test electron and the wakefield is constant in time. This approximation is realistic because an electron quickly attains $\gamma_z\gg\gamma_\phi$ and the transitory period in which the approximation fails is small compared to the dephasing time $L_d/c$. Hence $d\zeta/dt=v_z-v_\phi\simeq c-v_\phi \simeq c/(2\gamma_\phi^2)$ if $\gamma_\phi\gg1$. In agreement with the assumption used in the last expression of Eq. (\ref{chap4eq1}), we perform a perturbative treatment in the variable $p_{\bot}/p_z\ll1$. The variables with a hat are normalized by the choice $m=c=e=\omega_p=1$ and $\vec{\beta}=\vec{v}/c$ is the velocity normalized to the speed of light $c$. Equation (\ref{chap4eq1}) projected on each axis reads
\begin{align}
\label{chap4eq1a}
\frac{d(\gamma\beta_x)}{d\hat{t}}&=-\frac{\hat{x}}{2},\\
\label{chap4eq1b}
\frac{d(\gamma\beta_y)}{d\hat{t}}&=-\frac{\hat{y}}{2},\\
\label{chap4eq1c}
\frac{d(\gamma\beta_z)}{d\hat{t}}&=-\frac{\hat{\zeta}_i}{2}-\frac{\hat{t}}{4\gamma_\phi^2}.
\end{align}
At zero order ($p_{\bot}=0$), Eq. (\ref{chap4eq1c}) can be directly integrated to $\gamma(\hat{t})\simeq \hat{p}_z(\hat{t})=\gamma_{d}(1-\tau^2)$ with $\gamma_{d}=\gamma_{i}+\gamma_\phi^2\hat{\zeta}_i^2/2$, $\tau=(\hat{t}-\hat{t}_d)/\sqrt{\hat{t}_d^2+8\gamma_\phi^2\gamma_{i}}$ and $\hat{t}_d=-2\gamma_\phi^2\hat{\zeta}_i$ the dephasing time ($\hat{t}_d=k_pL_d$). This parabolic profile of $\gamma(\tau)$ can be inserted into Eqs. (\ref{chap4eq1a})-(\ref{chap4eq1b}) to obtain the first-order solution for $\hat{x}(\tau)$ and $\hat{y}(\tau)$. Note that motion in the $\vec{e}_x$ and $\vec{e}_y$ directions are decoupled at this order of calculation. Each equation takes the form of a Legendre differential equation,
\begin{align}
\label{chap4eq1d}
\frac{d}{d\tau}\Big[(1-\tau^2)\frac{d\hat{x}}{d\tau}\Big]+\nu(\nu+1)\hat{x}&=0,\\
\label{chap4eq1e}
\frac{d}{d\tau}\Big[(1-\tau^2)\frac{d\hat{y}}{d\tau}\Big]+\nu(\nu+1)\hat{y}&=0,
\end{align}
where $\nu(\nu+1)=4\gamma_\phi^2$. The solution space is generated by the Legendre functions of the first and second kinds $P_\nu(\tau)$ and $Q_\nu(\tau)$, which have the following asymptotic expressions for $\nu\gg1$ \cite{Math}:
\begin{align}
P_\nu(\tau)&=\sqrt{\frac{2}{\pi\nu(1-\tau^2)^{1/4}}}\cos[(\nu+1/2)\cos^{-1}\tau-\pi/4],\\
Q_\nu(\tau)&=\sqrt{\frac{\pi}{2\nu(1-\tau^2)^{1/4}}}\cos[(\nu+1/2)\cos^{-1}\tau+\pi/4].
\end{align}
Hence, the solutions for $\hat{x}(\tau)$ and $\hat{y}(\tau)$ can be written as
\begin{align}
\label{chap4eq1f}
\hat{x}(\tau)&=\frac{A_x}{(1-\tau^2)^{1/4}}\cos(2\gamma_\phi\cos^{-1}\tau+\varphi_x),\\
\label{chap4eq1g}
\hat{y}(\tau)&=\frac{A_y}{(1-\tau^2)^{1/4}}\cos(2\gamma_\phi\cos^{-1}\tau+\varphi_y),
\end{align}
where $(A_x,\varphi_x)$ and $(A_y,\varphi_y)$ are constants which have to be determined by initial conditions $(x_i,p_{xi})$ and $(y_i,p_{yi})$. They can be put into the simpler form
\begin{align}
\label{chap4eq1h}
\hat{x}(\hat{t})&=\hat{x}_\beta(\hat{t})\cos \left (\int_0^{\hat{t}}\hat{\omega}_\beta(\hat{t}^{\prime}) d\hat{t}^{\prime}+\varphi_x^{\prime} \right ),\\
\label{chap4eq1i}
\hat{y}(\hat{t})&=\hat{y}_\beta(\hat{t})\cos \left (\int_0^{\hat{t}}\hat{\omega}_\beta(\hat{t}^{\prime}) d\hat{t}^{\prime}+\varphi_y^{\prime} \right ),\\
\label{chap4eq1j}
\hat{\omega}_\beta(\hat{t})&=1/\sqrt{2\gamma(\hat{t})},\\
\label{chap4eq1k}
\hat{x}_\beta(\hat{t})&=A_{x}^{\prime}/\gamma(\hat{t})^{1/4},\\
\label{chap4eq1l}
\hat{y}_\beta(\hat{t})&=A_{y}^{\prime}/\gamma(\hat{t})^{1/4}.
\end{align}
The transverse motion consists of sinusoidal oscillations in each direction with time-dependent amplitude $\hat{x}_\beta(\hat{t}), \hat{y}_\beta(\hat{t})$, and frequency $\hat{\omega}_\beta(\hat{t})$. They depend on $\hat{t}$ only through $\gamma$. For an arbitrary $\gamma(\hat{t})$ profile, Eqs. (\ref{chap4eq1h})-(\ref{chap4eq1l}) can be derived from the adiabatic approximation and the WKB method [valid if $(1/\omega_\beta^2)d\omega_\beta/dt\ll 1$], using the conservation of the action variables \cite{PoP2004Kostyukov, PoP2010Thomas}. Thus,  Eqs. (\ref{chap4eq1j})-(\ref{chap4eq1l}) in terms of $\gamma$ are more general than the frame of the above calculation [\textit{i.e.}, the case of a parabolic $\gamma(\hat{t})$ profile]. The betatron amplitudes $x_\beta,y_\beta$ decrease as $\gamma^{-1/4}$ during acceleration, while the betatron frequency $\omega_\beta$ decreases as $\gamma^{-1/2}$. According to Eqs. (\ref{chap4eq1f})-(\ref{chap4eq1g}), the number of betatron oscillations between $\hat{t}=0$ and the dephasing time $\hat{t}=\hat{t}_d$ is approximately $\gamma_\phi/2$. Depending on the values of $\varphi_x^{\prime}$ and $\varphi_y^{\prime}$, the motion in each direction $\vec{e}_x$ and $\vec{e}_y$ can be in phase (motion confined in a plane) or not (helical motion). The latter case corresponds to an initial state with nonzero angular momentum $L_z=xp_y-yp_x$. Finally, the longitudinal motion $\hat{z}(\hat{t})$ can be obtained by integrating $\beta_z$ which is deduced from $\gamma^2=1/(1-\beta_x^2-\beta_y^2-\beta_z^2)$.

This trajectory is noticeably different from the one studied in Sec. \ref{chap2}. First, the betatron amplitude and frequency are time dependent. Nevertheless, in the wiggler regime, the resulting radiated energy $dI/d\omega$ can be seen as an integral over time of the radiated power $dP/d\omega(\omega_\beta, x_\beta, y_\beta)$ for each instantaneous amplitude and frequency if the characteristic time scale of acceleration is much longer than the betatron period. Second, $x(t)$ and $y(t)$ are sinusoidal functions of $t$ and not $z$. Even if the detail of the electron dynamics is slightly different from the case of Sec. \ref{chap2}, the betatron trajectory has the same properties. As in Sec. \ref{chap2}, the trajectory is a figure-eight motion in the electron local average rest frame. The radiation features derived in Sec. \ref{chap2} from the parameters $K$, $\lambda_u$, and $\gamma$ remain therefore valid. Assuming $dt\simeq dz/c$ in Eqs. (\ref{chap4eq1h})-(\ref{chap4eq1i}) and identifying the terms with the ones of Eq. (\ref{chap2eq2}), the local electron period $\lambda_u(t)$ and the local strength parameter $K(t)$ read in physical units
\begin{align}
\label{chap4eq2a}
\lambda_u(t)&=\sqrt{2\gamma(t)}\lambda_p,\\
\label{chap4eq2b}
K(t)&=r_\beta(t) k_p\sqrt{\gamma(t)/2},
\end{align}
in which we assume a motion confined in a plane with a betatron amplitude $r_\beta(t)=\sqrt{x_\beta(t)^2+y_\beta(t)^2}$. We recover the results obtained by \textcite{PRE2002Esarey,PoP2003Kostyukov} in the case of an ion channel without longitudinal acceleration, but in which the parameters  are time dependent through the variation of $\gamma$.

In the articles of \textcite{PRE2002Esarey, PoP2003Kostyukov}, the constant of motion $\gamma_z=\sqrt{1+\hat{p}_z^2}$ appears in Eqs. (\ref{chap4eq2a}) and (\ref{chap4eq2b}) instead of $\gamma$. The relative difference between $\gamma$ and $\gamma_z$ is of the order of $p_{\bot}^2/p_z^2$ and can be neglected at the order where the trajectory has been calculated. In this section, the definition $\gamma_z=\sqrt{1+\hat{p}_z^2}$ has not been used because it is different from the usual one $\gamma_z=1/\sqrt{1-\beta_z^2}$ (which is small compared to $\gamma$ for wigglers, as discussed in Sec. \ref{chap2secC}), commonly used in the synchrotron and FEL communities.\\
In practical units, Eqs. (\ref{chap4eq2a}) and (\ref{chap4eq2b}) read
\begin{align}
\label{chap4eq3b}
& \lambda_u[\mu \text{m}] = 4.72 \times 10^{10} \sqrt{ \gamma / n_e [\text{cm}^{-3}]},\\
\label{chap4eq3}
& K=1.33 \times 10^{-10} \sqrt{ \gamma n_e [\text{cm}^{-3}] }r_\beta[\mu\text{m}].
\end{align}

\bigskip
In the equation of motion (\ref{chap4eq1}), the electromagnetic fields correspond to an idealized cavitated regime. It assumes three conditions: (\textit{i}) the matching conditions in terms of focal spot size, laser pulse duration, and density are perfectly met; (\textit{ii}) the cavity is free of electrons; and (\textit{iii}) the trapped electrons are only submitted to the wakefields (there is no interaction with the electromagnetic fields of the laser pulse). In addition, in this approach, initial conditions have to be arbitrarily chosen since the injection mechanism is not taken into account. If the three above conditions are not met, the following effects occur.

First, the interaction regime is slightly different when the plasma wavelength is smaller than the focal spot size and the pulse duration (slightly off the matching conditions). In this regime, called the forced laser wakefield regime \cite{Science2002Malka}, the shape of the wake is not spherical anymore, the laser pulse does not necessarily expel all electrons from the focal spot. The wakefields are therefore different from the case of a cavity free of electrons. Nevertheless, the bubble model remains a good approximation to roughly determine the electron trajectories and the betatron radiation features.

Second, the cavity cannot be considered free of electrons when injection occurs. Trapped electrons can modify the fields of the cavity by the beam loading mechanism \cite{PRL2008Tzoufras, PoP2009Tzoufras, PRL2009Rechatin2}: they modify the motion of background electrons and hence the resulting wakefields. As a result the acceleration is reduced and, for massive electron injection, the term $\vec{F}_\parallel$ in (\ref{chap4eq1}) can be largely overestimated.

Third, electrons that have been accelerated can catch up with the back of the laser pulse. The interaction of an electron with the laser pulse can increase its oscillation amplitude $r_{\beta}$ in the direction of the laser polarization, enhancing the betatron radiation. The interaction is complex and strongly dependent on the phase velocity of the laser pulse. It corresponds to the mechanism of direct laser acceleration (DLA), in which some electrons can be in betatron resonance, if the induced transverse motion of the laser pulse fields is in the same direction as the betatron motion at all times \cite{RPP2003Pukhov, PRL2008Kneip, PoP1999Pukhov, PRL2005Mangles, PoP2000Tsakiris, PRL1999Gahn}. In the forced laser wakefield regime, electrons can catch up with the laser pulse more easily, resulting in a brighter betatron radiation.

Three-dimensional PIC simulations can be used to accurately describe al these effects, and to compute the electron orbits and the emitted radiation. Simple numerical simulation, integrating the equation of motion, can be used as well to provide the additional effects by modeling the laser pulse and by extracting effective acceleration and transverse fields from a PIC simulation. \textcite{PRL2008Nemeth} followed this approach and showed that the betatron motion can be driven by the laser pulse in the polarization direction if electrons interact with it. Such a model, without the laser pulse fields, has also been considered by \textcite{PoP2009Wu} who, for a given configuration with ultrahigh intensity ($a_0=20$), reduced the acceleration field by a factor of 0.35 and the transverse fields by a factor of 0.9 in order to fit the electron trajectories with the ones extracted directly from the PIC simulation. This highlights how the wakefields are overestimated when Eq. (\ref{chap4eq1}) is considered without any correction.

\subsection{Radiation properties}
\label{chap4secB}

The betatron motion derived in the above simple model leads to the emission of synchrotron radiation referred to as betatron radiation. It can be calculated within the general formalism of the radiation from a moving charge. The ion cavity acts as an undulator or a wiggler with a period $\lambda_u(t)$ and strength parameter $K(t)$ which depends on the electron initial conditions upon injection into the cavity. We first consider the radiation properties for constant $\lambda_u$, $K$, and $\gamma$ and then consider the radiation produced by the electron with acceleration.

\subsubsection{Without acceleration}

The spectrum of the emitted radiation depends on the amplitude of $K$. For a small amplitude of the betatron oscillation $K\ll1$, the radiation is emitted at the fundamental photon energy $\hbar\omega$ with a narrow bandwidth in the forward direction ($\theta=0$). As $K\rightarrow1$ harmonics of the fundamental start to appear in the spectrum, and for $K\gg1$ the spectrum contains many closely spaced harmonics and extends up to a critical energy $\hbar\omega_c$. These quantities are given by
\begin{align}
\nonumber
\hbar\omega &= (2 \gamma^2 hc/\lambda_u) / (1+K^2/2) \;\;\;\; \text{for} \: K <1,\\
\hbar\omega_c &=  \frac{3}{2} K \gamma^2 hc/\lambda_u \;\;\;\; \text{for} \: K\gg1,
\label{chap4eq4a}
\end{align}
which gives in practical units
\begin{align}
\nonumber
\hbar\omega[\text{eV}] &= 5.25 \times 10^{-11}\gamma^{1.5}\sqrt{n_e[\text{cm}^{-3}]} \;\;\;\; \text{for} \: K \ll1,\\
\label{chap4eq4b}
\hbar\omega_c[\text{eV}] &= 5.24 \times 10^{-21}\gamma^{2} n_e[\text{cm}^{-3}] r_\beta[\mu\text{m}] \;\;\;\; \text{for} \: K\gg1.
\end{align}

According to Sec. \ref{chap2}, the radiation is collimated within a cone of typical opening angle $\theta_{r}=1/\gamma$ in the undulator case. For a wiggler, the radiation is collimated within a typical opening angle $K/\gamma$ in the electron motion plane $(\vec{e}_x,\vec{e}_z)$ and $1/\gamma$ in the orthogonal plane $(\vec{e}_y,\vec{e}_z)$.

The x-ray pulse duration equals the electron bunch duration. Depending on the parameters and the electron injection process, it can be extremely short, a few fs \cite{NatPhys2011Lundh}, corresponding to a bunch length equal to a very small fraction of the bubble radius, or using massive self-injection at high density, the bunch length can be on the order of the bubble radius and $\tau_r \sim r_b/c$. 

The number of emitted photons can be estimated using the expressions of Sec. \ref{chap2}. In practical units, the number of photons emitted per period and per electron (at the mean photon energy $\langle\hbar\omega\rangle=0.3 \hbar \omega_c$ for the wiggler case) is
\begin{align}
\nonumber
N_\gamma  &= 1.53 \times 10^{-2} K^2 \;\;\;\; \text{for} \: K<1,\\
\label{chap4eq6}
N_\gamma  &= 3.31 \times 10^{-2}K \;\;\;\; \text{for} \: K\gg1.
\end{align}

From the above expressions, an estimation of the radiation properties for constant $\lambda_u$, $K$, and $\gamma$ can be obtained for a typical parameter regime. A 100 MeV electron ($\gamma\simeq 200$) undergoing betatron oscillations in a plasma of density $n_e=2\times 10^{19}$ cm$^{-3}$ is considered. The spatial period of the motion is $\lambda_u \simeq 150$ $\mu$m. For $K\ll1$, this electron will emit betatron radiation at a fundamental energy $\hbar\omega \simeq 650$ eV. In the same parameter regime but with $K \sim 10$, typical of our experimental conditions, the critical energy of the radiation is $\hbar\omega_c \sim 5$ keV. Along the interaction length $L_{\text{acc}}$, the electron accelerates and radiates but its main contribution in terms of energy to the betatron radiation comes from the part of the trajectory where its energy is maximal. If the electron is around its maximal energy during $\sim$ 3 betatron periods, the total number of photons emitted per electron at the mean energy $\hbar\omega \sim 1.5$ keV is $N_\gamma  \sim 1$. Considering that the number of electrons trapped into the ion cavity is on the order of $10^{8-9}$, the number of x-ray photons expected is in the range $10^{8-9}$ as well.
Finally, in a typical parameter regime where $K\sim 10$ and $\gamma \sim 200$, the betatron emission is collimated within a cone of typical solid angle of 50 mrad $\times$ 5 mrad for a single electron.

\subsubsection{With acceleration}

For an electron accelerating, the situation is more complex. In what follows, we consider the wiggler limit and we derive an approximation for the radiation spectrum taking into account the acceleration. This will be useful for discussing the scalings of betatron properties which will be presented in Sec. \ref{chap4secE}. The acceleration has a characteristic parabolic profile $\gamma(\tau)=\gamma_{d}(1-\tau^2)$, with $\tau=(\hat{t}-\hat{t}_d)/\sqrt{\hat{t}_d^2+8\gamma_\phi^2\gamma_{i}}$ and $\hat{t}_d=-2\gamma_\phi^2\hat{\zeta}_i$ the dephasing time ($\hat{t}_d=k_pL_d$), as shown in Sec. \ref{chap4secA}. Thus, from Eqs. (\ref{eq_chap2_power}), (\ref{chap4eq1k})-(\ref{chap4eq2b}), and (\ref{chap4eq4a}), we deduce
\begin{align}
\omega_\beta(\tau)&=\omega_{\beta,d}(1-\tau^2)^{-1/2},\\
K(\tau)&=K_{d}(1-\tau^2)^{1/4},\\
\hbar\omega_c(\tau)&=\hbar\omega_{c,d}(1-\tau^2)^{7/4},\\
\overline{P}_\gamma(\tau)&=\overline{P}_{\gamma,d}(1-\tau^2)^{3/2},
\end{align}
where $\omega_{\beta,d}$, $K_{d}$, $\hbar\omega_{c,d}$, and $\overline{P}_{\gamma,d}$ are the values of the parameters at the dephasing time $\tau=0$, and $\overline{P}_\gamma(\tau)$ is the radiated power averaged over one oscillation period [whose expression is given in Eq. (\ref{eq_chap2_power}) for a planar trajectory, and is increased by a factor of 2 for a helical trajectory]. We consider that the spectrum (integrated over angles) radiated per unit time is given by $dP/d\omega=(\overline{P}_\gamma/\omega_c) S(\omega/\omega_c)$, \textit{i.e.}, a synchrotron spectrum [see Eq. (\ref{chap2eq4})] in which $P_\gamma$ is replaced by $\overline{P}_\gamma$, which is exact for a helical trajectory (for which the radius of curvature $\rho$ does not depend on the phase of the oscillation) but only approximate for a sinusoidal trajectory (because $\rho$ varies during an oscillation and $\omega_c$ corresponds to the minimal value of $\rho$ in the oscillation). The radiation spectrum then reads
\begin{align}
\nonumber
\frac{dI}{d\omega}&\approx \int_{0}^{t_d}\frac{dP}{d\omega}(t)dt\\
\nonumber
&\approx t_d\int_{-1}^0d\tau\Bigg[\frac{\overline{P}_\gamma(\tau)}{\omega_c(\tau)}S(\omega/\omega_c(\tau))\Bigg]\\
\label{chap4eq6a}
&= \frac{\overline{P}_{\gamma,d}t_d}{\omega_{c,d}}S^{\prime}(\omega/\omega_{c,d}),
\end{align}
where the function $S^{\prime}$ is defined as
\begin{equation}
S^{\prime}(x)=\int_{-1}^0\frac{d\tau}{(1-\tau^2)^{1/4}}S[(1-\tau^2)^{-7/4}x].
\end{equation}
Equation (\ref{chap4eq6a}) highlights the fact that the radiation properties are encoded into the values of the parameters $\omega_{\beta,d}$, $K_{d}$, $\gamma_{d}$, $\hbar\omega_{c,d}$ and $t_d$. The shape of the spectrum is not described by the usual universal function $S$ of the synchrotron spectrum anymore, but by the function $S^{\prime}$ defined above. This latter function takes into account the photons emitted at low energies during the acceleration. Indeed, Eq. (\ref{chap4eq6}) shows that the number of emitted photons in one period has a weak dependence on $\gamma$ (only through $K\propto\gamma^{1/4}$), whereas the radiated energy is strongly dependent on $\gamma$ because photons are emitted at much higher energies for higher $\gamma$. As a result, the angular distribution of the radiated energy is dominated by the electron oscillations around maximal values of the parameters, \textit{i.e.}, around the dephasing time. Hence, to obtain rough estimations of the radiation properties, the formulas given above for constant parameters can be used, in which the values at the dephasing time are inserted. The information on the spectrum will be valid at high photon energies, but will underestimate the number of low-energy photons.

Finally, it is interesting to compare the scalings of $dI/d\omega$ and $N_\gamma $ for both cases of constant parameters and acceleration,
\begin{align}
\frac{dI}{d\omega}_{\big|\gamma=\text{constant}}&\propto N_\beta KS(\omega/\omega_c),\\
\frac{dI}{d\omega}_{\big|\text{acceleration}}&\propto \omega_{\beta,d}t_d K_dS^{\prime}(\omega/\omega_{c,d}),
\end{align}
which leads to, respectively,
\begin{align}
N_\gamma {_{\big|\gamma=\text{constant}}} & \propto N_\beta K,\\
N_\gamma {_{\big|\text{acceleration}}}&\propto \omega_{\beta,d}t_d K_d,
\end{align}
for the dependence of the total number of emitted photons, where $N_\beta$ is the number of betatron oscillations in the $\gamma=\text{const}$ case. The value $\omega_{\beta,d}t_d$ is also characteristic of the number of betatron oscillations for the acceleration case. Thus, we see that the scalings are identical for both situations (parametrized either by the constant parameters or by the parameters at the dephasing time). The only difference resides in the exact shape of the spectrum, given by the function $S^{\prime}$ instead of $S$.

\subsection{Numerical results}
\label{chap4secC}

In this section, the case of an electron experiencing the longitudinal acceleration force is treated numerically. The complete equation of the electron motion (\ref{chap4eq1}) is first integrated and then the features of the emitted radiation are calculated using the general formula of moving charge radiation (\ref{chap2eq1}). Finally, a particle in cell simulation of the betatron radiation taking into account all possible effects, including the injection process, is presented. This latter simulation is based on the extraction of the trapped electron trajectories from the PIC code and the calculation of the corresponding radiation via Eq. (\ref{chap2eq1}) in postprocessing.

\subsubsection{Test-particle simulation}
\label{chap4secCsec1}

\begin{figure}
\includegraphics[width=8.5cm]{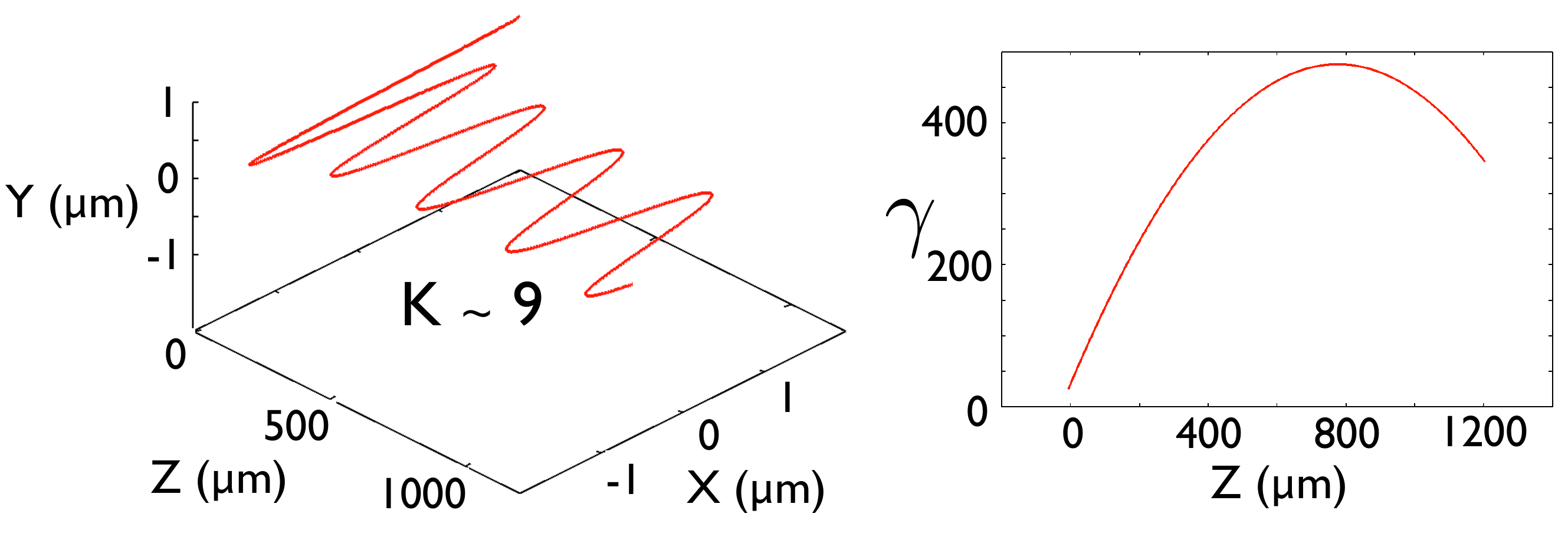}
\caption{Electron trajectory on the left and its gamma factor as a function of the longitudinal coordinate $z$ on the right. The set of initial conditions is  $x_i = 2$ $\mu$m, $y_i=0$, $z_i=\zeta_i=-r_b+0.3\:\mu\text{m}=-6.42\:\mu\text{m}$, $p_{xi}=0$, $p_{yi}=0$, $p_{zi}=25\:mc$ and the laser-plasma parameters are $n_e = 1 \times 10^{19}$ cm$^{-3}$ and $a_0=4$.}
\label{chap4fig2}
\end{figure}

An interaction regime accessible with tens of TW-class lasers and parameters typical of current experiments is considered. The left panel of Fig.  \ref{chap4fig2} represents the orbit of a test electron in the ion cavity obtained by integrating Eq. (\ref{chap4eq1}). The electron density is $n_e = 1 \times 10^{19}$ cm$^{-3}$, the propagation length is set at $1.2$ mm, and the laser strength parameter is $a_0 = 4$ (only used to calculate the bubble radius $r_b=6.72\:\mu\text{m}$). The test electron enters at the back of the cavity with $x_i = 2$ $\mu$m, $y_i=0$, $z_i=\zeta_i=-r_b+0.3\:\mu\text{m}=-6.42\:\mu\text{m}$ and $p_{xi}=0$, $p_{yi}=0$, $p_{zi}=25\:mc$.
\begin{figure}
\includegraphics[width=8.5cm]{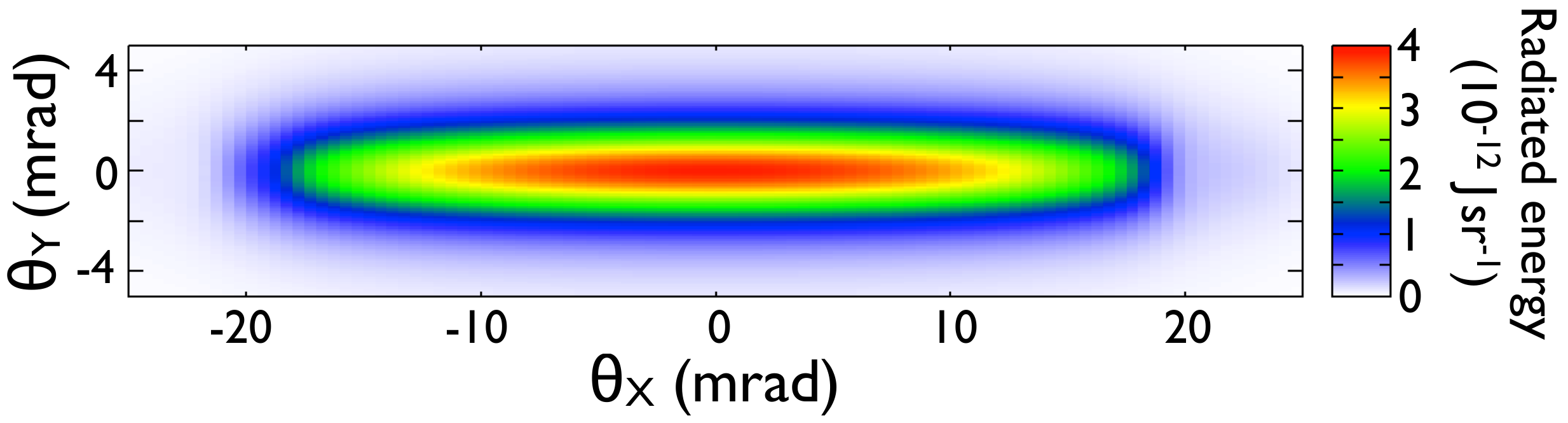}
\caption{X-ray angular distribution calculated from the trajectory of Fig. \ref{chap4fig2}. The color scale represents the radiated energy per unit solid angle.}
\label{chap4fig3}
\end{figure} 
The electron drifts along the longitudinal direction and oscillates in the transverse direction \cite{PoP2005TaPhuoc}. The betatron amplitude $r_{\beta}$ decreases as the electron gains energy ($r_\beta \propto \gamma^{-1/4}$) and presents a minimum at the maximum electron energy. The right panel of Fig.  \ref{chap4fig2} shows the $\gamma$ factor as a function of the longitudinal coordinate $z$. The electron is rapidly accelerated and acquires an energy up to $\sim 240$ MeV. It is decelerated after the dephasing length has been reached. As discussed in Sec. \ref{chap4secA}, the acceleration field may be largely overestimated and several effects not taken into account here can affect the acceleration process. Thus, the predicted maximum electron energy can be largely overestimated in this test-particle simulation. The shape of the trajectory is determined by the initial conditions $(x_i,y_i,z_i)$ and $(p_{xi},p_{yi},p_{zi})$ and can take various forms: planar as in the present simulation, circular, or  helical \cite{PRL2006TaPhuoc, PoP2008TaPhuoc1}.

\begin{figure}
\includegraphics[width=8.5cm]{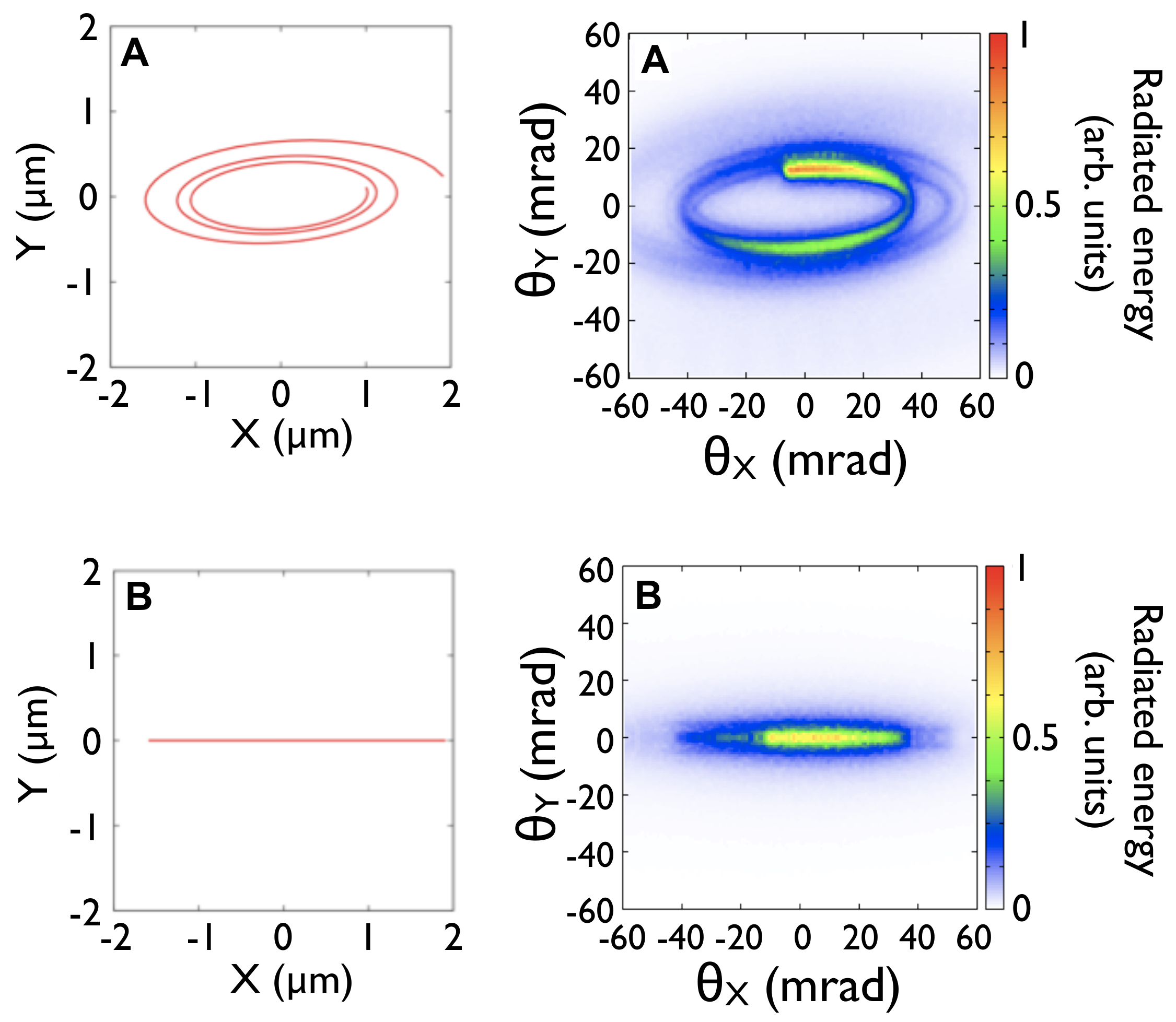}
\caption{Transverse trajectory (on the left) and angular profile of the corresponding emitted radiation (on the right, the color scale representing the radiated energy per unit solid angle) for two different cases. A: The trajectory is three dimensional and helical (elliptical in the transverse plane). B: The trajectory is planar. Both electrons are accelerated up to $\gamma\simeq380$ and the density is $2\times10^{19}$ cm$^{-3}$.}
\label{chap4fig3b}
\end{figure}

The radiation has been calculated using the general formula (\ref{chap2eq1}) in which the orbit calculated above is included. Figure \ref{chap4fig3} represents the spatial distribution of the radiated energy. The transverse profile of the emitted radiation has the same symmetry as the transverse orbit. In our particular case, the main divergence is in the ($\vec{e}_x,\vec{e}_z$) plane and the radiation beam is confined within a typical angle $1/\gamma$ in the perpendicular direction. The typical opening angles in each direction are $\theta_{Xr}\sim 18$ mrad and $\theta_{Yr} \sim 2$ mrad.
The spatial distribution is a direct signature of the transverse electron orbits in the cavity \cite{PRL2006TaPhuoc, PoP2008TaPhuoc1}. As an example, the spatial distributions of the radiation produced by an electron undergoing two types of transverse orbits are represented in Fig. \ref{chap4fig3b}. For minimum radius of curvature along the helical orbit (case A), the transverse electron momentum is along $\vec{e}_y$, so the radiation is emitted up and down and the radiated energy is maximal. For maximum radius of curvature, the transverse momentum is along $\vec{e}_x$, so the radiation is emitted left and right and the radiated energy is minimal. Therefore, the x-ray beam profile is mapping out $p_x/p_z$ and $p_y/p_z$. Also note that the increase of the radiated energy with the electron energy appears in the radiation angular profile. Measurement of the spatial distribution in the x-ray range can therefore provide information on the transverse electron momentum during their oscillation in the wakefield cavity, provided that the energy and $p_z$ are known. It can also provide some insight on the initial conditions of electron injection that determines the type of trajectory. 

\begin{figure}
\includegraphics[width=8.5cm]{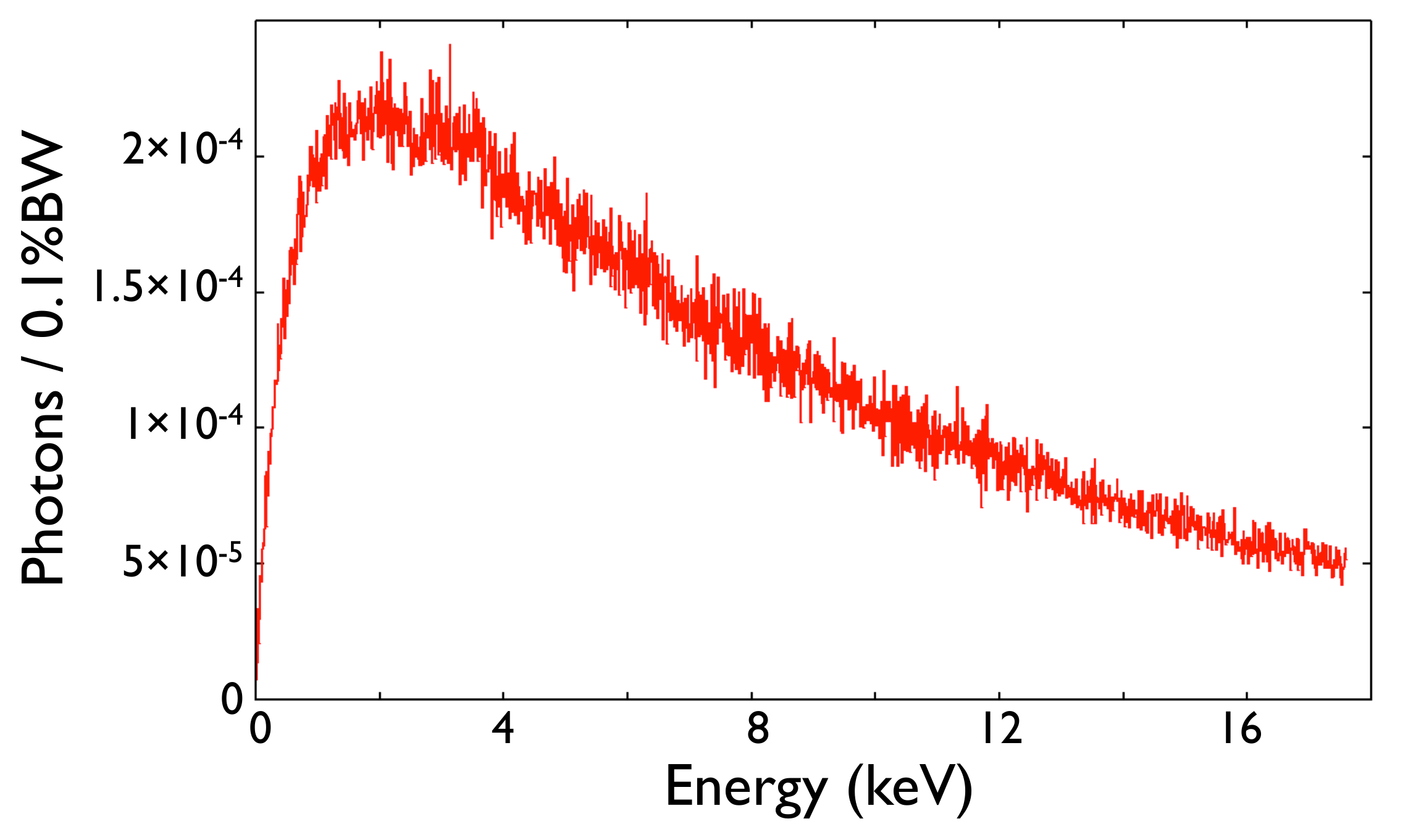}
\caption{Betatron radiation spectrum, integrated over angles and corresponding to the trajectory of Fig. \ref{chap4fig2}, in number of photons per 0.1\% bandwidth and per electron.}
\label{chap4fig4}
\end{figure}

Figure \ref{chap4fig4} presents the spectrum of the radiation integrated over the spatial distribution of Fig. \ref{chap4fig3}.  Instead of the radiated energy primarily given by Eq. (\ref{chap2eq1}), Fig. \ref{chap4fig4} shows the number of emitted photons per 0.1\% bandwidth and per electron. The spectrum extends well above the keV range with a significant number of x-ray photons per electron. An estimate of the total number of photons can be obtained on the basis of the number of electrons that are currently accelerated in a laser-plasma accelerator. Assuming $10^{9}$ electrons gives $\sim 10^{5}\ \text{photons}/0.1\%\ \text{bandwidth}\ (\text{BW})$ between 1 and 10 keV. The total photon number is on the order of $10^9$.

\subsubsection{Particle In Cell simulation}
\label{chap4secCsec2}

We now look at a 3D PIC simulation in a typical experimental parameters range. We are using the CALDER-Circ code \cite{calder-circ}, which, for the cost of a few 2D simulations, can provide the fully 3D trajectories of particles in the plasma. Indeed, this model uses Fourier decomposition of the electromagnetic fields in the azimuthal direction. The axially symmetric (or cylindrical) assumption consists in computing only the single mode $m=0$. In order to be able to account for planar fields, such as the linearly polarized laser, we assume the so-called quasicylindrical geometry and compute only one additional mode, $m=1$. The linearly polarized laser's wakefield in a ultrahigh-intensity regime is well described by those first two modes since fields never stray far from the axial symmetry. The macroparticles, each representing an assembly of electrons, evolve in a full 3D space in which the 3D fields have been reconstructed from the two known Fourier modes. Monitoring macroparticle trajectories gives us the same information as a test-particle code but with realistic wakefield, energies, and injection properties in space and time. The injected particles in the code are detected by their energy. As soon as a particle reaches 45 MeV, it is considered as being injected and we keep track of its trajectory for the rest of the simulation time. These trajectories thus begin after the injection time of each single particle and finish at simulation end, after the particles come out of the plasma. Approximately $10^5$ macroparticles are tracked. Their three position coordinates and their $\gamma$ factor are recorded every $\Delta t_{\text{traj}}=100\: dt_{\text{comput}}\simeq 5$ fs, where $dt_{\text{comput}}$ is the simulation time step. Their contribution to the radiation is computed similarly to the test-particle code [\textit{i.e.}, using Eq. (\ref{chap2eq1})]. We note, however, that to reduce the computing time, it is possible to use the synchrotron radiation formulas instead of Eq. (\ref{chap2eq1}), since betatron radiation occurs in the wiggler regime. For each electron, the angle-integrated spectrum can be calculated by $dI/d\omega=\int dP/d\omega(t)dt$, where $dP/d\omega$ is the synchrotron spectrum given in Eq. (\ref{chap2eq4}) and depends on the radius of curvature which can be evaluated numerically along the arbitrary electron trajectory. The angle-dependent spectrum $d^2I/d\omega d\Omega$ can also be obtained in the wiggler limit by using the ``saddle point'' method~\cite{PoP2003Kostyukov}. The saddle points are defined as the points in the electron trajectories where the velocity vector $\vec{v}$ is parallel to a given observation direction $\vec{n}$. The total radiation spectrum is then obtained as the sum of the synchrotronlike bursts of all saddle points.

\begin{figure}
\includegraphics[width=8.5cm]{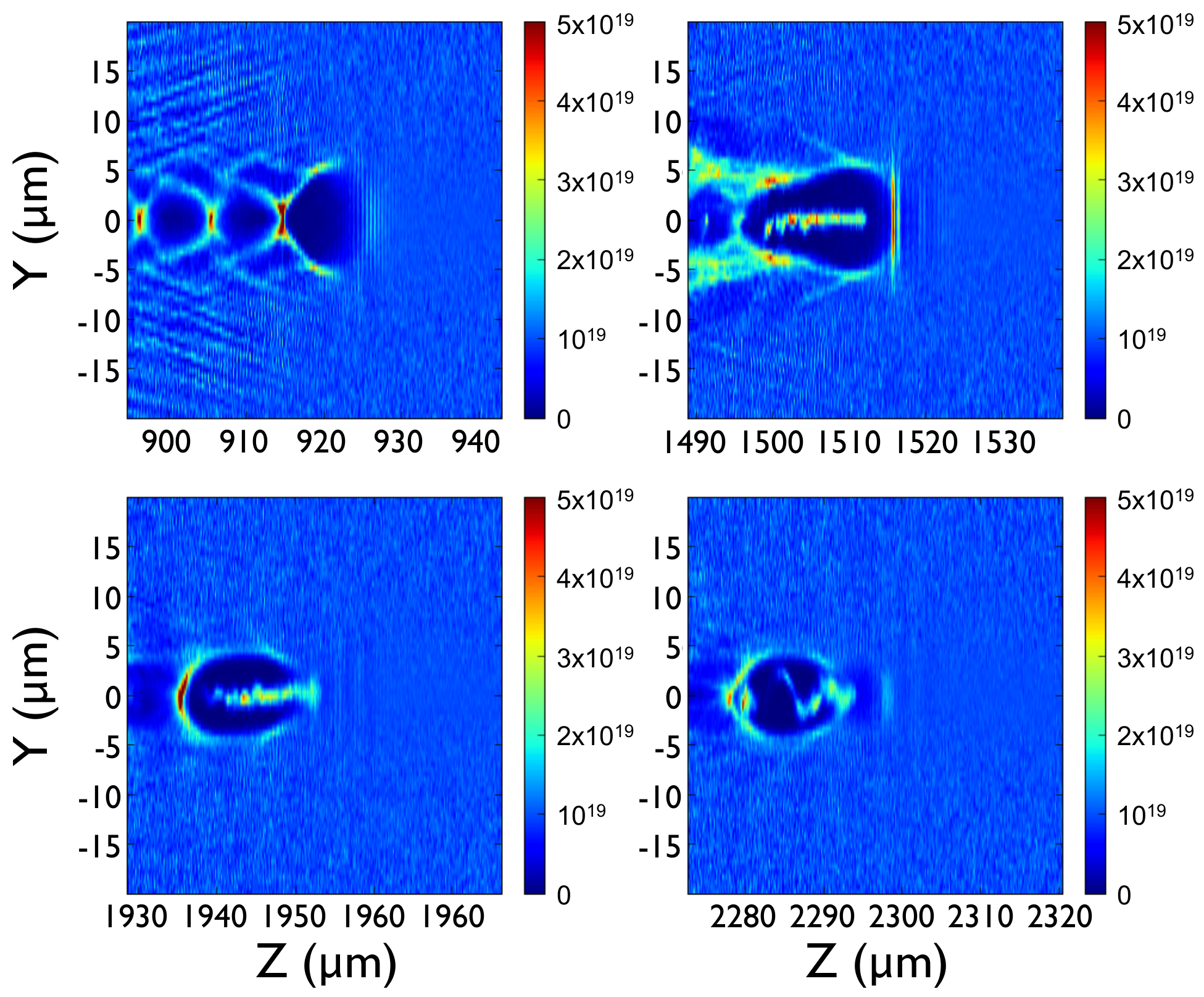}
\caption{Electronic density in the $z-y$ plane averaged over $5\ \mu \text{m}$ in the $x$ direction. Densities are given in $\text{cm}^{-3}$. $z$ is the distance from the beginning of the first density ramp. Different steps of the simulation are shown. First the bubble forms, then injection starts, a whole bunch of accelerated electrons is created inside the bubble and finally the bunch comes out of the bubble and is scattered. The dynamics of the bunch is such that the trailing electrons have a much higher oscillation amplitude than leading electrons.}
\label{injection_phases}
\end{figure}

A 30 fs (FWHM) laser pulse is focused at the beginning of the density plateau of a 4.2 mm wide gas jet, to a peak normalized electric field of $a_0=1.2$. It is linearly polarized in the $\vec{e}_x$ direction and the focal spot size is 18 $\mu$m (FWHM). The gas jet density profile consists in a density plateau of 3 mm at $n_e=1.5\times 10^{19}\ \text{cm}^{-3}$ and the edges are described by ramps varying linearly from 0 to $n_e=1.5\times 10^{19}\ \text{cm}^{-3}$ over $600\ \mu \text{m}$. The simulation describes all the successive steps of the bubble regime (see Fig. \ref{injection_phases}): penetration of the laser pulse in the plasma, formation of the bubble, injection of electrons at the back of the bubble and acceleration, wiggling of the electron bunch in the bubble, laser energy depletion and scattering of the bunch in the plasma when it comes out of the focusing transverse electric field of the bubble. 

\begin{figure}
\includegraphics[width=8.5cm]{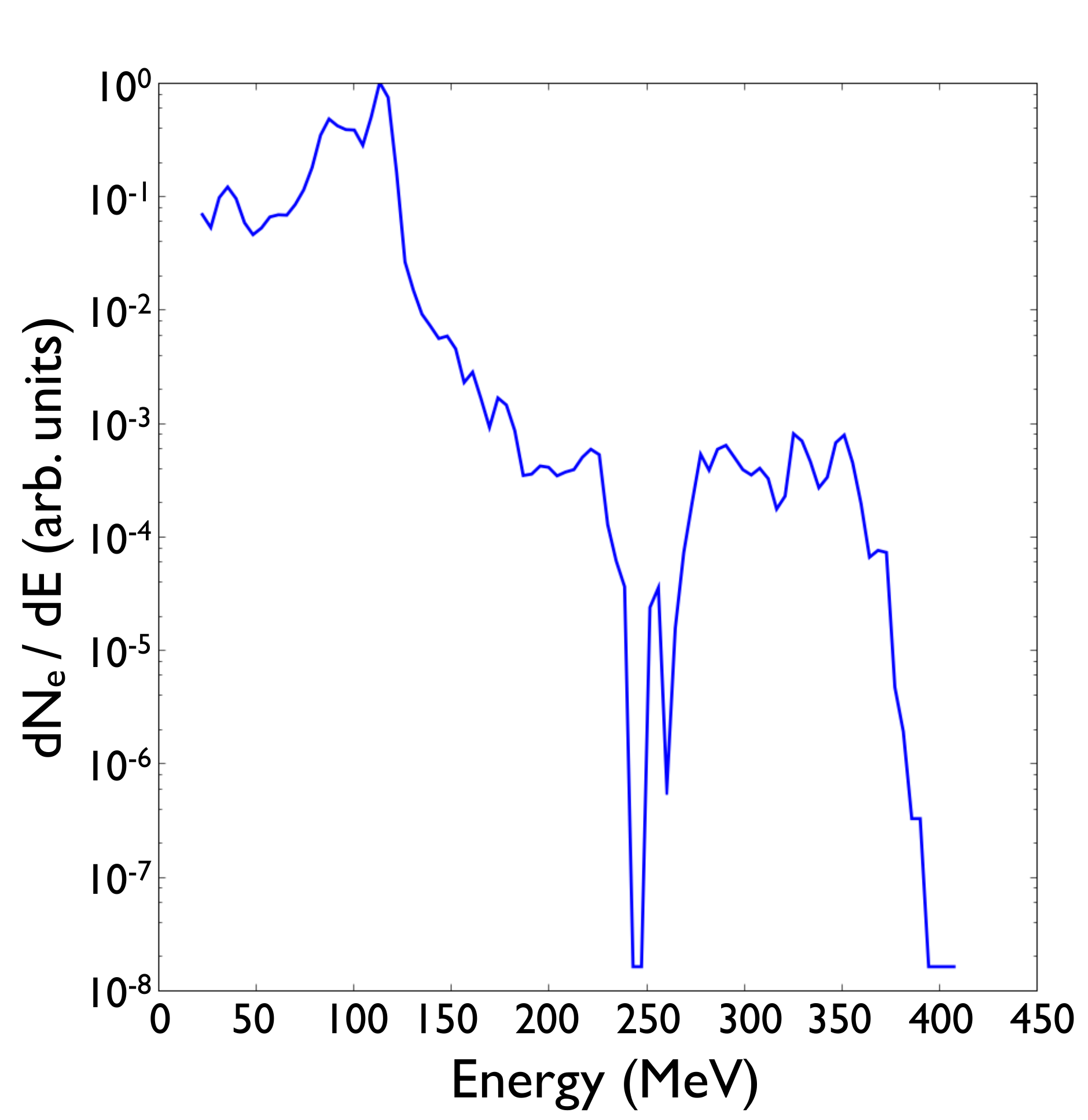}
\caption{Normalized electron beam spectrum before scattering at $z\simeq 1800$ $\mu$m.}
\label{electron_distribution}
\end{figure}

Figure \ref{electron_distribution} shows the electron beam spectrum before scattering. Most of the electrons are around 100 MeV but the tail extends up to 350 MeV. The total charge of the bunch is 640 pC, \textit{i.e.}, approximately $4\times10^9$ electrons.

\begin{figure}
\includegraphics[width=8.5cm]{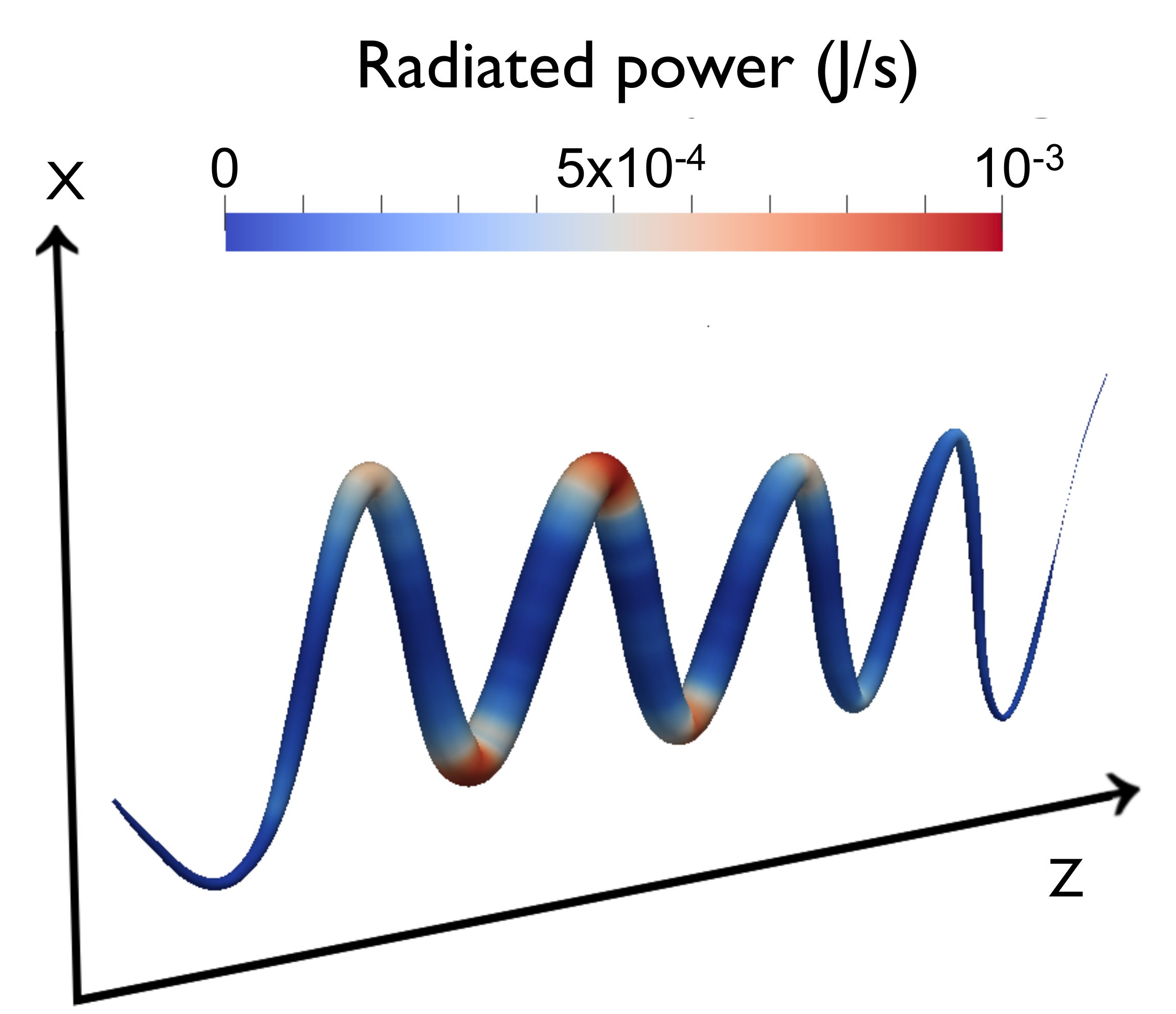}
\caption{Trajectory, energy and radiated power of a single electron before scattering. The tube shows the particle's trajectory. Its radius is proportional to the particle's energy and the color measures the instantaneous radiated power of the particle. The orbit of this particle is in the polarization plan.}
\label{single_radiation}
\end{figure}

Figure \ref{single_radiation} focuses on a single particle.  It illustrates the fact that a particle in the bunch, first accelerated and then dephased, follows a sinusoidal orbit around the propagation axis and radiates the most when it has both a high $\gamma$ and a strong perpendicular acceleration (when the curvature radius is small). 

\begin{figure}
\includegraphics[width=8.5cm]{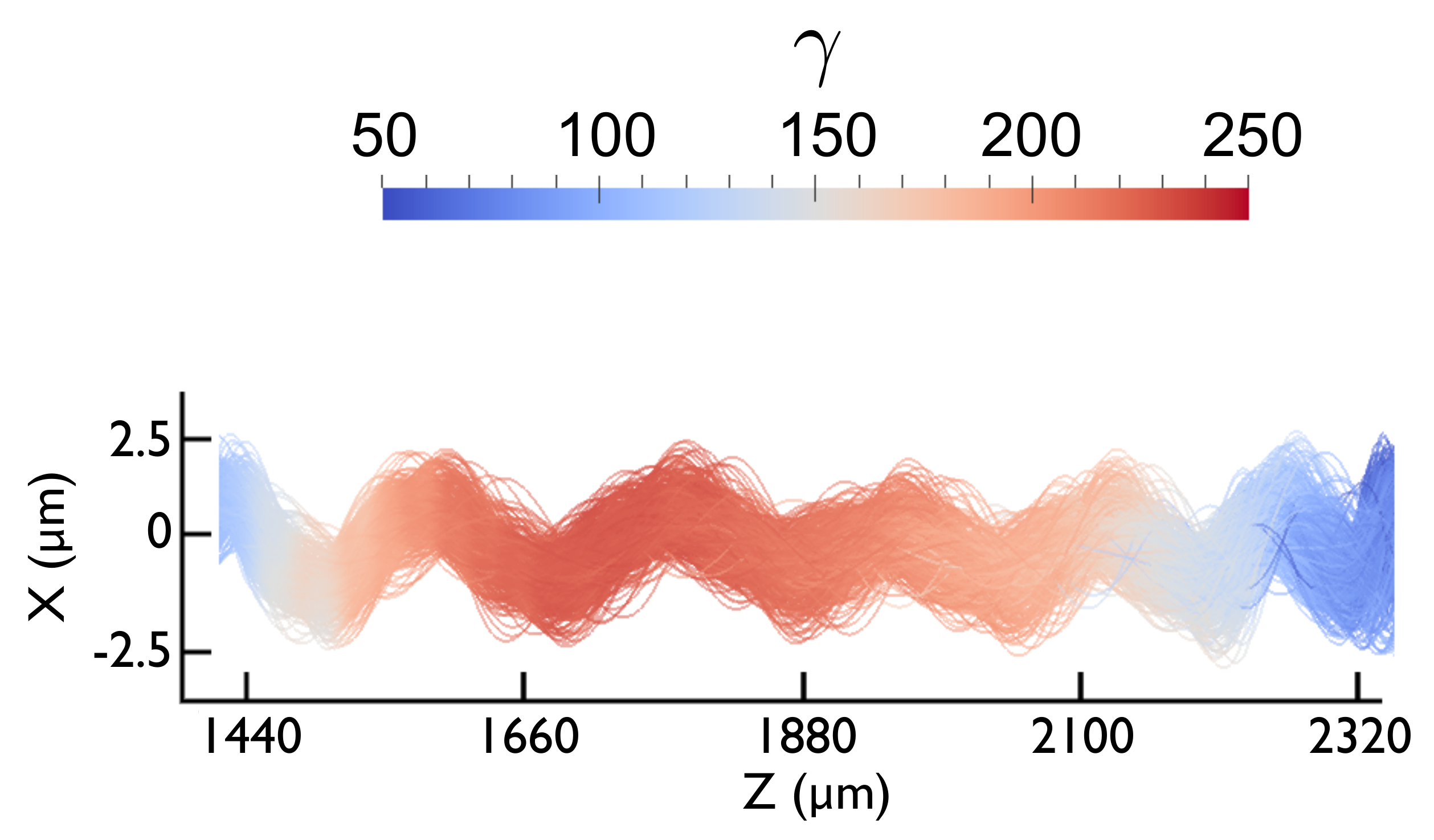}
\caption{Projection in the $z-x$ plane of trajectories and energies of 500 particles from injection until scattering. The laser is polarized in the $\vec{e}_x$ direction and modulates the particle trajectories. Note the different scales in $x$ and $z$ axes.}
\label{trajectories_gamma}
\end{figure}

In a more global point of view, Fig. \ref{trajectories_gamma} shows typical trajectories of electrons during their acceleration inside the bubble in the polarization plane. Each line represents the trajectory of one macroparticle in the code. The color of a line changes as the macroparticle is accelerated to higher energy (represented here by its relativistic $\gamma$ factor). The 500 trajectories plotted in Fig. \ref{trajectories_gamma} are computed from macroparticles injected around the same instant and thus in the same conditions. As the behavior of a particle strongly depends on its injection conditions, it is not surprising to see that, in majority, the represented particles follow very close orbits during the acceleration phase. Moreover, the oscillation is modulated and slightly amplified by the interaction with the laser pulse. This interaction was neglected in the test-particle simulation.

\begin{figure}
\includegraphics[width=8.5cm]{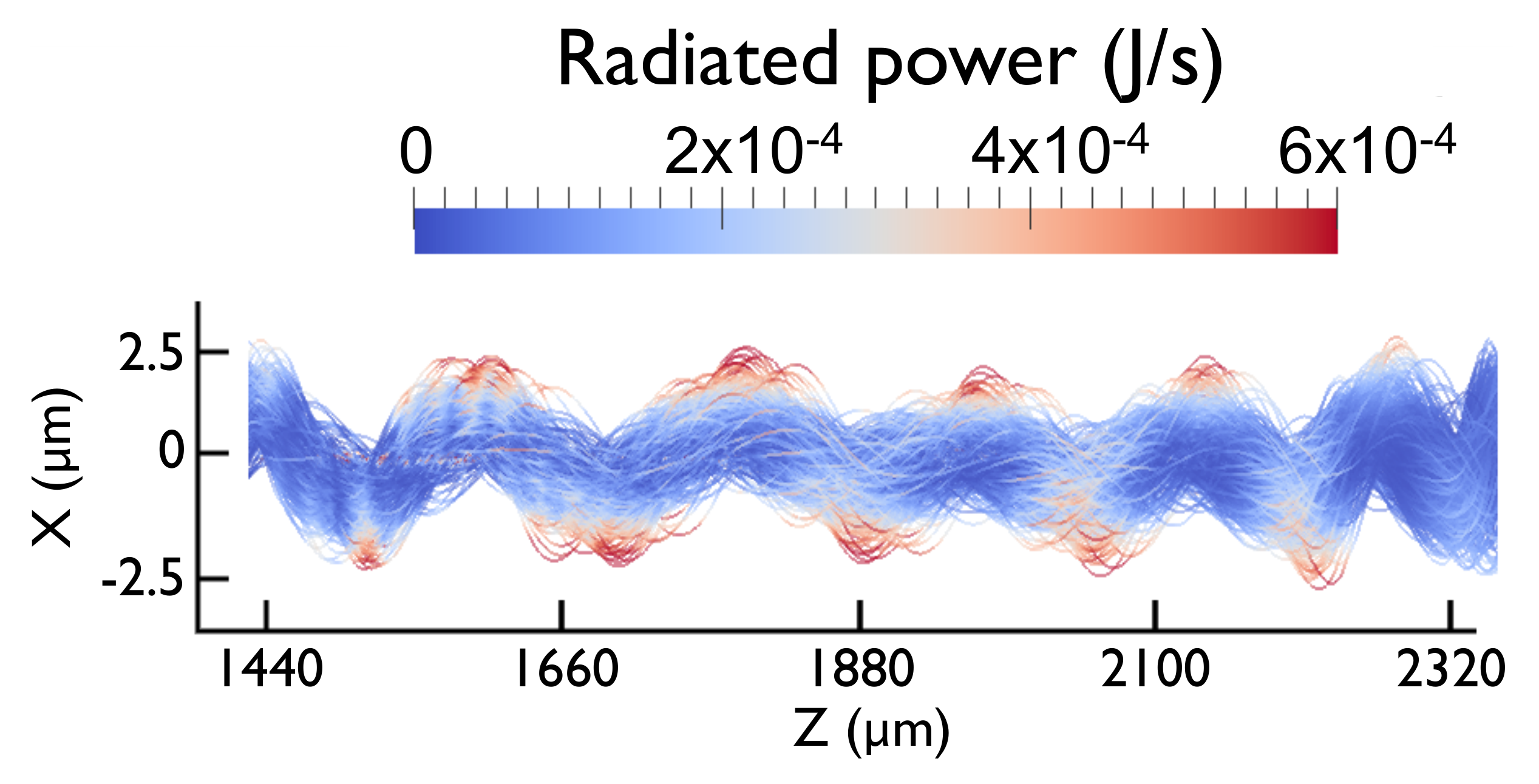}
\caption{Trajectories and radiated power of the same 500 particles as Fig. \ref{trajectories_gamma}. The laser is polarized in the $\vec{e}_x$ direction.}
\label{bunch_radiation}
\end{figure}

Figure \ref{bunch_radiation} shows trajectories and radiated power of the same particles. It can be directly compared to Fig. \ref{trajectories_gamma}. As expected, most of the radiation is produced inside the bubble at high electron energies and far from the axis, where the transverse acceleration is the highest. 

\begin{figure}
\includegraphics[width=8.5cm]{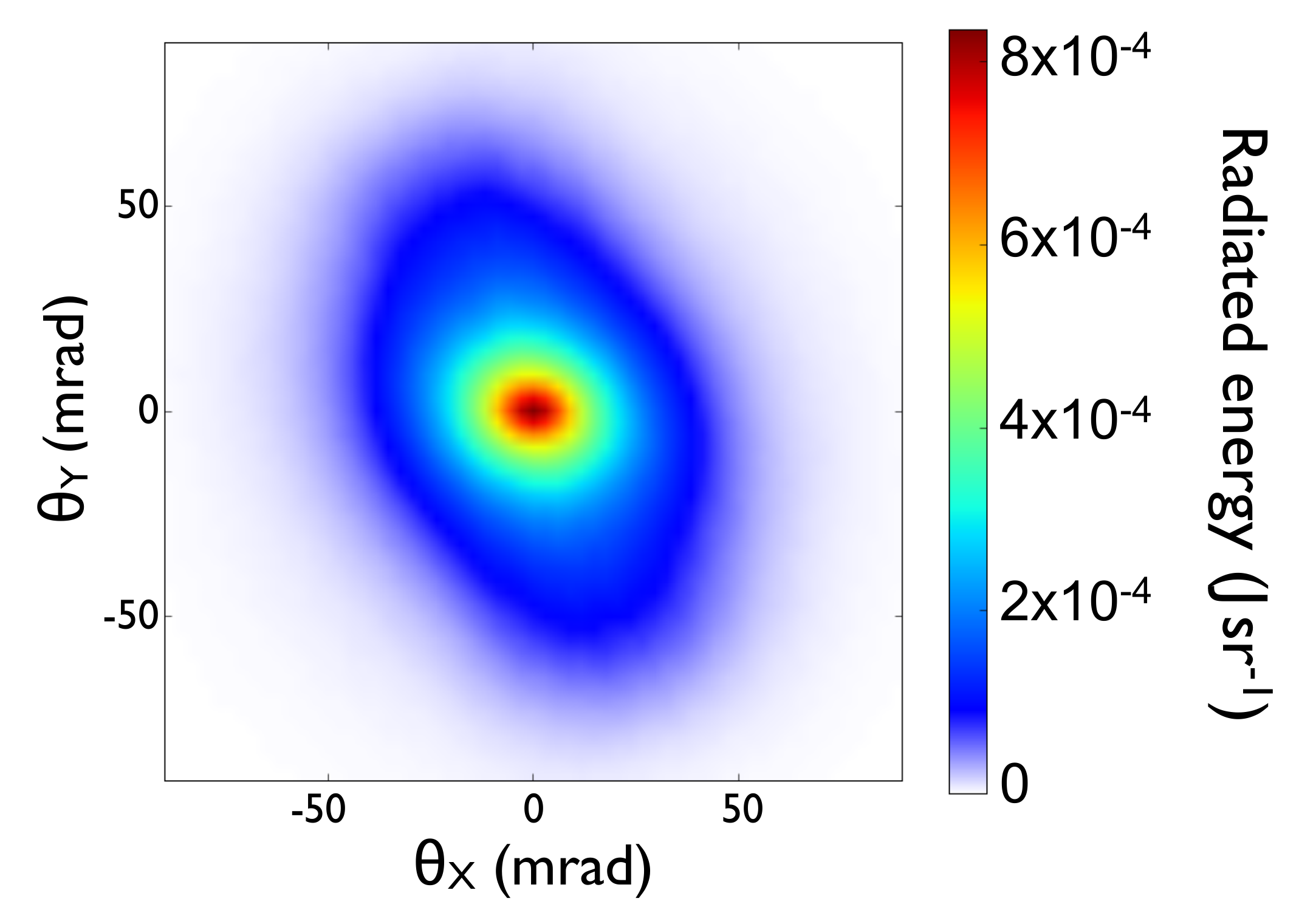}
\caption{Angular profile of the emitted radiation.}
\label{radiated_energy}
\end{figure}

As shown in Sec. \ref{chap2}, the radiation is emitted in the direction of the electron velocity. Figure \ref{radiated_energy} shows the angular distribution of the radiated energy. It is very well centered on the propagation axis and has an angular width of 30 mrad (FWHM).

The frequency distribution of the radiated energy integrated over all angles also yields interesting information.
\begin{figure}
\includegraphics[width=8.5cm]{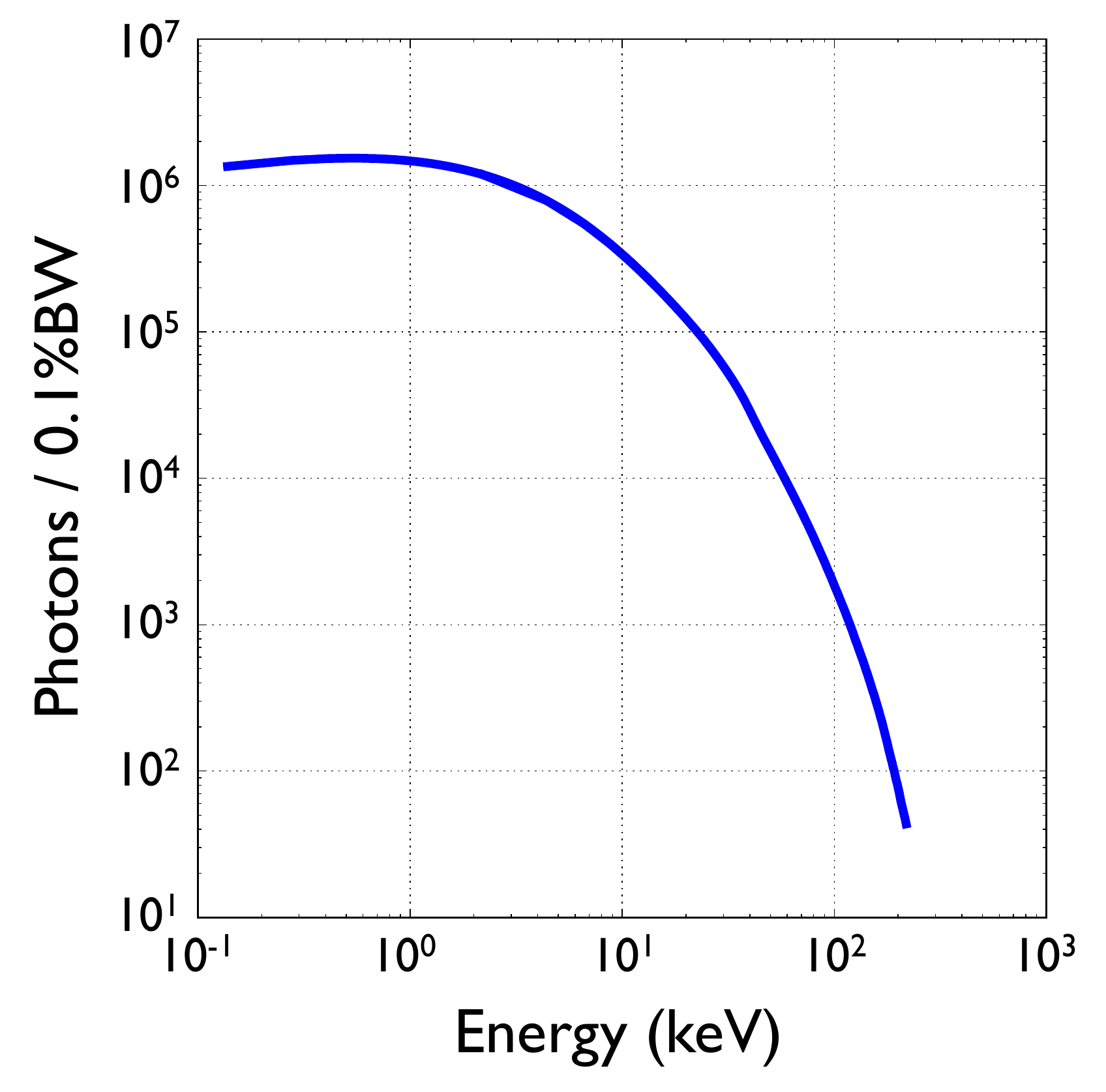}
\caption{Spectrum of the x-ray radiation emitted by all injected macroparticles.}
\label{bunch_spectrum}
\end{figure}
Figure \ref{bunch_spectrum} represents the calculated x-ray spectrum emitted by all injected macroparticles. The spectral distribution is given by the number of emitted photons per 0.1\% bandwidth. The total number of emitted photons is $N_\gamma \simeq 4.5\times10^9$ and we have $\approx 10^6\ \text{photons}/0.1\%\text{BW}$ at 1 keV and $\approx 3\times 10^5\ \text{photons}/0.1\%\text{BW}$ at 10 keV. This simulation reproduces the global behavior observed in the experiments (see Sec. \ref{chap4secD}) but tends to produce more photons with a spectrum extending to higher energies.

\subsection{Experimental results}
\label{chap4secD}

\begin{figure}
\includegraphics[width=8.5cm]{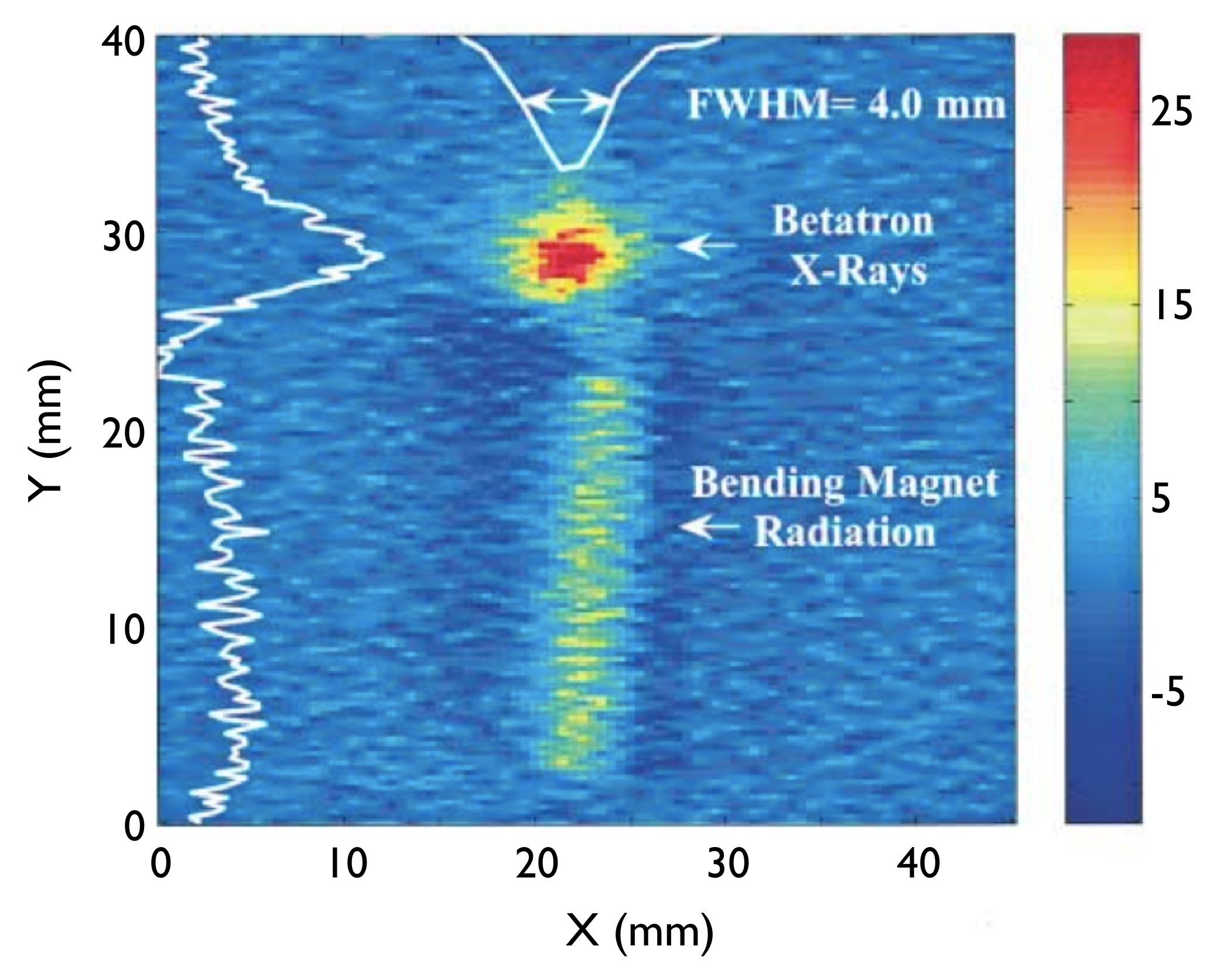}
\caption{Betatron x-ray beam (round spot), produced by the 28.5 GeV SLAC electron beam propagating through a plasma of density $\sim2\times 10^{13}$ cm$^{-3}$, and recorded by a fluorescent screen imaged onto a visible CCD camera. From \textcite{PRL2002Wang}.}
\label{chap4fig5b}
\end{figure}
The production of betatron radiation requires a relativistic electron bunch and an ion cavity or channel. The laser-plasma interaction is not the only mechanism that can create this ion cavity. Another approach was investigated first experimentally: an electron beam  propagating in a plasma can expel background electrons via the repulsing electric space-charge force and create an ion cavity if its own density is larger than the plasma density. The cavity being created by the front edge of the bunch, its main body propagates in the self-produced cavity and experiences its fields. This scheme of betatron x-ray emission was demonstrated by \textcite{PRL2002Wang}. The experiment, carried out at the SLAC National Accelerator Laboratory, was based on the use of the 28.5 GeV SLAC electron beam, containing $1.8\times10^{10}$ electrons per bunch and focused near the entrance of a preformed lithium plasma of length 1.4 m. For a plasma density of $1.7\times 10^{14}$ cm$^{-3}$, the strength parameter of betatron oscillations was $K = 16.8$, and the number of emitted photons and the peak spectral brightness were estimated to be respectively $6\times10^5\ \text{photons}/0.1\%\text{BW}$ and  $7\times10^{18}\ \text{photons}/(\text{s}\: \text{mrad}^2\: \text{mm}^2\: 0.1\%\text{BW})$, at 14.2 keV. The betatron x-ray beam (shown in Fig. \ref{chap4fig5b} for a smaller plasma density, $\sim2\times 10^{13}$ cm$^{-3}$) was found to be collimated in the divergence angle of $\sim (1-3)\times10^{-4}$ rad. The expected quadratic density dependence of the spontaneously emitted betatron x-ray radiation [$\overline{P}_\gamma\propto\gamma^2K^2/\lambda_u^2\propto n_e^2$ at constant $\gamma$ and $r_\beta$; see Eqs. (\ref{eq_chap2_power}) and (\ref{chap4eq3b})-(\ref{chap4eq3})] was also observed, with a detection sensible in the spectral range 5 - 30 keV. These features were found to be in good agreement with theoretical predictions.

\bigskip
\begin{figure}
\includegraphics[width=8.5cm]{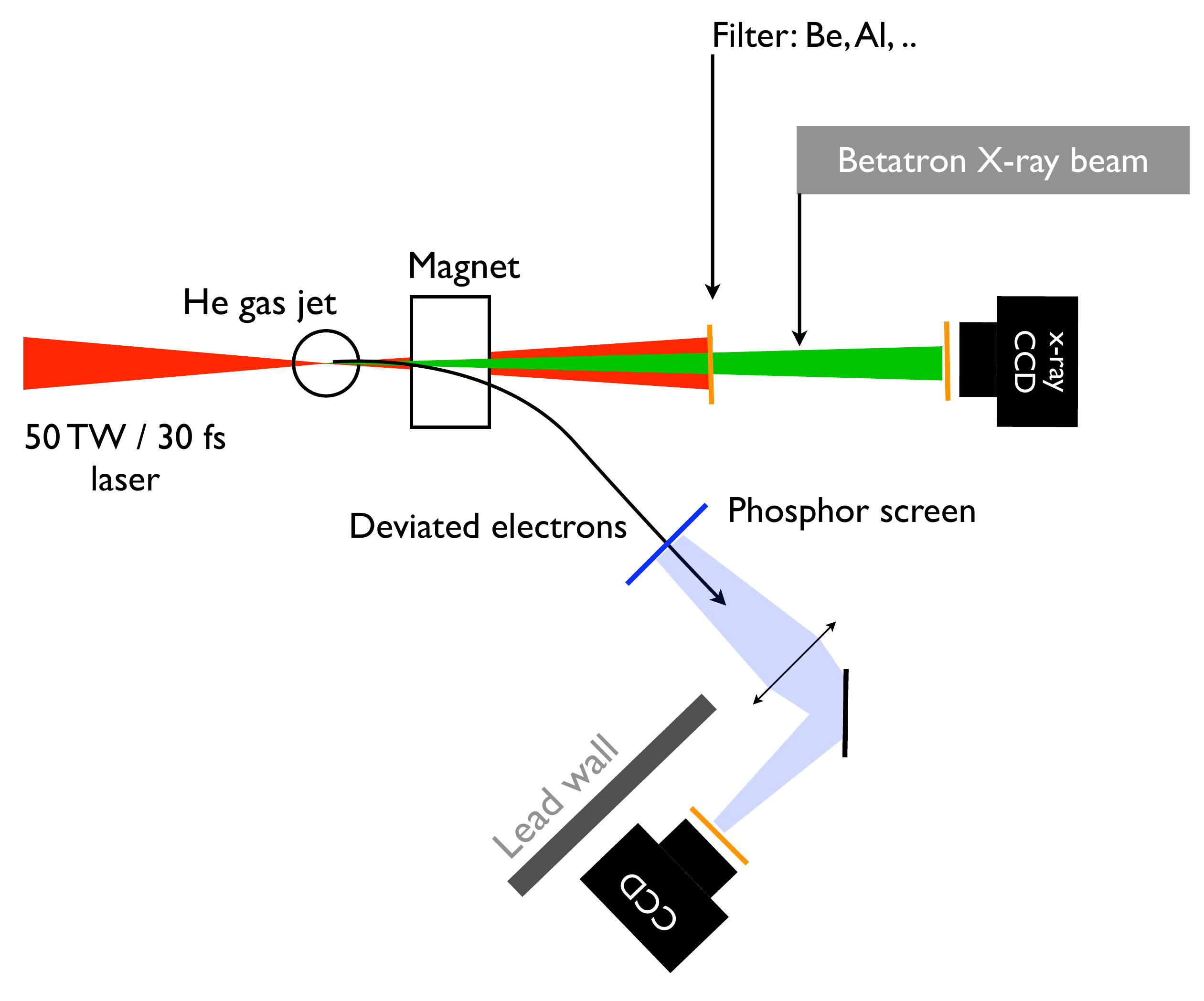}
\caption{Experimental setup for the generation and observation of betatron radiation. A 50 TW / 30 fs class laser is focused onto the front edge of a gas jet. Electron spectra are obtained by deflecting electrons with a magnet and using a LANEX phosphor screen imaged onto a CCD. Betatron radiation is recorded on axis using a charged coupled device sensitive up to 10 keV (CCD X). A filter (Be, Al, etc.) is placed before the CCD to block infrared light.}
\label{chap4fig5}
\end{figure}
The first experiment based on a fully laser-plasma approach discussed above was performed by \textcite{PRL2004Rousse} using a 50 TW / 30 fs laser focused onto a 3 mm helium gas jet at an electron density of $\sim 10^{19}$ cm$^{-3}$. A typical betatron experimental setup is shown in Fig. \ref{chap4fig5}. An intense femtosecond laser pulse is focused onto a supersonic helium gas jet. Electrons accelerated during the interaction are deflected toward a phosphor screen using a permanent magnet with a field approaching 1 T. Because the magnet bends the electron trajectories in only one direction, the image on the screen reveals the electron spectrum and the beam divergence. The x-ray radiation is observed using an x-ray CCD placed on axis. With the camera sensitive in a broad spectral range, all radiation from the laser and the plasma below the keV range is blocked using filters (beryllium and aluminum).

\begin{figure}
\includegraphics[width=8.5cm]{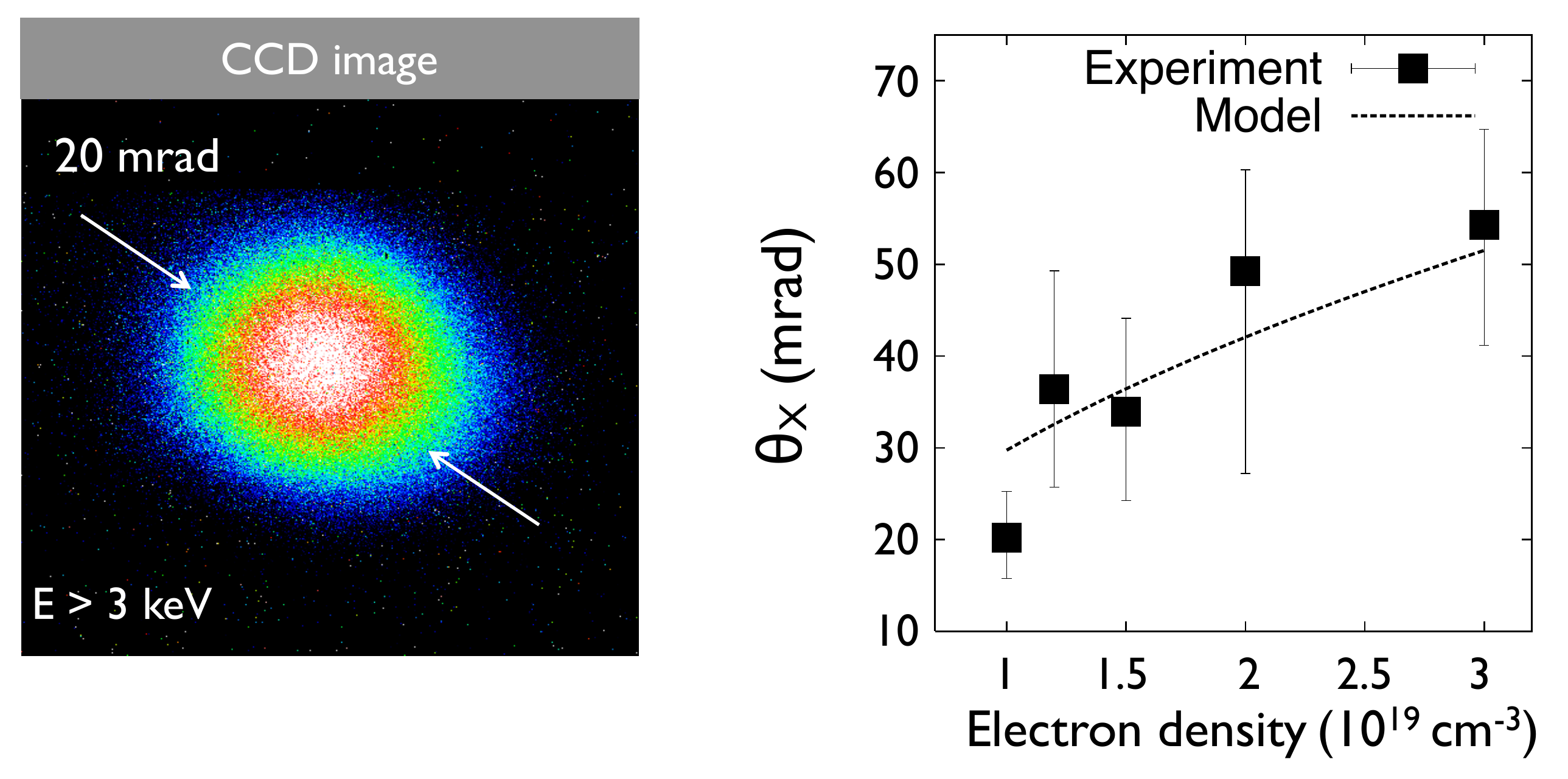}
\caption{Left: A raw image of the angular distribution of the betatron radiation beyond 3 keV recorded by the CCD in a single shot. Right, divergence (full width at half maximum) of radiation as a function of density. The dotted line shows the dependence of $\theta$ on $n_e$ for constant $r_\beta$ and $\gamma$ according to the equations of Secs. \ref{chap4secA} and Sec. \ref{chap4secB}.}
\label{chap4fig6}
\end{figure}
The left part of Fig. \ref{chap4fig6} shows the x-ray beam profile observed on the camera (at energy $>$ 3 keV as a $500$ $\mu$m beryllium filter was placed in front of the camera). The right part of the figure represents the evolution of the beam divergence (full width at half maximum) as a function of the electron density. The radiation is collimated on axis with a divergence in the range $[10:100]$ mrad. The divergence increases with the electron density due to the reductions of both the betatron period and the electron energy as observed in the experiment. These experimental observations are in good agreement with the numerical predictions presented above \cite{PoP2005TaPhuoc,PRL2006TaPhuoc,PoP2008TaPhuoc1,PRE2008Albert} and with PIC simulations \cite{PRL2004Kiselev, PRL2004Rousse,PoP2005TaPhuoc,NF2004Pukhov}.

\begin{figure}
\includegraphics[width=8.5cm]{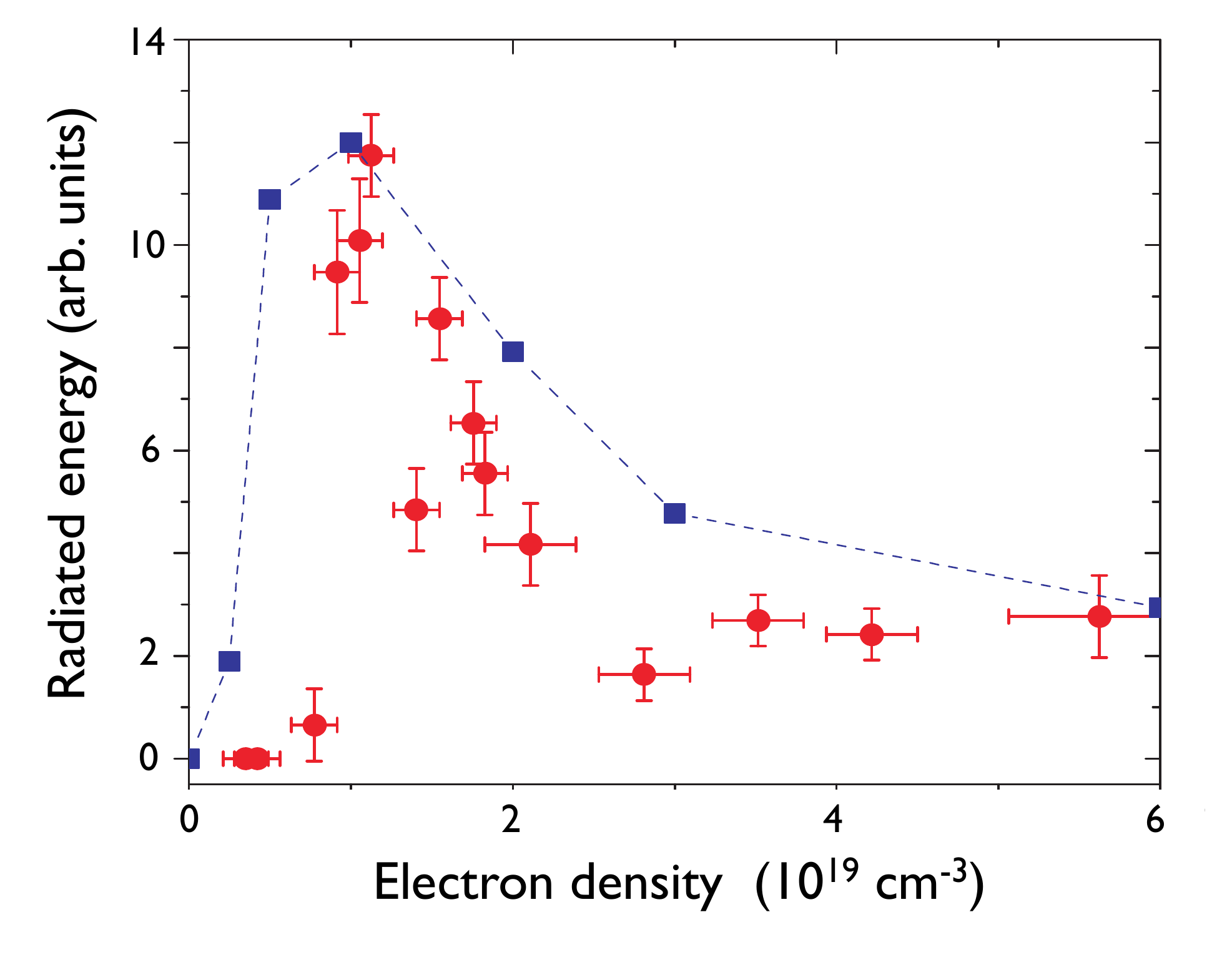}
\caption{X-ray signal (proportional to radiated energy beyond 1 keV per unit solid angle) as a function of plasma electron density. Each point corresponds to an average value over ten shots. The dotted line corresponds to the results obtained using 3D PIC simulations. From \textcite{PRL2004Rousse}.}
\label{chap4fig7}
\end{figure}
\begin{figure}
\includegraphics[width=8.5cm]{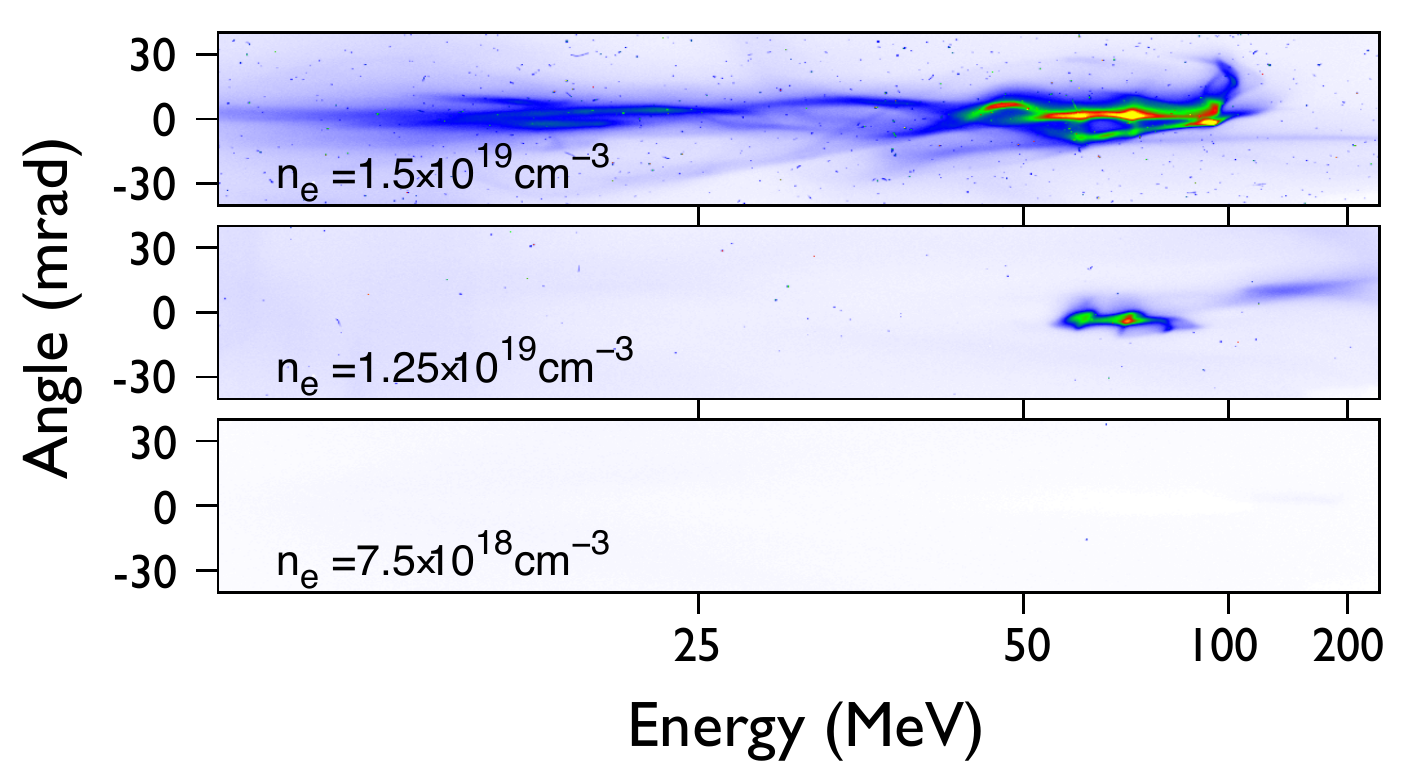}
\caption{Raw electron spectra for three different densities and fixed laser parameters. Horizontal axis, electron energy; vertical axis, exit angle; and color scale, number of counts. This latter gives an indication of the beam charge. No electron is observed below the trapping threshold ($n_e=7.5\times10^{18}~\text{cm}^{-3}$), a quasimonoenergetic electron beam is produced just above ($n_e=1.25\times10^{19}~\text{cm}^{-3}$), and a broadband spectrum is recorded for a higher density ($n_e=1.5\times10^{19}~\text{cm}^{-3}$). The threshold density is not the same as in Fig. \ref{chap4fig7} because the laser parameters are different.}
\label{chap4fig7b}
\end{figure}

The betatron radiation has a specific dependence with plasma density (for given laser parameters) \cite{PRL2004Rousse,PoP2005TaPhuoc}. Figure \ref{chap4fig7} shows the maximum of the radiated energy per unit solid angle beyond 1 keV (in arbitrary units), as a function of the electron density, obtained in the experiment and in PIC simulations. The observed x-ray emission is peaked at $n_e \sim 1.1 \times 10^{19}{\text{cm}}^{-3}$. Below $5\times10^{18}$cm$^{-3}$, the x-ray signal vanishes simply because no electron is trapped (the laser intensity is too low for such densities in order to ensure electron trapping). This is confirmed in the experiment for which no electron was detected by the spectrometer. Just above this threshold, electrons are trapped in the bubble regime and monoenergetic electron bunches as well as betatron radiation of relatively low intensity can be obtained. When the density is higher, the plasma wavelength is smaller and the focal spot size and pulse duration are not exactly matched. It first corresponds to the FLWF regime \cite{Science2002Malka, PoP2003Najmudin}, and then for even higher density to SM-LWFA \cite{Nature1995Modena, Science1996Umstadter, PRL2001Santala, PRL1997Wagner, PoP2003Najmudin, PoP1997Ting, PRL1995Coverdale}. This results in acceleration of electron bunches of poor quality compared to the bubble regime (broad spectrum, high divergence, and large shot-to-shot fluctuations). The counterpart of this poor quality electron beam is that the number of electrons and the amplitude of motion $r_{\beta}$ are higher and the betatron period is smaller, such that the betatron radiation is brighter even if the mean electron energy is smaller. An additional effect which can occur is the interaction of the electron bunch with the laser pulse. From $n_e \sim 1.1 \times 10^{19}$ to $2.7 \times 10^{19}\:\text{cm}^{-3}$, the x-ray signal drops down and a plateau is reached. The PIC numerical simulations \cite{PRL2004Rousse,PoP2005TaPhuoc} shown in Fig. \ref{chap4fig7} reproduce this experimental behavior: a sharp increase of the x-ray intensity followed by a smoother decrease of the signal. Three typical electron spectra which can be obtained simultaneously with the measurement of betatron radiation are represented in Fig. \ref{chap4fig7b}. The first shows a broadband spectrum obtained at relatively high density, characteristic of the forced laser wakefield regime, and the second corresponds to a monoenergetic electron bunch (signature of the bubble regime) obtained for a density just above the trapping threshold. Below this threshold, no electron is observed (third spectrum). For some broadband spectra, a correlation of the electron output angle with the electron energy can be observed. It is attributed to off-axis injection of the electron beam due to asymmetric laser pulse intensity profile or tilted energy front. For these shots, direct measurement of the betatron motion can be obtained from the oscillation observed in the spectrum \cite{EPL2008Glinec}.

\begin{figure}
\includegraphics[width=8.5cm]{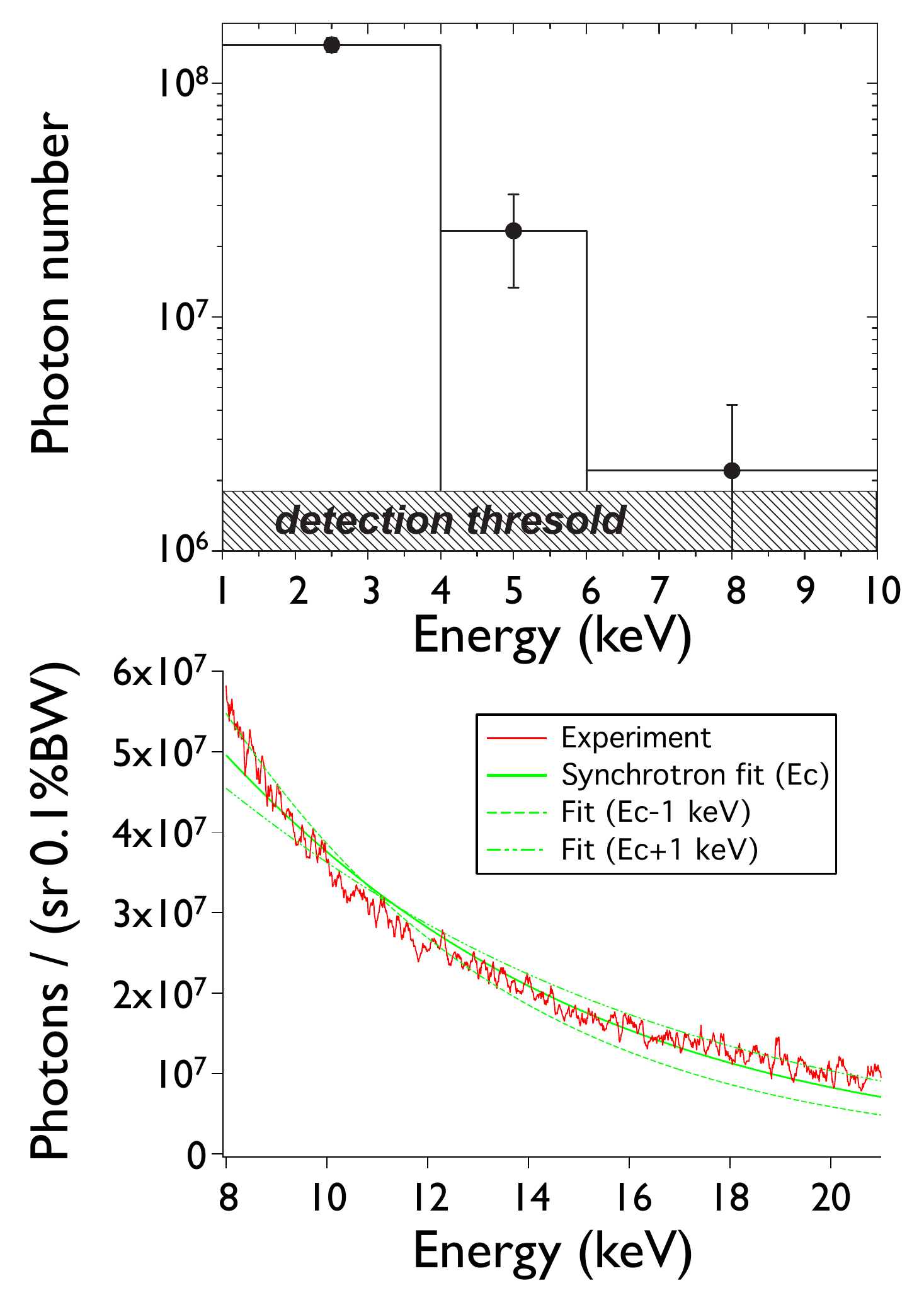}
\caption{Top: Total number of photons obtained experimentally within the spectral bandwidths 
determined by filters: 25 $\mu$m Be $(1 < E < 10$ keV), 25 $\mu$m Be + 40 $\mu$m Al $(4  < E < 10$ keV), and 25 $\mu$m Be + 25 $\mu$m Cu $(6 < E < 10$ keV). From \textcite{PRL2004Rousse}. Bottom: Averaged spectrum (in red line) of the radiation obtained by photon counting, with a fit (in green line) by the function $S(\omega/\omega_c)$ with $\hbar\omega_c=5.6\pm1$ keV. From \textcite{NJP2011Fourmaux}.}
\label{chap4fig8}
\end{figure}
The spectrum of the betatron radiation was first measured using a set of filters \cite{PRL2004Rousse,PoP2005TaPhuoc}, for an electron density $n_e=1\times10^{19}$ cm$^{-3}$. The top panel of Fig. \ref{chap4fig8} presents the number of x-ray photons obtained experimentally, integrated over the beam divergence and over the spectral bandwidths determined by (1) 25 $\mu$m Be ($1< E<10$ keV),  (2) 25 $\mu$m Be + 40 $\mu$m Al ($4 < E < 10$ keV), and (3) 25 $\mu$m Be $+$ 25 $\mu$m Cu ($6 < E < 10$ keV). The bottom part of the figure represents a measurement of the spectrum performed by photon counting, at higher resolution (350 eV), in the range 8 to 21 keV. In this experiment, betatron radiation was produced with a 80 TW laser system (2.5 J and 30 fs) at the Advanced Laser Light Source \cite{NJP2011Fourmaux}, for an electron density $n_e=5.4\times10^{18}$ cm$^{-3}$. The experimental spectrum (averaged over 10 shots) was well described by a synchrotron function at $\hbar\omega_c=5.6\pm1$ keV in the range 8 to 21 keV, demonstrating the synchrotron-type spectrum of betatron radiation \cite{NJP2011Fourmaux}. Depending on the experiment and available laser energy in the focal spot, the total number of photons range from $10^8$ \cite{PRL2004Rousse} to $10^9$ with $2.2\times10^8\ \text{photons}/(\text{sr}\: 0.1\%\text{BW})$ at 10 keV \cite{OL2011Fourmaux}. The latter result also showed a critical energy above 10 keV \cite{OL2011Fourmaux}.

\begin{figure}
\includegraphics[width=8.5cm]{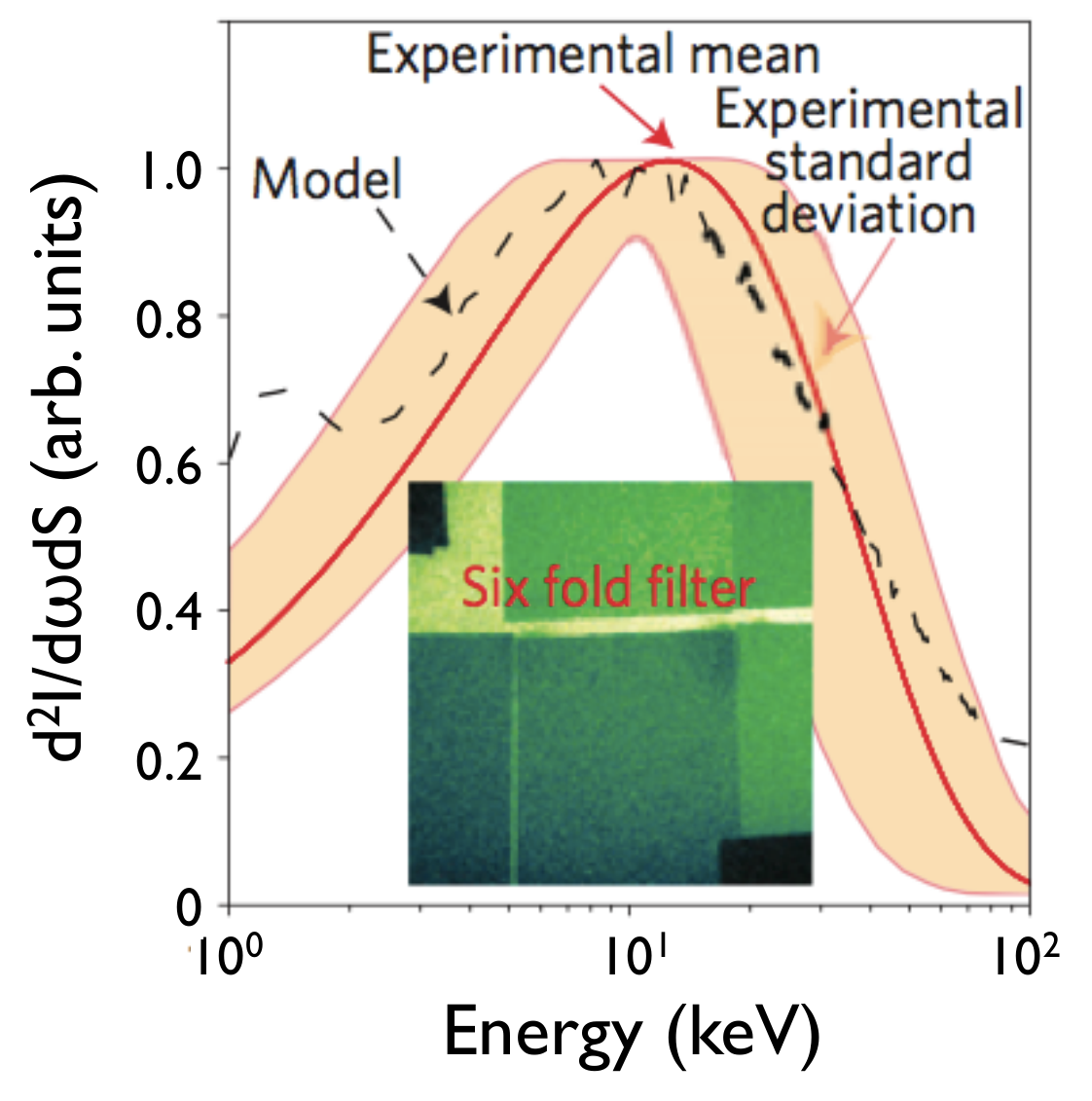}
\caption{Synchrotron spectrum fitted from transmission values through six different filters. The fit uses the synchrotron function $(\omega/\omega_c)^2K^2_{2/3}(\omega/\omega_c)$ instead of $S(\omega/\omega_c)$, and the fitted critical energy is $29\pm13$ keV. From \textcite{NatPhys2010Kneip}.}
\label{chap4fig8b}
\end{figure}

Betatron radiation has also been observed in a recent experiment performed on the Michigan Hercules laser system at the University of Michigan \cite{NatPhys2010Kneip}. The results demonstrated high photon energies and small x-ray beam divergence. The betatron radiation had a broadband spectrum, characterized using a set of filters. For an electron density $n_e=1\times10^{19}$ cm$^{-3}$, the transmission values through the six filters were fitted with the synchrotron function $(\omega/\omega_c)^2K^2_{2/3}(\omega/\omega_c)$ [which differs from $S(\omega/\omega_c)$] with a critical energy $\hbar\omega_c\sim29\pm13$ keV (see Fig. \ref{chap4fig8b}). The divergence angle was found to be $\sim 10$ mrad (FWHM), and the electron spectrum, simultaneously recorded, indicated electron acceleration up to 400 MeV. The peak spectral brightness was estimated to be $\sim 10^{22}\ \text{photons}/(\text{s}\ \text{mrad}^2\: \text{mm}^2\: 0.1\%\text{BW})$, which is comparable to currently existing third-generation conventional light sources.

The duration of short-pulse betatron x rays was first estimated using time resolved x-ray diffraction. In this experiment \cite{PoP2007TaPhuoc}, an ultrafast phase transition (nonthermal melting of InSb) was used as a Bragg switch to sample the x-ray pulse duration. The dynamics of this specific transition is well known. The InSb is initially crystalline and diffracts the x-ray radiation according to the Bragg law. However, when irradiated by a femtosecond laser pulse (at a fluence in the 100 mJ/cm$^2$ range), the surface of the InSb is melted and disorders appears in a few tens of femtoseconds. As a consequence, x-ray radiation can no longer be diffracted. The duration of this phase transition has been determined in several experiments based on the use of femtosecond synchrotron radiation \cite{Science2005Lindenberg} and femtosecond $K_\alpha$ radiation \cite{Nature2001Rousse}. Here, assuming the dynamics of the phase transition is known, the duration of the betatron x-ray pulse was estimated to be less than 1 ps with a best fit below 100 fs \cite{PoP2007TaPhuoc}. Recently, the electron bunch duration was measured \cite{NatPhys2011Lundh} using spectral characterization of coherent transition radiation (CTR), in a laser-plasma accelerator working with external colliding pulse injection \cite{Nature2006Faure}. The CTR result demonstrates a root-mean-square duration of 1.5 fs for the electron bunch. Betatron radiation was shown to be strongly correlated with the properties of these electron bunches \cite{PRL2011Corde2}, showing that the radiation is effectively produced by these ultrashort electron bunches and that it inherits its duration of a few femtoseconds.

The source size has been determined with two different methods. The first uses the Fresnel edge diffraction of the x-ray beam and is based on the spatial coherence of the radiation \cite{PRE2006Shah}. Because electrons emit x rays incoherently (see Sec. \ref{chap2secF}) in the betatron mechanism, the source size determines the degree of spatial coherence which results in interference fringes in the Fresnel diffraction experiment. This latter experiment corresponds to the measurement of the intensity profile of the shadow made by a knife edge (here the edge of a cleaved crystal) placed in the x-ray beam onto a detector at a distance sufficient to ensure an appropriate resolution. Using this technique, it was demonstrated that the source size was less than 8 $\mu$m (FWHM). A knife-edge technique was also used in betatron experiments with the Michigan Hercules laser system at the University of Michigan \cite{NatPhys2010Kneip}, and with the Advanced Laser Light Source laser system \cite{OL2011Fourmaux}, showing an x-ray source size of $\sim 1-2$ $\mu$m (FWHM). The second method relies on the information contained in the spatial distribution of the radiation. Indeed, because the betatron radiation is emitted in the direction of the electron velocity, its profile represents a direct signature of the electron orbits in the cavity. In addition, the divergence and the structure of the distribution depends on the amplitude of the electrons orbits. This method, discussed in the previous section, is presented in references \textcite{PRL2006TaPhuoc, PoP2008TaPhuoc1}. It is shown that the spatial distribution of the radiation implies a limited choice of electron orbits. Therefore, by determining the electron orbits, the source size was deduced and the result was in the range of $\sim 1-2$ $\mu$m (FWHM).

\begin{figure}
\includegraphics[width=8.5cm]{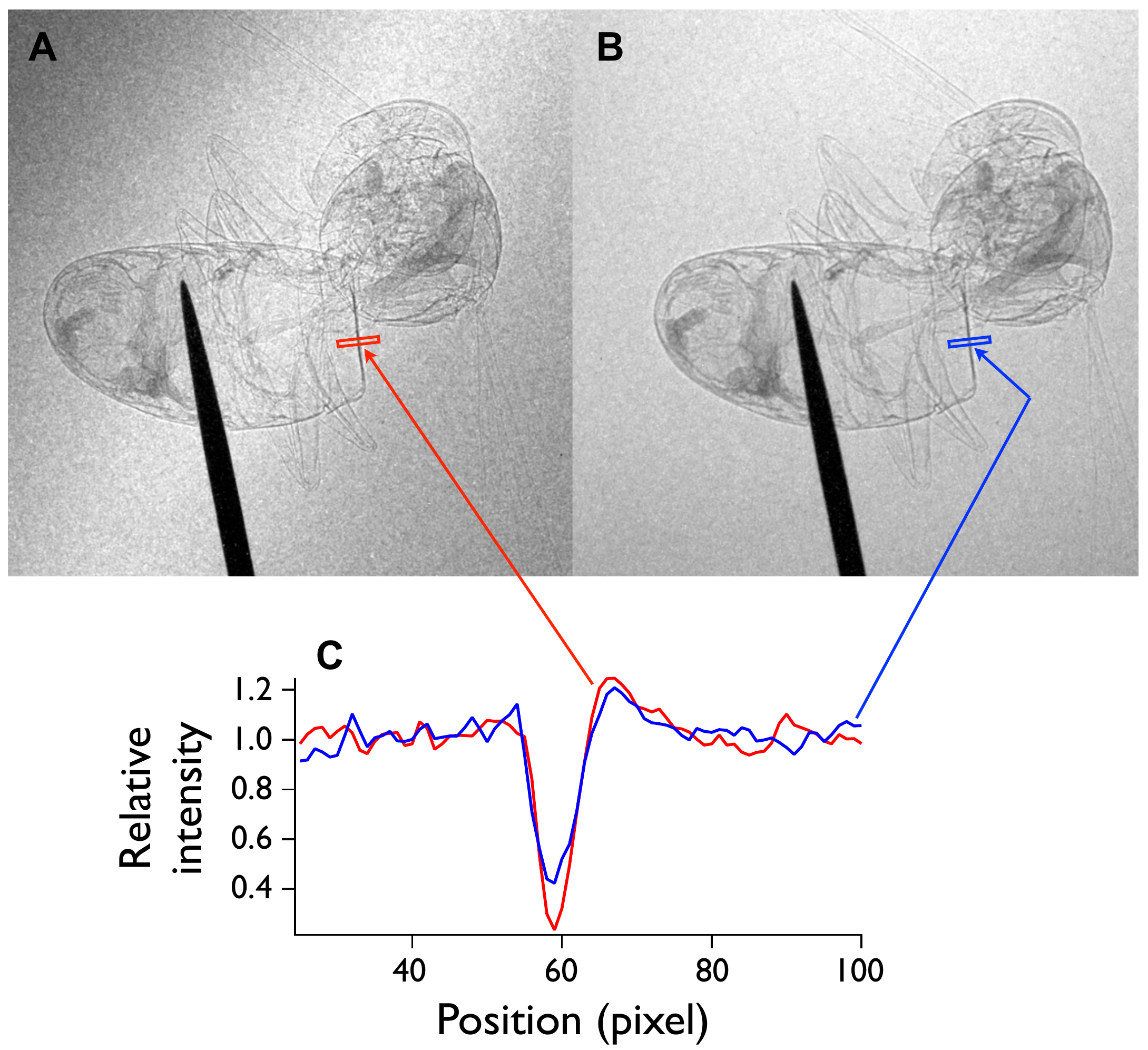}
\caption{Bee imaged with the betatron x-ray beam with an edge line out indicated by the rectangular area: (a) one laser shot; (b) 13 laser shots. (c) Line out of the images taken at the indicated rectangular area. From \textcite{OL2011Fourmaux}.}
\label{chap4fig8c}
\end{figure}
Betatron x-ray radiation has interesting properties for application experiments: high peak spectral brightness, ultrashort duration, very small source size, as well as femtosecond time-scale synchronization in pump-probe experiments. The potential of betatron radiation was demonstrated with the example of single shot phase contrast imaging of biological samples \cite{OL2011Fourmaux, APL2011Kneip}, where the experiment takes advantage of the very small source size and high spatial coherence of the x-ray beam. For an x-ray source size of $\sim1-2$ $\mu$m, the coherence length at 1 m from the source is $\sim 10$ $\mu$m, which allows to perform phase contrast imaging with a compact setup. Figure \ref{chap4fig8c} shows a phase contrast of a bee in single shot [Fig. \ref{chap4fig8c}(a)] as well as with accumulation over 13 shots [Fig. \ref{chap4fig8c}(b)] and the contrast is found to be 68 \% in single shot for the indicated line out  \cite{OL2011Fourmaux}. Betatron x-ray radiation is also a powerful tool to study laser-plasma accelerator physics and for noninvasive measurement of these properties. A novel method has shown the possibility to measure the longitudinal profile of the x-ray emission region, giving insight into the history of the laser-plasma interaction \cite{PRL2011Corde1}. Betatron radiation from optically injected, tunable, and monoenergetic electron bunches have demonstrated the strong correlation between electron and x-ray properties \cite{PRL2011Corde2}, showing that betatron radiation can be used to perform noninvasive measurements of some of the electron beam parameters, such as the normalized transverse emittance of the electron beam \cite{PRSTAB2012Kneip}.

\bigskip
Betatron radiation was also observed in the petawatt regime using a laser delivering 300 J within 600 fs pulses \cite{PRL2008Kneip}. In this parameter regime, the laser pulse is much longer than the plasma period and the physical mechanisms are different. For a long laser pulse ($\tau_L \gg 1/\omega_p$), electrons can be accelerated and wiggled in the SM-LWFA and/or DLA regimes, depending on the laser strength parameter $a_0$. Electrons experience both the transverse focusing force of the channel and the electromagnetic field of the laser. As a consequence, the transverse amplitude of the electron orbits can be significantly increased due to betatron resonance.
The results showed the possibility to produce betatron radiation with a spectrum fitted by the synchrotron function $S(\omega/\omega_c)$ with a critical energy up to $\hbar\omega_c=14.5$ keV [note that in \textcite{PRL2008Kneip}, their convention for $\hbar\omega_c$ leads to a value of 29 keV] and with a divergence of $\sim 1$ rad. In this experiment, the total number of photons was found to be more than $5\times10^8\ \text{photons}/(\text{mrad}^2\: 0.1\%\text{BW})$ in the spectral range $[7$ keV$;12$ keV$]$ and a typical electron charge of 2 nC was observed with electron energies in the 50 MeV range. From these properties, the amplitude of motion and the strength parameter $K$ were deduced to be $\sim30$ $\mu$m and $\sim130$ respectively, in agreement with PIC simulations.

\subsection{Scalings and perspectives}
\label{chap4secE}

The betatron source offers several routes of development in terms of number of photons, spectral range and divergence. According to Sec. \ref{chap4secB}, these quantities scale in the wiggler limit as \cite{EPJD2007Rousse}
\begin{align}
\label{chap4eq7}
&N_\gamma \propto N_eN_\beta K\propto N_eN_\beta\sqrt{\gamma n_e}r_{\beta},\\
\label{chap4eq8}
&\hbar\omega_c\propto \gamma^{1.5}\sqrt{n_e}K\propto \gamma^2n_er_{\beta},\\
\label{chap4eq9}
&\theta_r\propto K/\gamma\propto \sqrt{n_e}r_{\beta}/\sqrt{\gamma},
\end{align}
where $N_\beta$ is the number of betatron periods and $N_e$ is the number of electrons. As discussed in Sec. \ref{chap4secB}, these formulas are valid for time-constant parameters, but can be used with the values of the parameters at the dephasing time for the realistic case of an electron being accelerated in the ion cavity. In addition to these formulas, the scalings of the acceleration process are required. Considering the phenomenological model of the bubble regime described by \textcite{PRL2006Lu, PoP2006Lu, PRSTAB2007Lu}, higher electron energy gain requires lower plasma density. Its scaling is given by $\gamma\simeq\Delta \mathcal{E}/mc^2=(2\omega_L^2/3\omega_p^2)a_0\propto a_0/n_e$. The matched spot size condition reads $k_pw_0=2\sqrt{a_0}$ and the bubble radius should equal the laser pulse waist $r_b=w_0\propto a_0^{1/2}n_e^{-1/2}$. The number of trapped electrons, obtained from an energy balance between the electron beam and the bubble electromagnetic fields, scales as the ionic charge of the bubble: $N_e \propto r_b^3 n_e\propto a_0^{3/2}n_e^{-1/2}$. The dephasing length scales as $L_d\propto a_0^{1/2}n_e^{-3/2}$ and the betatron period as $\lambda_u\propto a_0^{1/2}n_e^{-1}$, such that the number of betatron oscillations obeys $N_\beta\propto n_e^{-1/2}$. Different electron injection mechanisms can lead to different scalings for $r_\beta$, but betatron radiation is optimized when $r_\beta$ is maximum, and this latter is limited by $r_b$. We therefore assume that the transverse size of the electron bunch roughly scales as the bubble radius for optimized conditions for betatron radiation: $r_\beta\propto r_b\propto a_0^{1/2}n_e^{-1/2}$.

The scalings of Eqs. (\ref{chap4eq7}), (\ref{chap4eq8}) and (\ref{chap4eq9}) can then be reformulated as functions of $a_0$ and $n_e$ as
\begin{align}
\label{chap4eq10}
&N_\gamma \propto a_0^{5/2}n_e^{-3/2},\\
\label{chap4eq11}
&\hbar\omega_c\propto a_0^{5/2}n_e^{-3/2},\\
\label{chap4eq12}
&\theta_r\propto n_e^{1/2},
\end{align}
while the needed laser energy $E_{\text{laser}}\propto a_0^2 w_0^2 \tau$ obeys
\begin{equation}
\label{chap4eq13}
E_{\text{laser}}\propto a_0^{7/2}n_e^{-3/2}.
\end{equation}
This expression uses the optimal pulse duration $c\tau=2w_0/3$ corresponding to the situation where the dephasing length $L_d$ matches the pump depletion length $L_{pd}$. These scalings take into account neither the possible interaction between the electron beam and the laser pulse nor beam loading effects, and describe the betatron perspectives in the purely bubble regime. However, they highlight the route toward harder x rays and higher radiated energies: increasing the laser pulse strength parameter and/or decreasing the density. Equation (\ref{chap4eq13}) indicates that it is more efficient in terms of laser energy to decrease the plasma density and not to increase the laser strength parameter. The efficiency $\eta_X$ of the conversion of laser energy into x rays, defined as the ratio of the radiated energy by the laser energy, scales as $a_0^{3/2}n_e^{-3/2}$. The betatron mechanism will become more and more efficient when going to higher and higher photon energies.

Petawatt class lasers and capillaries both allow one to decrease the plasma density. With petawatt lasers producing sub-$100$ fs pulses, the higher laser energy can be focused into a larger focal spot matched to the lower plasma density according to $k_pw_0=2\sqrt{a_0}$. For self-injection to occur, $a_0\sim4-5$ is needed, while the self-guided propagation requires $a_0\gtrsim(n_c/n_e)^{1/5}$, where $n_c=m\epsilon_0\omega_L^2/e^2$ is the critical density (this condition expresses that energy loss at the front of the laser pulse due to pump depletion is higher than the loss due to diffraction) \cite{PRSTAB2007Lu}. Capillaries allow one to guide the laser pulse over distances much larger than the Rayleigh length such that no condition apply on $a_0$. This technique has already been used to accelerate electrons up to 1 GeV in a few centimeters capillary \cite{NatPhys2006Leemans}. Combined with external injection, a strength parameter as small as $a_0=2$ can be used, which permits one again to focus the laser energy into a larger focal spot and to decrease the plasma density. A drawback of these natural scalings is the increase of the x-ray source size and pulse duration as $r_\beta, \tau_r\propto a_0^{1/2}n_e^{-1/2}$. As a consequence, the peak spectral brightness (number of photons divided by the pulse duration, by the solid angle, and by the source size in 0.1\% spectral bandwidth) scales as $a_0/n_e$. Nevertheless, an external injection could provide small duration and therefore better time resolution and brightness for application experiments. 

Using 15 J of laser energy focused into a gas jet of density $n_e=1.1\times10^{18}\:$cm$^{-3}$ with a waist $w_0=21\: \mu$m and compressed to a duration $\tau=48\:$fs (300TW on target), the laser strength parameter is $a_0=4.4$, ensuring self-guiding and self-injection, and trapped electrons reach the maximum energy of 2.4 GeV after $23\:$mm of plasma \cite{PhDRechatin}. Assuming a reasonable value for the betatron amplitude $r_\beta=3\:\mu$m, we expect $K\sim 30$ and x-ray photons emitted up to $\hbar\omega_c \sim 400$ keV, collimated within a typical opening angle of 6 mrad, and containing $N_\gamma  \sim$ 1 photon per electron and per betatron period. On the other hand, a configuration with 3 J of laser energy, $n_e=5.1\times10^{17}\:$cm$^{-3}$, $w_0=21\:\mu$m, $\tau=47\:$fs (60TW on target) corresponding to $a_0=2$ and assisted by external guiding (capillaries) and external injection [colliding pulse injection \cite{Nature2006Faure} or density gradient injection \cite{PRL2008Geddes}] provides the same output electron energy of 2.4 GeV after 52 mm of plasma. With the same value of betatron amplitude $r_\beta=3\:\mu$m, x rays are emitted with the following properties: $K\sim20$, $\hbar\omega_c\sim200$ keV, $\theta_r\sim4\:$mrad, and $N_\gamma \sim$ 0.7 photon per electron and per betatron period. Compared to the self-guided self-injected situation, the number of trapped electrons $N_e$ is 2 times smaller, the number of betatron periods $N_\beta$ is 1.4 times higher (see their respective scalings above), and the total number of emitted photons is 2 times smaller, but within a better divergence ($\theta_r$ is 1.4 times smaller) and for a needed laser power 5 times smaller. For such configurations, the number of betatron photons could reasonably reach the level of $10^{10}$ photons.

\bigskip
Another route toward brighter betatron radiation and higher photon energies makes use of density-tailored plasmas \cite{PRL2007Layer} as recently suggested \cite{PoP2008TaPhuoc2}. This method relies on the control of the electron orbits by varying the density, and therefore the forces acting on the electrons along the propagation. For appropriately chosen density modulations, numerical simulations show that the amplitude of the betatron oscillations $r_\beta$ can be significantly increased. The basic idea is to freeze the wakefield during a short duration in order to let the electron beam defocus and acquire larger betatron amplitude. An additional effect is that, for large betatron amplitude, the mean longitudinal velocity is lowered and electrons dephase slower and can reach higher energies. In an optimistic configuration described by \textcite{PoP2008TaPhuoc2}, the critical energy of the radiation can be increased by a factor of $\approx9$,  the number of photons by a factor of $\approx5$, but with an increase of the divergence by a factor $\approx4$.

\section{Plasma Accelerator and Conventional Undulator: Synchrotron Radiation}
\label{chap5}

\begin{figure}
\includegraphics[width=8.5cm]{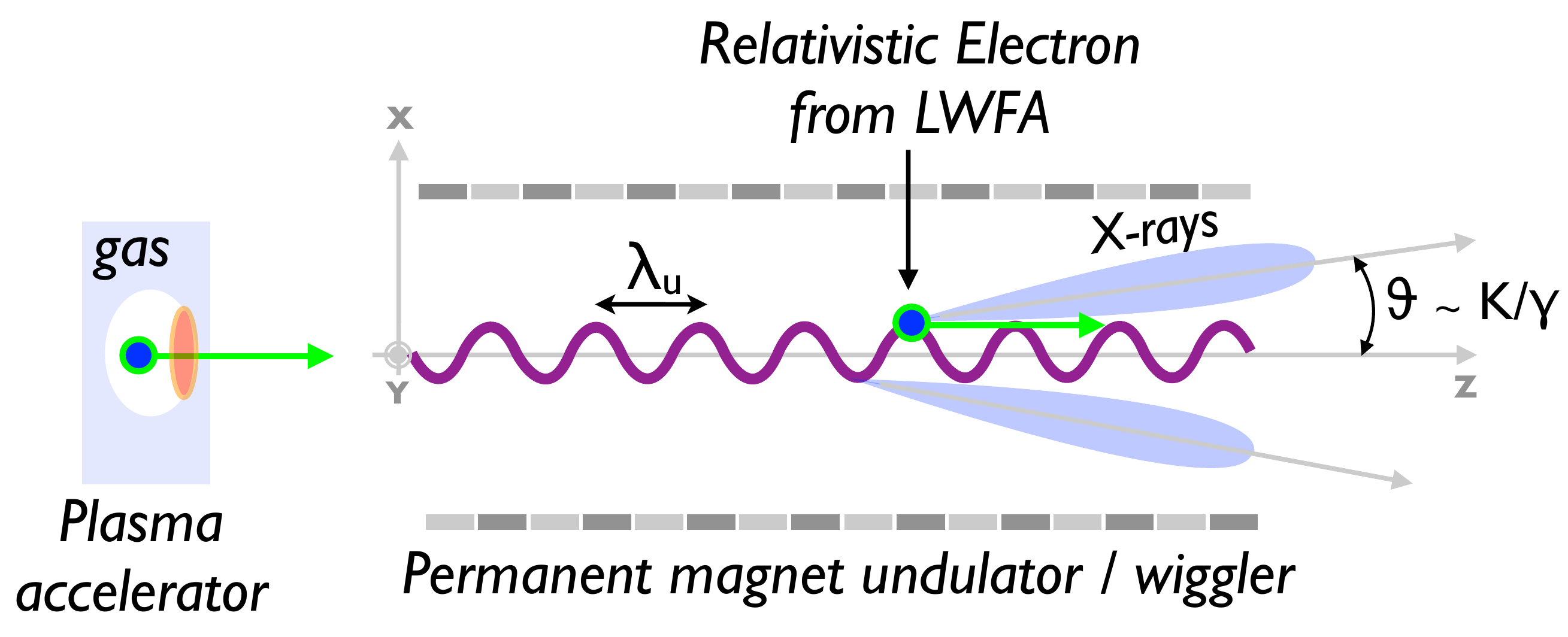}
\caption{Schematic of the undulator source. A laser-plasma accelerated electron beam is injected in a conventional undulator. Electrons oscillate transversally and emit x rays due to this motion.}
\label{chap5fig1}
\end{figure}
Synchrotron radiation can be produced from laser-plasma accelerated electrons propagating and oscillating in a conventional undulator or wiggler. A conventional undulator or wiggler is a periodical structure of magnets generating a periodic static magnetic field \cite{Clarke}. Figure \ref{chap5fig1} represents the principle of the scheme. Electrons are first accelerated in a plasma accelerator (gas jet, capillary or steady-state-flow gas cell) and then transported into an undulator or wiggler. The size of such a device is on the meter scale, with a typical magnetic period in the centimeter range. Moreover, the values of $\lambda_u$, technologically limited, imply that GeV electrons are necessary to produce radiation in the x-ray range. These represent the disadvantages of this scheme, which however is of crucial interest because it represents a possible route toward the production of a compact free-electron laser based on a laser-plasma accelerator \cite{APB2007Gruner, NatPhys2008Nakajima, NatPhys2008Malka} (see Sec. \ref{chap7}).
In the following sections, the electron orbits in undulators and the main properties of synchrotron radiation produced from laser-plasma electron bunches are described in a presently realizable case. After a summary of the experimental results, we conclude with the short term developments foreseen. For detailed properties of undulator radiation, such as the angular distribution of individual undulator harmonics, the different undulator types (planar or helical), the polarization properties and the spatial or temporal coherence properties, see \textcite{Wiedemann, Clarke, Accelerator}.

\subsection{Electron motion}
\label{chap5secA}

We consider a laser wakefield accelerator producing a monoenergetic electron bunch of energy $\mathcal{E}=\gamma_imc^2$ with velocity directed in the $\vec{e}_z$ axis. The simplest model to describe the electron dynamics in the undulator is to approximate the static magnetic field near the undulator axis by
\begin{equation}
\vec{B}=B_0\sin(k_uz)\vec{e}_y,
\end{equation}
where $k_u=2\pi/\lambda_u$ with $\:\lambda_u$ the magnetic period of the undulator. The equation of motion for a test electron in this idealized model is given by
\begin{equation}
\frac{d\vec{p}}{dt}=-e\vec{v}\times\vec{B}.
\end{equation}
The normalized Hamiltonian describing the electron dynamics is given by
\begin{equation}
\hat{\mathcal{H}}(\hat{\vec{r}},\hat{\vec{P}},\hat{t})=\gamma=\sqrt{1+(\hat{\vec{P}}+\vec{a})^2},
\end{equation}
where $\hat{\vec{P}}$ is the normalized canonical electron momentum and $\vec{a}=-a_0\cos(k_uz)\vec{e}_x$ is the normalized vector potential verifying $\vec{B}=(mc/e) \nabla \times \vec{a}$. $\hat{\mathcal{H}}$  does not depend on $\hat{x}$, $\hat{y}$, and $\hat{t}$. The transverse canonical momentum $\hat{\vec{P}}_{\bot}$ and the normalized energy $\gamma$ are therefore conserved. Integrating these constants of motion with the initial conditions $\hat{\vec{P}}_{\bot}=0$ and $\gamma=\gamma_i$ leads to, in physical units,
\begin{align}
&\gamma=\text{cte}=\gamma_i,\\
&x(z)\simeq -\frac{K}{\gamma k_u}\sin(k_uz),\\
&K=a_0=\frac{eB_0}{k_umc}.
\end{align}
The general description developed in Sec. \ref{chap2} is recovered and the practical expression of $K$ in terms of experimental parameters and fundamental constants has been obtained. In particular, $K$ does not depend on $\gamma$ and the radiation regime (undulator or wiggler) is the same for all electron energies. The strength parameter $K$ is determined by the amplitude of the magnetic field $B_0$ and the period $\lambda_u$ and is, in practical units, given by
\begin{equation}
K=0.934 \lambda_u[\text{cm}]B_0[\text{T}]. 
\end{equation}

\subsection{Radiation properties}
\label{chap5secB}

The radiation properties are the same as in Sec. \ref{chap2}. The spectrum of the emitted radiation depends on $K$. For a small amplitude of oscillation $K\ll1$, radiation is emitted at the fundamental photon energy $\hbar\omega$ with a narrow bandwidth in the forward direction ($\theta=0$). As $K\rightarrow1$ harmonics of the fundamental start to appear in the spectrum, and for $K\gg1$ the spectrum contains many closely spaced harmonics and extends up to a critical energy $\hbar\omega_c$. These quantities are given by
\begin{align}
\nonumber
\hbar\omega &= (2 \gamma^2 hc/\lambda_u) / (1+K^2/2) \;\;\;\; \text{for} \: K <1,\\
\hbar\omega_c &=  \frac{3}{2}K \gamma^2 hc/\lambda_u \;\;\;\; \text{for} \: K\gg1.
\end{align}
In practical units, this gives
\begin{align}
\nonumber
\hbar\omega[\text{eV}] &= 2.48\times10^{-4}\gamma^2/\lambda_u\text{[cm]} \;\;\;\; \text{for} \: K \ll1,\\
\hbar\omega_c[\text{eV}] &= 1.74\times10^{-4}\gamma^2B_0\text{[T]} \;\;\;\; \text{for} \: K\gg1.
\end{align}

The radiation is collimated within a cone of typical opening angle $\theta_{r}=1/\gamma$ in the undulator case. For a wiggler, the radiation is collimated within a typical opening angle $K/\gamma$ in the electron motion plane $(\vec{e}_x,\vec{e}_z)$ and $1/\gamma$ in the orthogonal plane $(\vec{e}_y,\vec{e}_z)$.

The radiation temporal profile is simply given by the convolution between the electron bunch temporal profile and the radiation profile from a single electron. For x rays, the radiation length from a single electron $l_r{_{|_{N_e=1}}}=N\lambda$ is in the nanometer range, much smaller than the typical electron bunch length which is in the micron range, and therefore $\tau_r{_{|_{N_e}}}\simeq\tau_b$. However, for UV and XUV radiation, the slippage between the radiation and the electron bunch is not necessarily negligible, and the radiation duration can be stretched by the single electron radiation duration. The electron bunch duration depends on the acceleration mechanism and on its transport from its source to the undulator. Presently, the shortest electron bunches are obtained using a laser-driven plasma-based accelerator with an external optical injection mechanism \cite{Nature2006Faure, PPCF2007Faure, PoP2008Davoine, PoP2009Malka, PRL2009Rechatin1}. In this scheme, electron bunches with rms (root-mean-square) duration as small as 1.5 fs can be produced \cite{NatPhys2011Lundh}. 
However, the necessary transport of the electron bunch in the undulator can deteriorate the duration (and the transverse emittance), because different electrons in the bunch travel different distances. For example, considering usual transport devices (quadrupoles), placed at $\sim$ 1 m from the laser-plasma accelerator, an initially 1-$\mu$m-long bunch has its length increased to $\sim 10\;\mu$m during its transport to an undulator placed several meters after the laser-plasma accelerator. An ultracompact transport system, placed as closed as possible to the plasma source, is required to avoid such a deterioration.

The number of emitted photons follows the expressions given in Sec. \ref{chap2}.  In practical units, the number of photons emitted per period and per electron (at the mean photon energy $\langle\hbar\omega\rangle=0.3\hbar \omega_c$ for the wiggler limit) is given by
\begin{align}
\nonumber
N_\gamma  &= 1.53 \times 10^{-2} K^2 \;\;\;\; \text{for} \: K<1,\\
N_\gamma  &= 3.31 \times 10^{-2}K \;\;\;\; \text{for} \: K\gg1.
\end{align}

From these expressions the features of the produced radiation can be obtained for a realizable case: an electron bunch at an energy of 1 GeV containing 100 pC and an undulator of 100 periods with $\lambda_u = 1$ cm and $K=0.5$. The emitted radiation has a narrow band spectrum centered at $E_{X} \sim 1 $ keV in the forward direction ($\theta=0$), is collimated within a typical opening angle of $\sim 500$ $\mu$rad, and contains $\sim 2 \times 10^8$ photons.

\subsection{Numerical results}
\label{chap5secC}

\begin{figure}
\includegraphics[width=8.5cm]{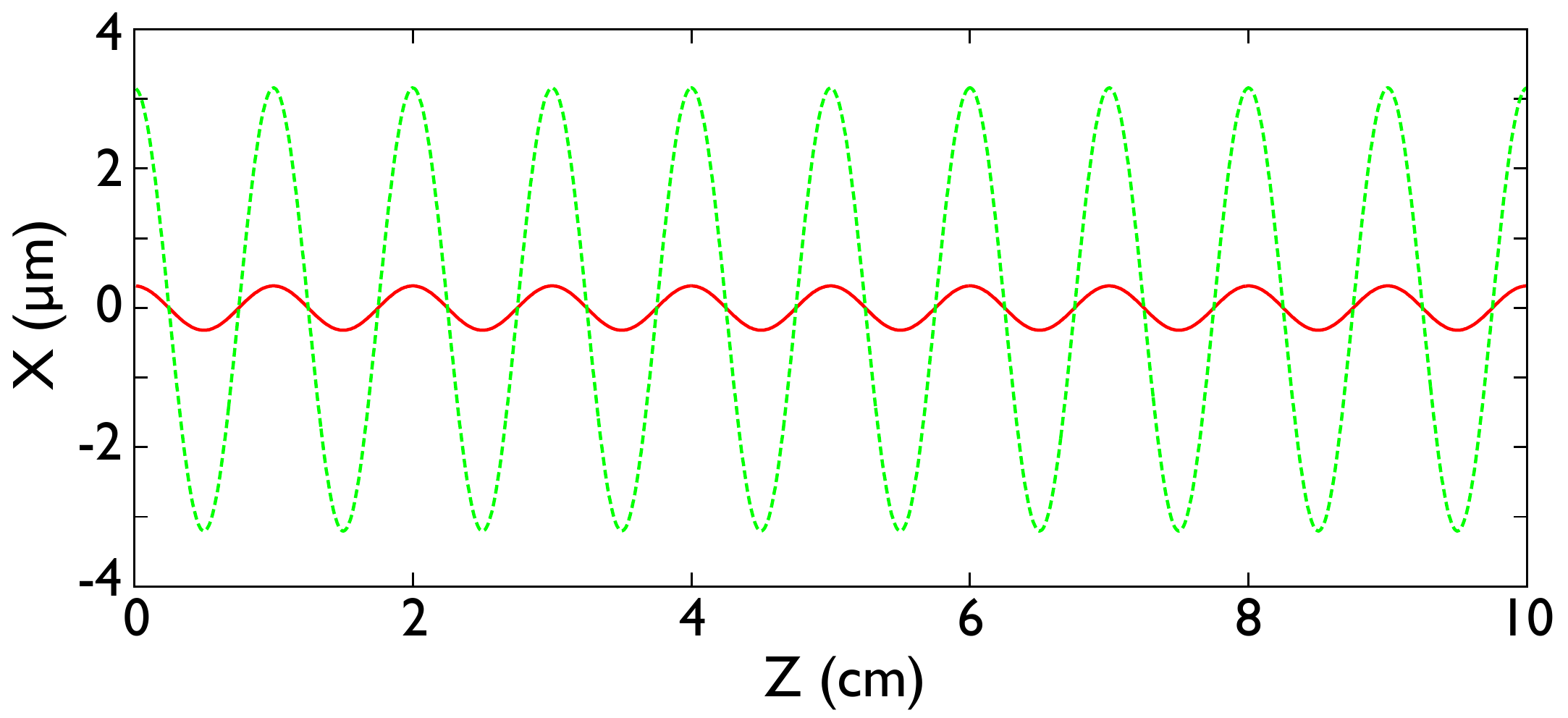}
\caption{Electron trajectories. The orbit is plotted as a red solid line for the undulator case ($K=0.2$) and green dashed line for the wiggler case ($K=2$). At the entrance, the electron has $\gamma=1000$ and $\vec{p}=\sqrt{\gamma^2-1}\vec{e}_z\:$ in both situations. The undulator or wiggler is ten periods long.}
\label{chap5fig2}
\end{figure}
In this section, the electron motion and the emitted radiation are calculated numerically. Both the cases of an undulator and a wiggler are considered, with strength parameter of $K=0.2$ and 2, respectively. The magnetic structure has a period $\lambda_u=1$ cm and the test electron propagates in the $\vec{e}_z$ direction with $\gamma = 1000$ ($\simeq 500$ MeV). Figure \ref{chap5fig2} shows the electron motion along $N=10$ oscillation periods (the number of periods of an undulator or wiggler is usually on the order of 100). The motion, with $\gamma$ constant, consists of a transverse oscillation at a period $\lambda_u$ combined with a longitudinal drift. The amplitude of the transverse motion is given by $\lambda_u K/2\pi\gamma$ (here $\simeq 1.6 \times K[\mu\text{m}]$). For $K=0.2$ and 2 the transverse excursions are $0.32$ and $3.2$ $\mu$m, respectively.

\begin{figure}[!t]
\includegraphics[width=8.5cm]{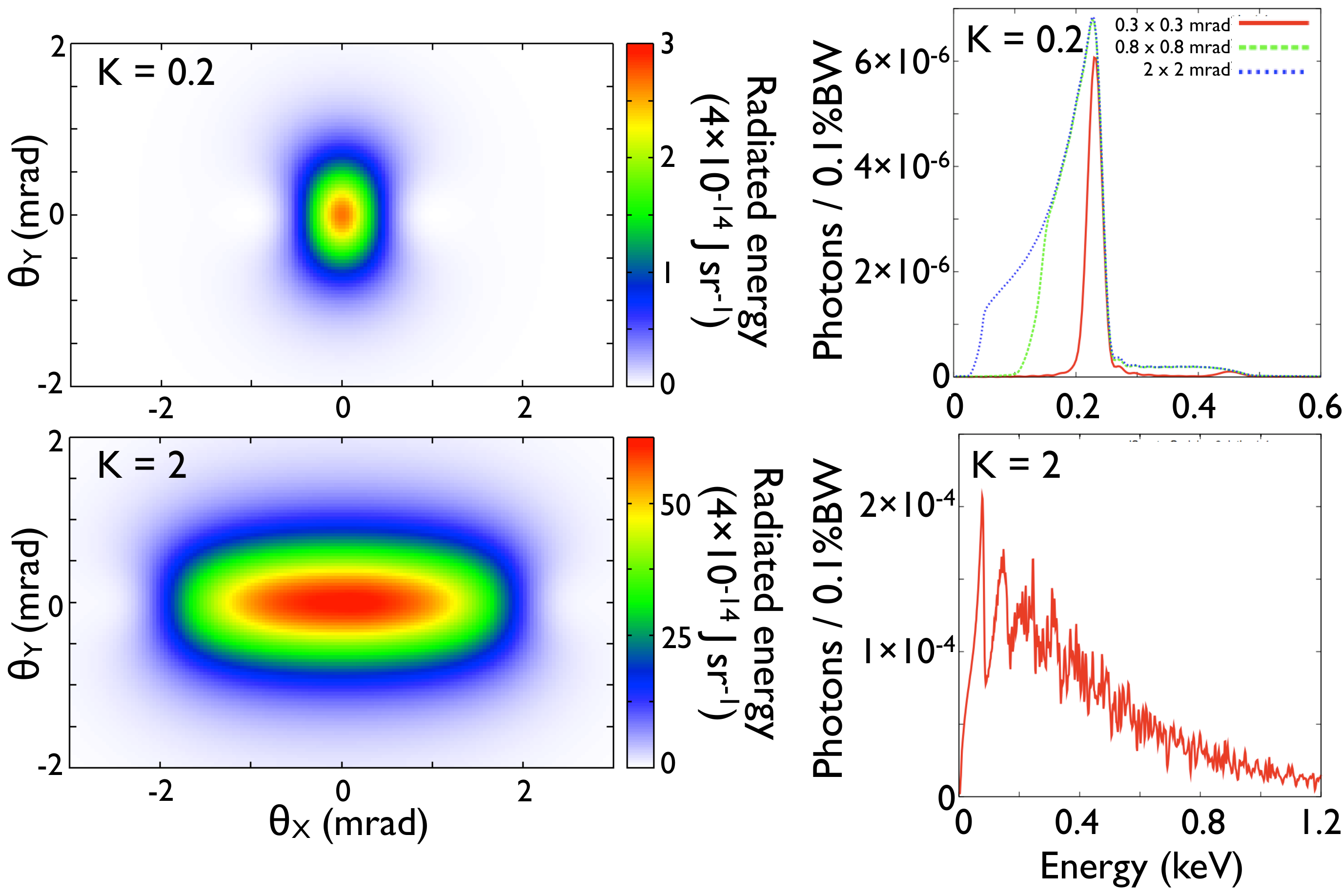}
\caption{On the left is represented the angular distributions of the radiation (radiated energy per unit solid angle) and the corresponding spectra are shown on the right (in number of photons per 0.1\% bandwidth, per electron, and per shot). The undulator case, $K=0.2$, is on the top whereas the wiggler case, $K=2$, is on the bottom. In the first situation, three different solid angles of integration have been considered to highlight the stretching of the spectrum. In the wiggler situation, the integration is performed over the total angular distribution.}
\label{chap5fig3}
\end{figure}
The spatial distributions of the radiation produced by the electron undergoing these trajectories are presented on the left of Fig. \ref{chap5fig3}. For $K=0.2$ (top), the radiation has a divergence of typical opening angle $1/\gamma \sim 1$ mrad. For $K=2$ (bottom), the radiation has a divergence of typical opening angle $K/\gamma \sim 2$ mrad in the plane of the electron motion and $1/\gamma \sim 1$ mrad in the orthogonal direction. The right part of Fig. \ref{chap5fig3} represents the spectra of the emitted radiation for each situation (undulator on the top and wiggler on the bottom). For the undulator case, the spectrum is plotted for three different solid angles of integration ($0.3\times0.3$, $0.8\times0.8$, and $2\times2$ mrad$^2$), whereas for the wiggler case it is integrated over the overall spatial distribution. For $K=0.2$ the spectrum is nearly monochromatic at the energy $\hbar\omega \sim 200$ eV in the forward direction, with a width of the emission line given by $\Delta \lambda = \lambda / N$ (where $N=10$ is the number of oscillation periods), and it is stretched after integration over the emission angles due to the angular dependence of the radiated wavelength $\lambda \simeq (\lambda_u/2\gamma^2)(1+\frac{K^2}{2}+\gamma^2\theta^2)$ [see Sec. \ref{chap2}, Eq. (\ref{chap2eq3})]. For $K=2$ the spectrum becomes broadband with a critical energy of around 300 eV. An estimate of the total photon number can be obtained on the basis of the number of electrons that are currently accelerated in a laser-plasma accelerator. Assuming $1 \times 10^{8}$ electrons gives $\sim 10^{5}$ photons at 200 eV within a spectral bandwidth of 10\% in the undulator case (selection of the $0.3\times0.3$ mrad$^2$ solid angle). In the wiggler case there are $\sim 5 \times 10^{3}\ \text{photons}/0.1\%\text{BW}$ at 600 eV and a total photon number in the range $10^{7-8}$.

\subsection{Experimental results}
\label{chap5secD}

Synchrotron facilities have been developed worldwide and have been providing x-ray radiation to users for several decades. These installations are very robust and the x-ray beams have high quality. They are based on conventional accelerator technology.

The first demonstration of the production of synchrotron radiation from laser-accelerated electrons was performed by \textcite{NatPhys2008Schlenvoigt, IEEE2008Schlenvoigt}.
In this experiment, laser-produced electron bunches between 55 and 75 MeV were injected, without transport,  into a 1-m-long undulator having a period $\lambda_u=2\:\text{cm}$ and a strength parameter $K=0.6$. They obtained synchrotron radiation in the visible and infrared part of the spectrum (the wavelength was in the range $700-1000$ nm) and estimated a peak spectral brightness of $6.5\times10^{16}\ \text{photons}/(\text{s}\: \text{mrad}^2\: \text{mm}^2\: 0.1\%\text{BW})$ and a total number of photons of $2.8\times 10^{5}$. The radiation wavelength was observed to scale with the electron bunch energy as expected.

More recently, synchrotron radiation from laser-plasma accelerated electrons was produced up to $\sim 130$ eV \cite{NatPhys2009Fuchs}. This experiment represents an important advance compared to the first demonstration. First, a stable electron beam was produced by the interaction of a laser at $\sim2\times10^{18}$ W.cm$^{-2}$ with a 15-mm-long hydrogen-filled steady-state-flow gas cell \cite{PRL2008Osterhoff}. The electron bunch was in the 150 - 220 MeV energy range and contained 30 pC. Second, a pair of miniature permanent-magnet quadrupole lenses was implemented in order to make the electron beam collimated at a selected energy and to transport it through a 30-cm-long undulator having a period $\lambda_u=5$ mm and a strength parameter $K=0.55$ \cite{PRSTAB2007Eichner}. Thanks to the high energy of the electron and the short period of the undulator, the radiation was produced down to the XUV range with a tunable wavelength. Note that this tunability resides in the possibility to select, with the magnetic lenses, a specific electron energy in the 150 - 220 MeV range (electrons with different energies are not collimated and will not produce a collimated radiation beam). The radiation was analyzed using a spectrometer based on a transmission grating. The fundamental wavelength ($\lambda=18$ nm) as well as the second harmonic were measured. In addition, the parabolic dependence of the radiation wavelength on the angular direction $\lambda = (\lambda_u/2\gamma^2)(1+K^2/2+\gamma^2\theta^2)$ was observed with good agreement. This soft-x-ray undulator source delivers $\sim 10^5\ \text{photons}$ per shot integrated over the fundamental within a detection cone of  $\pm$0.7 mrad, corresponding to a relative bandwidth of 30\% (FWHM). The peak spectral brightness was estimated to be $\sim1.3\times10^{17}\ \text{photons}/(\text{s}\: \text{mrad}^2\: \text{mm}^2\: 0.1\%\text{BW})$.

\subsection{Perspectives}
\label{chap5secE}

The experiments performed open perspectives for the production of a tunable synchrotron-type x-ray source. However, further progress is necessary to produce tunable and reproducible monoenergetic electron bunches in the GeV range. The foreseen possibilities to achieve these features rely on the use of higher power lasers (petawatt) and/or capillaries. This gain in laser power compared to actual systems would allow one to decrease the plasma density, increase the acceleration length, and as a consequence increase the electron energy. The use of an external injection will be required as well to easily control the electron energy (optical injection \cite{Nature2006Faure} or plasma density gradient injection \cite{PRL2008Geddes}, as presented in Sec. \ref{chap3}). Finally, a compact electron transport system from the plasma source to the undulator has to be implemented in order to avoid deterioration of the electron beam parameters and to reduce the radiation source size.

In the short term, this approach will allow university-scale laboratories to access x-ray sources with angstrom wavelengths and sub-10-fs pulse durations for four-dimensional imaging with atomic resolution.
Besides this future development, the main and more challenging perspective relies on the possibility to produce a free-electron laser based on a laser-plasma accelerator combined with an undulator. This radiation source has a brightness orders of magnitude higher than a synchrotron, thanks to the coherent addition of radiation from each electron. Electron bunches produced with laser-plasma accelerators are interesting candidates for a FEL source since such bunches are naturally femtosecond and have high current. In Sec. \ref{chap7}, we discuss the principle of the free-electron laser and its possible realization using electrons from laser-plasma accelerators which could have revolutionary impacts in many fields of science, technology, and medicine.  

\section{Electromagnetic Wave Undulator: Nonlinear Thomson Scattering and Thomson Backscattering}
\label{chap6}

In this section, the production of femtosecond x-ray and gamma-ray beams from electrons oscillating in an electromagnetic wave is reviewed. The involved radiative mechanism is referred to as Compton scattering. In the framework of quantum electrodynamics, it corresponds to the absorption of one (linear) or several (nonlinear) photons by an electron and to the emission of a single photon. When, in the rest frame, the electron experiences a negligible recoil (which happens when $\hbar\omega\ll mc^2$ in the rest frame), the mechanism is referred to as Thomson scattering, the low-energy limit of Compton scattering. In the following, we consider Thomson scattering and use a classical description, assuming that quantum and radiation reaction effects are negligible (see Sec. \ref{chap2secRR}).

At the laser-plasma interaction, two schemes of femtosecond x-ray or gamma-ray sources based on Thomson scattering have been proposed and demonstrated. Based on the same principle, the two methods differ by the initial energy of the electron and the intensity of the laser pulse scattering off the electrons.

In the first scheme, electrons are initially at rest and laser wakefield acceleration is not invoked. Here, Thomson scattering occurs in a highly nonlinear regime. Indeed, for $a_0\gg1$, electrons have a highly non linear motion and the emitted radiation consists of high-order harmonics. The harmonic spectrum can extend up to the x-ray range. This scheme offers simplicity and the possibility to produce a large x-ray photon flux because the number of electrons participating in the emission is on the order of the number of electrons in the focal volume. However, reaching the keV energy range requires very high laser strength parameters, typically $a_0 >10$.

The second scheme relies on Thomson scattering an intense laser pulse off a counterpropagating relativistic electron beam. Here the x-ray range is reached even with modest-energy electrons thanks to two successive Doppler shifts. In the average rest frame of the electrons, moving with a relativistic factor $\bar{\gamma}$ with respect to the laboratory frame, the frequency of the incident electromagnetic wave (the laser pulse) becomes $\omega^{\prime}_i=2\bar{\gamma}\omega_i$. The light is scattered by the electrons at the same frequency $\omega^{\prime}_r=\omega^{\prime}_i$ (for a low intensity electromagnetic wave corresponding to the linear Thomson scattering and the undulator regime) or at harmonics $\omega^{\prime}_r=n\omega^{\prime}_i$ (for high intensity corresponding to nonlinear Thomson scattering and the wiggler regime). When observed in the laboratory frame, this reflected wave is Doppler shifted and its frequency becomes $\omega_r=4\bar{\gamma}^2\omega_i$ (linear Thomson backscattering) or its harmonics (nonlinear Thomson backscattering). This scheme offers the possibility to produce x rays in the keV range even with modest energy electrons (tens of MeV) or gamma rays if 100 MeV range electrons are used. 

In the following sections the two schemes are discussed. For both, electron orbits, radiation properties, numerical, and experimental results are presented. 

\subsection{Nonlinear Thomson scattering}

\begin{figure}
\includegraphics[width=8.5cm]{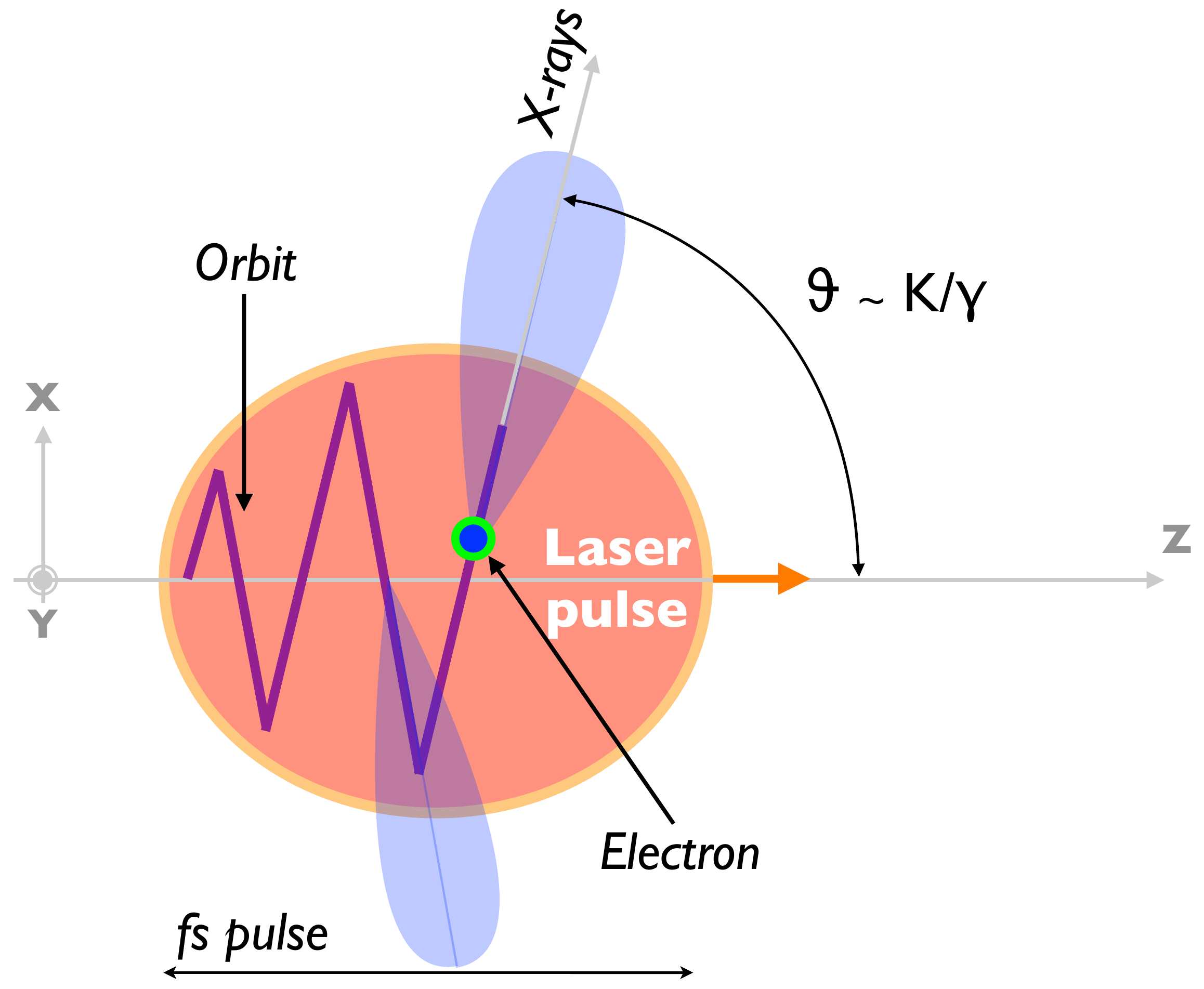}
\caption{Schematic of the nonlinear Thomson scattering source. A free electron is submitted to a relativistic laser pulse ($a_0\gg1$). It is simultaneously accelerated and wiggled, and emits XUV radiation due to this motion. This illustrates the motion of an electron in the case of a linearly polarized laser pulse.}
\label{corde_fig33}
\end{figure}

In this section, the production of femtosecond x-ray radiation from electrons initially at rest is discussed. In this scheme, Thomson scattering occurs in a strongly nonlinear regime and electrons are directly accelerated and wiggled within an intense laser field. This scheme offers the possibility to produce a large photon flux because the number of electrons participating in emission can be much larger than the number of electrons trapped and accelerated in a laser-plasma accelerator. It is of the order of the number of electrons in the focal volume. However, producing x-ray radiation requires  a laser strength parameter $a_0$ much greater than unity. Indeed, in that regime, the term $-e\vec{v}\times\vec{B}$ of the Lorentz force, negligible at low intensity, becomes comparable to the $-e\vec{E}$ component and the motion is a nonlinear function of the driving field in addition to becoming relativistic. The electron motion is no longer harmonic and the radiation emitted consists of high-order harmonics forming a broadband spectrum that extends up to the x-ray range. The spectrum shifts to higher energies as $a_0$ increases. This radiative mechanism is commonly called nonlinear or relativistic Thomson scattering. The principle of this source is displayed in Fig. \ref{corde_fig33}. It simply consists of focusing an intense laser on a target (here an underdense plasma is considered).

In the following sections, the properties of nonlinear Thomson scattering \footnote{For an overview of the subject see \textcite{PRE1993Esarey, PR1962Vachaspati, PR1964Brown, PRD1970Sarachik, PRA1989Bardsley, IEEE1993Esarey, IEEE1993Castillo, PRA1996Salamin, IEEE1997Gibbon, OC1997Shen, LPB1999Ueshima, PoP2002He, PoP2003Lau, PoP2006Tian}.} are reviewed. The electron orbit is derived in the case of a circularly polarized laser pulse, because it leads to a situation where the concept of wiggling, developed in Sec. \ref{chap2}, can be applied. For a linearly polarized laser pulse, analytical results can be easily obtained for the electron orbit, but not for the radiation features. In that latter case, the electron movement does not correspond to a transverse wiggling around a relativistic drift motion along the propagation axis. The electron performs violent accelerations from $\gamma=1$ to $\gamma=1+a_0^2/2$ and decelerations to $\gamma=1$ for each half period of the movement. The emitted radiation comes from the longitudinal acceleration, and not from the transverse one anymore. Hence, in the Secs. \ref{chap6secAsec1} and \ref{chap6secAsec2} analytical expressions are derived for the circular polarization but the linear polarization will be discussed in the numerical Sec. \ref{chap6secAsec3}. After a brief summary of the experimental results obtained, we finally conclude with the short-term developments foreseen.

\subsubsection{Electron orbit in an intense laser pulse}
\label{chap6secAsec1}

We consider a test electron initially at rest ($\gamma_i=1$) submitted to an intense laser pulse modeled by a circularly polarized plane electromagnetic wave propagating along the $\vec{e}_z$ axis, with a wave vector $\vec{k_i}=2\pi/\lambda_L\:\vec{e}_z$ and a frequency $\omega_i=2\pi c/\lambda_L$, and with a normalized vector potential given by
\begin{equation}
\vec{a}=a_0\left [ \frac{1}{\sqrt{2}}\cos(\omega_it-k_iz)\vec{e}_x+\frac{1}{\sqrt{2}}\sin(\omega_it-k_iz)\vec{e}_y \right ].
\end{equation}
The equation of motion of the test electron is
 \begin{equation}
\label{chap6eq1}
\frac{d\vec{p}}{dt}=-e(\vec{E}+\vec{v}\times\vec{B}).
\end{equation}
All quantities with a hat are normalized by the following choice of units: $m=c=e=\omega_i=1$. The Hamiltonian describing the test electron dynamics in the electromagnetic wave is written as
\begin{equation}
\label{chap6eq2}
\hat{\mathcal{H}}(\hat{\vec{r}},\hat{\vec{P}},\hat{t})=\gamma=\sqrt{1+\hat{\vec{p}}\:^2}=\sqrt{1+(\hat{\vec{P}}+\vec{a})^2}.
\end{equation}
This system of an electron in a electromagnetic plane wave is integrable and action-angle variables can be found to solve the electron motion. Here, for simplicity, we analyze the symmetries of the system to directly yield the conserved quantities according to Noether theorem.
The Hamiltonian $\hat{\mathcal{H}}$ depends on the canonical momentum $\hat{\vec{P}}$ and on the potential vector $\vec{a}(\varphi)$ through the variable $\varphi=\hat{t}-\hat{z}$. Hence $\hat{\mathcal{H}}$ is independent of $\hat{x}$ and $\hat{y}$, which implies that the transverse canonical momentum is a constant of motion. The first constant of motion reads
\begin{equation}
\hat{\vec{P}}_{\bot}=\hat{\vec{p}}_{\bot}-\vec{a}=\vec{0}.
\end{equation}
In addition, $\hat{\mathcal{H}}$ depends on $\hat{t}$ and $\hat{z}$ only through $\varphi=\hat{t}-\hat{z}$. Thus $\partial\hat{\mathcal{H}}/\partial \hat{t}=-\partial\hat{\mathcal{H}}/\partial \hat{z}$, which leads to the second constant of motion $\text{C}$,
\begin{equation}
\gamma-\hat{p}_z=\text{C}.
\end{equation}
For an electron initially at rest, $\text{C}=1$, and the test electron orbit is obtained by integrating the constants of motion,
\begin{align}
\hat{x}(\varphi) &= \frac{a_0}{\sqrt{2}}\sin(\varphi),\\
\hat{y}(\varphi) &= -\frac{a_0}{\sqrt{2}}\cos(\varphi),\\
\hat{z}(\varphi) &= \frac{a_0^2}{4}\varphi,\\
\gamma &= 1+\frac{a_0^2}{4}.
\end{align}
The electron motion in the laser field consists of a drift in the longitudinal direction combined with a transverse oscillation. The electron orbit is mainly longitudinal and relativistic for $a_0>1$ while it is essentially transverse and nonrelativistic for $a_0<1$. For $a_0<1$, the test electron motion is a simple nonrelativistic dipole motion which is not relevant for the production of x rays since radiation is emitted at the same wavelength as the incident field. For $a_0>1$, the longitudinal motion becomes relativistic and nonlinear effects and Doppler shift occur. In the following, the interesting nonlinear case $a_0>1$, which results in the generation of smaller wavelength radiation than the incident laser pulse, is studied.

The period of the motion is given by
\begin{equation}
\lambda_u = \frac{a_0^2}{4}\lambda_L.
\end{equation}
The electron orbit is similar to the standard orbits discussed in Sec. \ref{chap2} since the normalized energy $\gamma$ is constant. However, the trajectory takes place in three dimensions and is helical. The qualitative regime and the angular distribution can be derived from the parameters $K_X=\gamma\psi_X$ and $K_Y=\gamma\psi_Y$ which are given by
\begin{equation}
K_X = K_Y = \frac{4}{\sqrt{2}a_0}(1+a_0^2/4)\simeq a_0/\sqrt{2},
\end{equation}
where $\psi_X$ and $\psi_Y$ are the maximum deflection angles of the orbits in the $\vec{e}_x$ and $\vec{e}_y$ directions.

\subsubsection{Radiation properties}
\label{chap6secAsec2}

For $a_0>1$, the relativistic electron motion described above leads to the emission of nonlinear Thomson scattering radiation whose features can be calculated within the frame of the general formalism of the radiation from a moving charge. As seen in Sec. \ref{chap2}, the spectrum of the produced radiation depends on the parameter $K$. According to the above expressions of $K_X$ and $K_Y$, the motion is driven in the wiggler regime since the nonlinear case $a_0>1$ is considered. The spectrum is broadband and extends up to the critical energy
\begin{equation}
\hbar\omega_c = \frac{3}{2}\gamma^3 \hbar \frac{c}{\rho},
\end{equation}
where $\rho= (\lambda_L/2\pi)(a_0/\sqrt{2}+\sqrt{2}a_0^3/16)\simeq(\lambda_L/2\pi)\times\sqrt{2}a_0^3/16$ is the instantaneous radius of curvature of the electron orbit obtained from the trajectory. Basically, the critical energy grows as $a_0^3$ for $a_0 \gg 1$. In practical units, it reads for $a_0 \gg 1$
\begin{equation}
\hbar\omega_c\text{[eV]}= 0.3 \frac{a_0^3}{\lambda_L[\mu\text{m}]}.
\end{equation}

Being an emission from a relativistically moving electron, nonlinear Thomson scattering radiation is emitted in the direction of the electron velocity and has the same symmetry as the electron orbit. For circular polarization, the radiation is emitted symmetrically around the laser propagation axis at an angle $\theta\simeq2\sqrt{2}/a_0$, within a typical angular width $\Delta \theta = 1/\gamma \simeq 4/a_0^2$.

The number of photons can be estimated by integrating, over one oscillation period, the expression of the radiated power $P(t)=(e^2/6\pi\epsilon_0 c)\gamma^2[(d\hat{\vec{p}}/dt)^2-(d\gamma/dt)^2]$ in which the trajectory calculated above is inserted. In practical units, the number of photons emitted at the mean energy $\langle\hbar\omega\rangle=0.3 \hbar \omega_c$  per one electron undergoing one oscillation period is, for $a_0 \gg 1$,
\begin{equation}
N_\gamma = 4.68\times 10^{-2} a_0.
\end{equation}

From the above expressions an estimation of the radiation properties can be obtained for a typical parameter regime. The laser pulse is focused along a Rayleigh length $z_r = 500$ $\mu$m with $a_0 = 10 $, and has a duration of 20 fs FWHM and a wavelength $\lambda_L=800$ nm. The period of the electron motion is $\lambda_u \sim 20$ $\mu$m and the electron executes about $\sim 8$ oscillations within the laser pulse (FWHM). The critical energy of the radiation is $\hbar\omega_c \sim 400$ eV. The emission is emitted at an angle $\theta \sim 280$ mrad with respect to the laser propagation axis. For a plasma density of $1 \times 10^{18}$ cm$^{-3}$, and if we consider that the number of electrons participating to the emission is the number of electrons in the focal volume $\sim 10 \times 10 \times 1000$ $\mu$m$^3$ (which is equal to $10^{11}$ electrons), the number of x-ray photons per shot is $\sim4\times 10^{11}$.

\subsubsection{Numerical results}
\label{chap6secAsec3}

The features of the radiation have been described so far for a free electron submitted to a circularly polarized plane wave for which the concept of wiggling applies and analytical solutions are straightforward. In this section, the case of a linearly polarized laser field is studied using numerical simulations. Electron orbits and radiation features are presented.

\begin{figure}
\includegraphics[width=8.5cm]{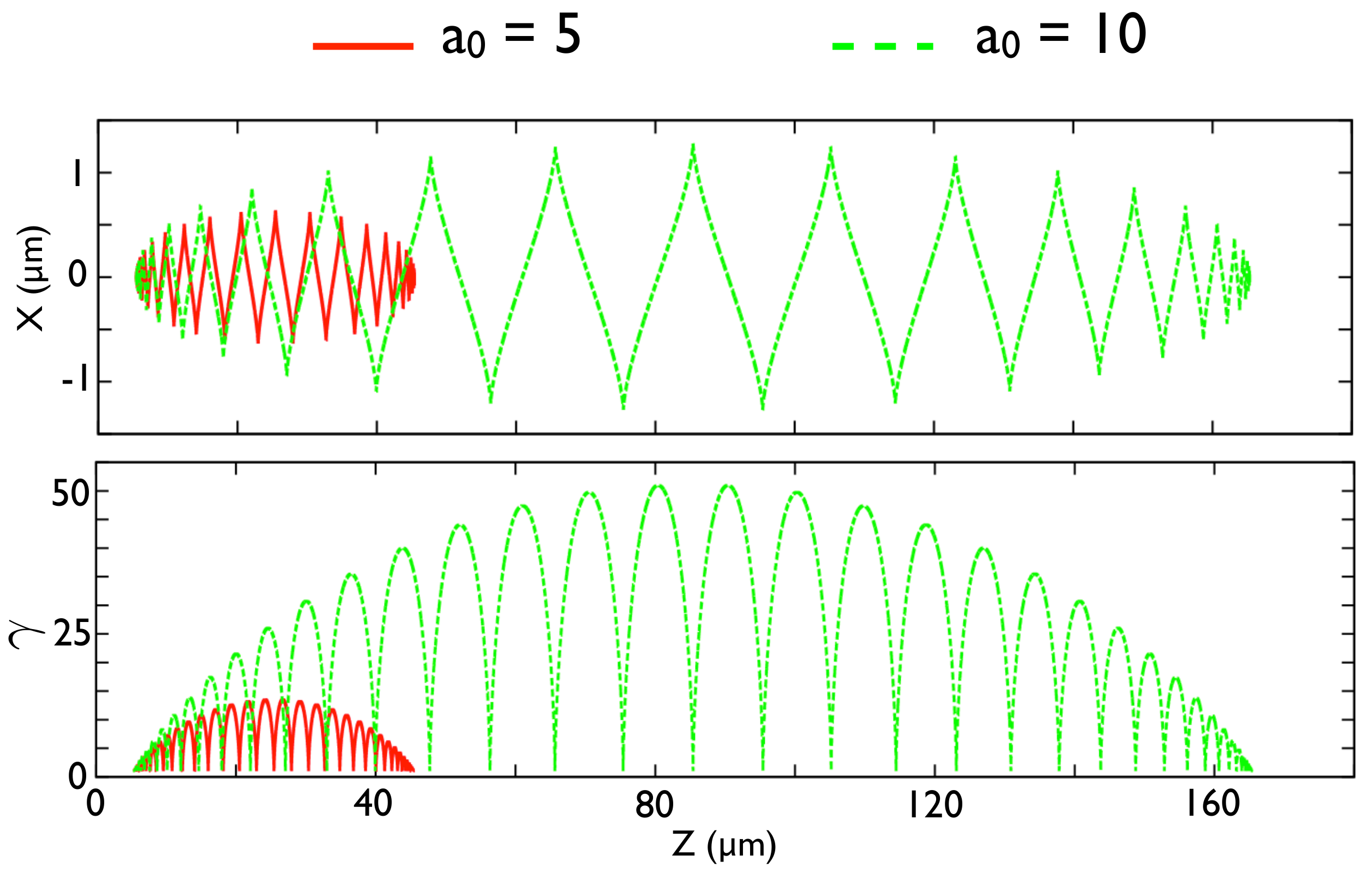}
\caption{Top: The electron orbit in an intense linearly polarized laser pulse with $a_0=5$ (red solid line) and $a_0=10$ (green dashed line). Bottom: The $\gamma$ factor as a function of the longitudinal position for each case. The electron is initially at rest and is submitted only to the linearly polarized laser pulse whose temporal profile is Gaussian with 20 fs duration (FWHM).}
\label{corde_fig34}
\end{figure}

We consider a laser pulse with a Gaussian temporal profile, whose normalized vector potential reads
\begin{align}
\vec{a}=a_0\exp[-(2\ln2)(z-ct)^2/(c^2\tau^2)]\cos(\omega_it-k_iz)\vec{e}_x.
\end{align}
The laser pulse duration (FWHM) is $\tau = 20$ fs and the polarization is linear along the $\vec{e}_x$ direction. The test electron is initially at rest. Figure \ref{corde_fig34} displays the electron orbits for $a_0 = 5$ and $10$ and the evolution of the relativistic factor of the electron $\gamma$ along its motion.
The trajectory consists in successive straight lines with violent longitudinal acceleration or deceleration, the velocity vanishing at each corner. The electron acquires energy in the tens of MeV range while oscillating in the laser pulse, and is again at rest once the laser has passed. There is no net energy gain.

\begin{figure}
\includegraphics[width=8.5cm]{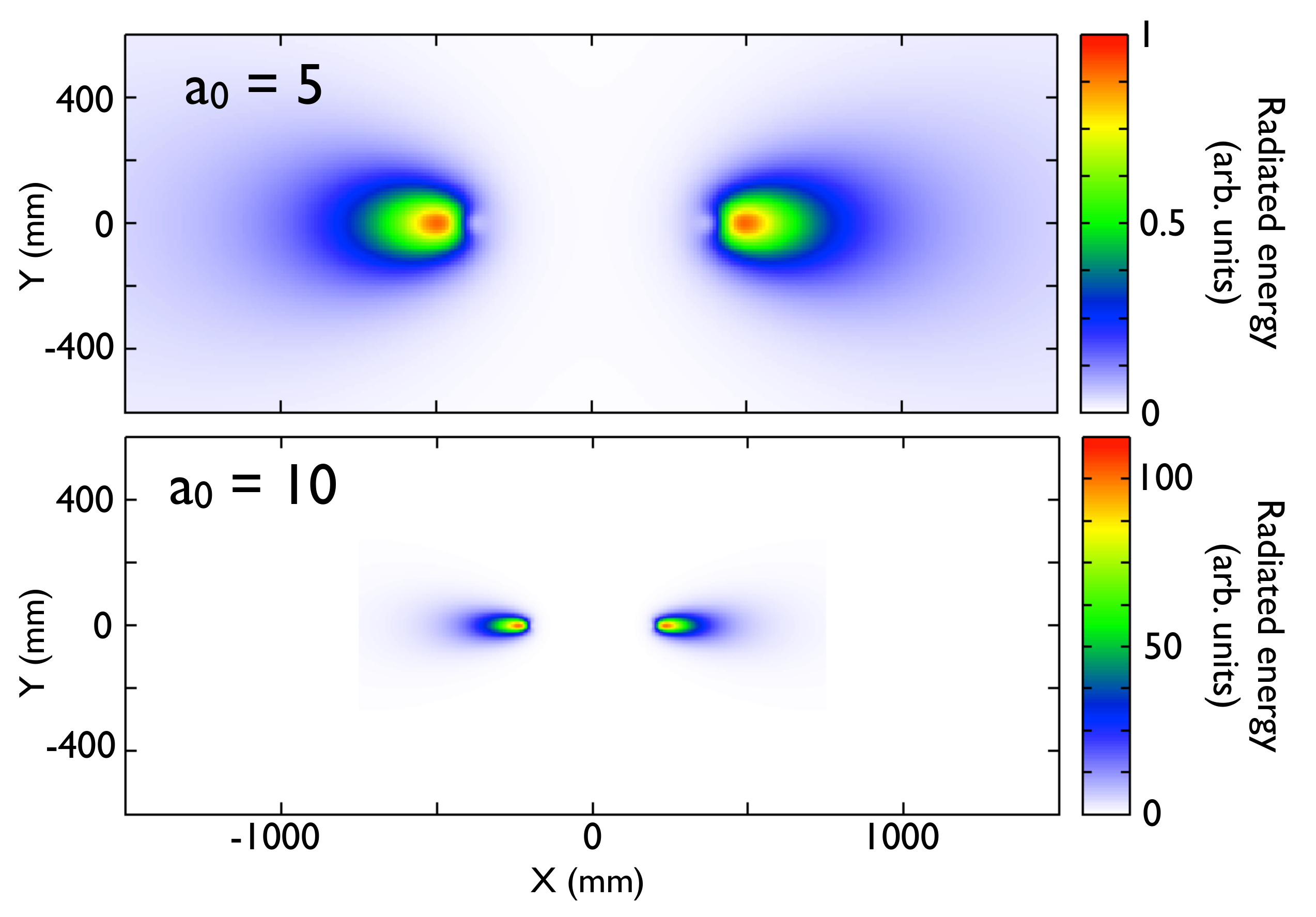}
\caption{Spatial distribution of the nonlinear Thomson scattering, corresponding to trajectories of Fig. \ref{corde_fig34}. The radiated energy per unit surface in the plane situated at 1 m from the source and perpendicular to the propagation axis is represented. On the top is the case $a_0=5$ and on the bottom the case $a_0=10$.}
\label{corde_fig35}
\end{figure}
\begin{figure}
\includegraphics[width=8.5cm]{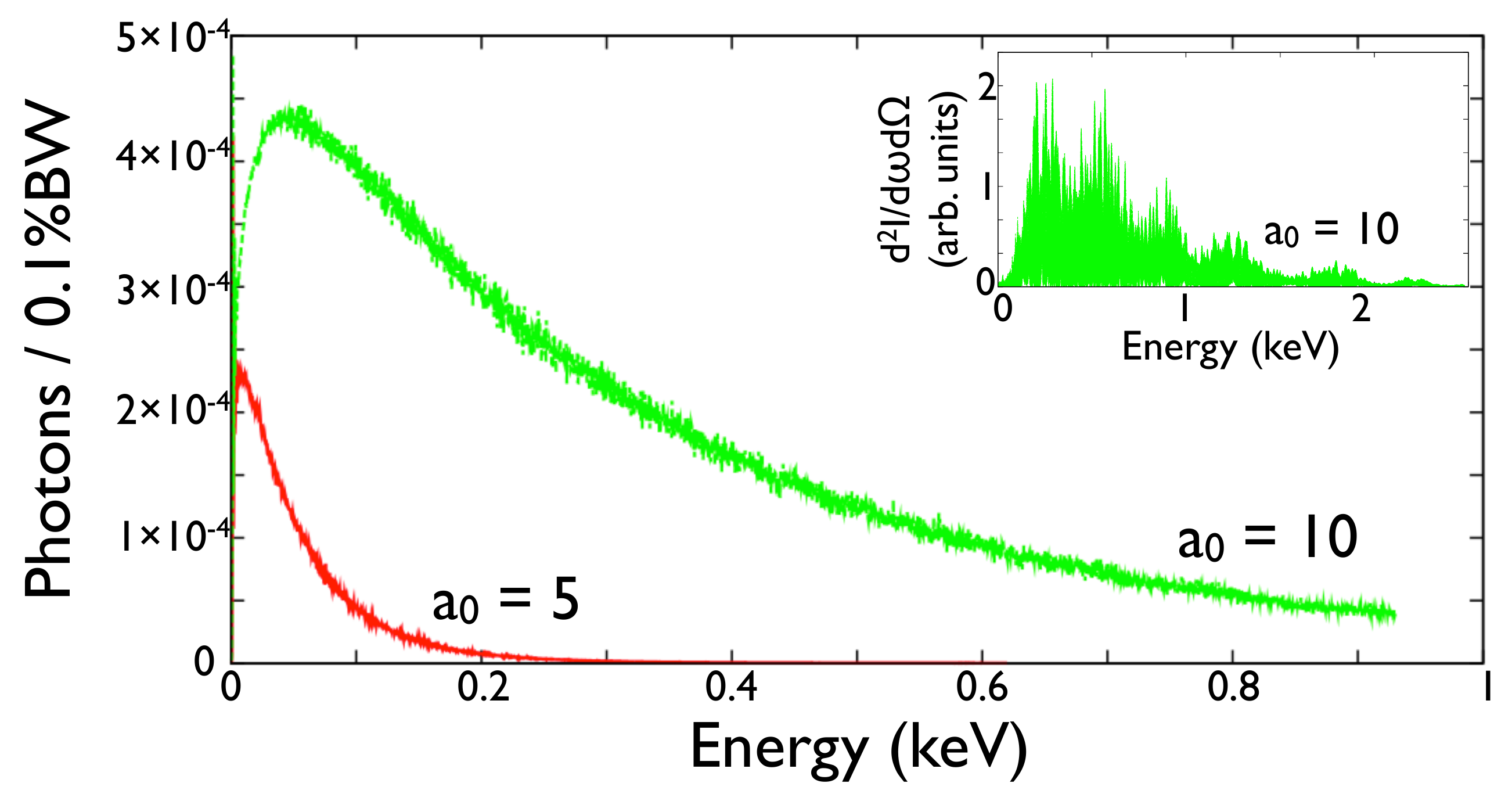}
\caption{Nonlinear Thomson scattering spectrum, integrated over angles and corresponding to the trajectories of Fig. \ref{corde_fig34}. The number of photons per 0.1\% bandwidth for one electron is displayed for the two cases: $a_0=5$ in red solid line and $a_0=10$ in green dashed line. The spectrum near the maximum angle of emission is represented in the inset for the case $a_0=10$.}
\label{corde_fig36}
\end{figure}

The radiation has been calculated using the general formula (\ref{chap2eq1}) in which the orbits calculated above are inserted. In Fig. \ref{corde_fig35}, the spatial distribution of the radiated energy is represented. Because the polarization is linear, the radiation consists in two lobes corresponding to the two directions pointed by the straight lines of the orbit. These two lobes form, with respect to the laser propagation axis, an angle $\theta_X \sim 450 $ mrad for $a_0=5$ and $\theta_X \sim 230 $ mrad for $a_0=10$. As expected, the radiation is more collimated for stronger $a_0$. Figure \ref{corde_fig36} displays the spectra of the radiation integrated over the spatial distribution (the photon number per 0.1\% bandwidth per electron and per shot). An estimate of the total photon number can be obtained assuming that all electrons in the focal volume participate to the emission. Assuming a focal volume of dimension $10 \times 10 \times 1000$ $\mu$m$^3$ and a density of $10^{18}$ cm$^{-3}$ gives $10^{11}$ electrons which radiate $\sim 10^7$ photon/0.1\%BW at 50 eV for $a_0 = 5$ and $\sim 1.5 \times 10^7$ photon/0.1\%BW at 400 eV for $a_0 = 10$.
Comparing these numerical results with the analytical expressions obtained for the circular polarization, it appears that the spectral properties depend on the observation angle (see the inset of Fig. \ref{corde_fig36}). The shape of the spectrum at the maximum angle of emission is not synchrotronlike, because it does not correspond to a wiggling motion. The radiation for linear polarization is due to the longitudinal acceleration from $\gamma=1$ to $\gamma=1+a_0^2/2$ and deceleration to $\gamma=1$  occurring for each straight line section of the motion. The time profile of the radiated field in a single straight line section consists of a double peak structure: the first corresponds to the moment when the acceleration is maximal, and the second to the moment when the deceleration is maximal. This complex behavior is at the origin of the spectral modulation observed at the maximum angle of emission in the inset of Fig. \ref{corde_fig36} \cite{PRE2003Lee}.

It is important to note that the most simple case of nonlinear Thomson scattering has been considered. For a more accurate description of the mechanism, several effects must be taken into account. 
In the tightly focused case, the exact laser field can strongly modify the electron orbit. In particular, the electron is expelled from the high-intensity region. This results in a radial drift and limits the number of electron oscillations in the high-intensity regions. Therefore, the radiated energy is smaller than in the ideal case and the divergence of the x-ray beam is higher and broadened. The presence of the ions from the plasma, or more generally, the collective fields of the plasma, can as well have a significant impact on the electron orbit. They can reduce the longitudinal drift or accelerate the electrons. In the first case, the longitudinal drift being reduced, the divergence is increased. Oppositely, if the electron gains energy from the plasma, the radiation becomes more collimated. Finally, the laser pulse propagates in the plasma with a group velocity $v_g$ smaller than $c$, which can strongly modify the electron orbit if the electron longitudinal velocity $v_z$ becomes comparable or greater than $v_g$.

\subsubsection{Experimental results}
\label{chap6secAsec4}

While nonlinear Thomson scattering was anticipated in the 1960s \cite{PR1962Vachaspati, PR1964Brown, PRD1970Sarachik}, the first experimental demonstration of nonlinear Thomson scattering was performed in 1998 only \cite{Nature1998Chen}. Using a 4 TW laser pulse with a duration of 400 fs, $\lambda_L = 1$ $\mu$m, and $a_0 \sim 2$, focused into a helium gas jet with a density of a few $10^{19}$ cm$^{-3}$, nonlinear Thomson scattering radiation was observed for the first time. The measured radiation followed the specific features of nonlinear Thomson scattering and other radiative processes were ruled out. In particular, the spatial distributions measured were consistent with the nonlinear Thomson scattering properties obtained from the picture of the single free electron submitted to a plane wave. In this experiment, radiation was detected at wavelengths up to the third harmonic of the laser (few eV range). The efficiency was estimated to be of the order of a few $10^{-4}$ photons per electron per pulse. In latter experiments \cite{PoP2002Banerjee, JOSAB2003Banerjee}, the nonlinear Thomson scattering radiation was observed up to the 30th harmonic with a spatial distribution collimated in the direction of the laser propagation axis (and maximum on axis, in contradiction with the single free-electron model).
\begin{figure}
\includegraphics[width=8.5cm]{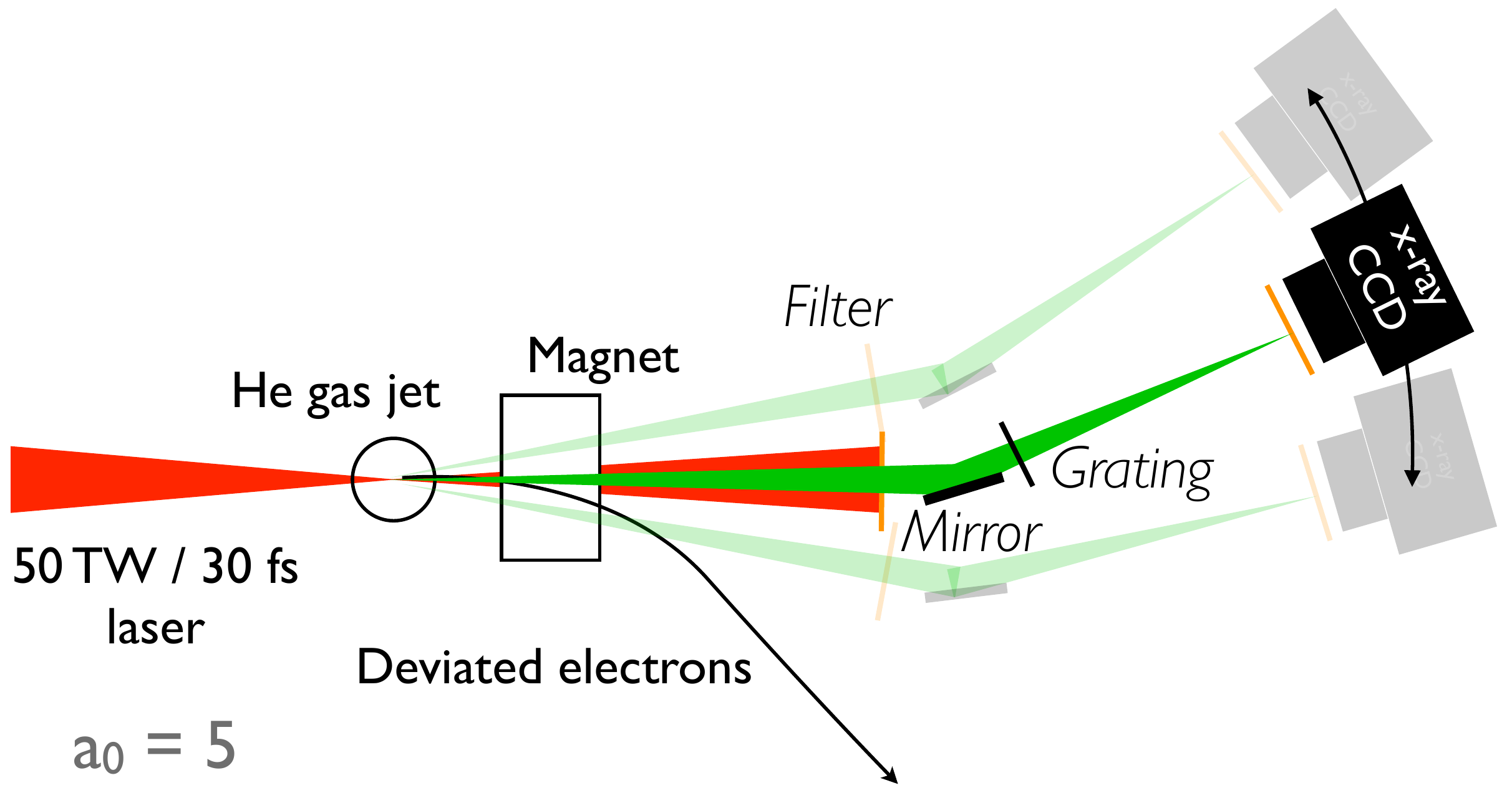}
\caption{Experimental setup for the nonlinear Thomson scattering experiment. Radiation is produced from the interaction of a laser pulse with $a_0\sim5$ and a helium gas jet. Radiation is collected using a grazing incidence spherical mirror and spectrally analyzed using a transmission grating. Filters are used to block the infrared light and to select a specific spectral range. To scan the angular distribution of the radiation, the whole system is mounted on a rotation stage centered on the gas jet.}
\label{corde_fig37}
\end{figure}
In 2003, thanks to the advent of high-intensity lasers, nonlinear Thomson scattering was demonstrated in the XUV range \cite{PRL2003TaPhuoc,EPJD2005TaPhuoc,JOSAB2003TaPhuoc}. This experiment was based on the interaction of a 50 TW laser (with $a_0 \sim 5$, 30 fs, 800 nm) and a helium gas (density in the range $10^{18} - 10^{19}$ cm$^{-3}$). The experimental setup is presented in Fig. \ref{corde_fig37}. The laser was focused within a 6 $\mu$m focal spot (FWHM) onto the front edge of a supersonic helium gas jet (3 mm in diameter). Radiation was collected using a grazing incidence spherical mirror and spectrally analyzed using a transmission grating. The whole system was mounted on a rotation stage centered on the gas jet in order to measure the spatial distribution of the radiation. Nonlinear Thomson scattering radiation was observed in the few tens of eV range up to 1 keV using a set of filters (Al, Zr, Ni, and Be). As at previous experiments, radiative mechanisms others than nonlinear Thomson scattering were ruled out by the variations of the XUV signal as a function of experimental parameters. The signal was found to increase linearly with the plasma density, the radiation was anisotropic and the energy range fitted what is expected for nonlinear Thomson scattering.

\begin{figure}
\includegraphics[width=8.5cm]{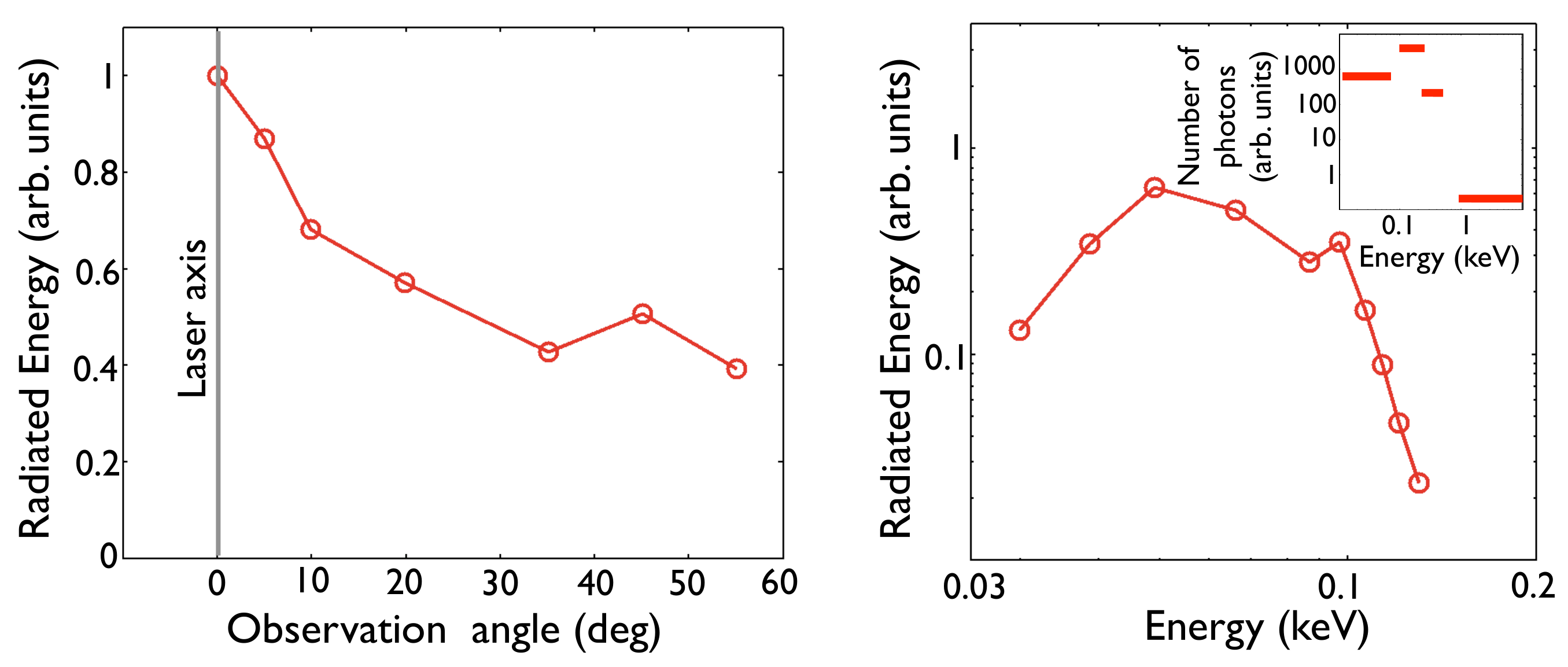}
\caption{On the left is represented the angular dependence of the radiated energy recorded experimentally by rotating the whole detection system. The spectrum of the radiation on axis is given in logarithmic scales on the right of the figure and the inset shows the total number of photons integrated over several spectral bands determined by the choice of filters (the horizontal extension of the lines indicates the spectral band).}
\label{corde_fig38}
\end{figure}

However, the results differ from the expected emission with the model of free electrons in a plane electromagnetic wave. The radiation spatial distribution, presented in Fig. \ref{corde_fig38} (left), was collimated within a large angle (within a cone of half angle $\sim 30$ deg) and was maximum on axis instead of consisting in two lobes centered at $+23$ deg and $-23$ deg. This result can be explained considering the multitude of electron orbits, the focusing of the laser and the effect of plasma fields. This was supported by the fact that electrons at relativistic energies were observed during the experiment. In addition, laser filamentation was observed \cite{PoP2002Faure, PRL2007Thomas, PPCF2009Thomas} indicating a complex propagation of the laser.

The spectrum, measured on axis in a limited spectral range (30 - 120 eV), is presented in Fig. \ref{corde_fig38} (right). The measurement was performed using a transmission grating. In the inset is presented the spectrum measured with filters (Al, Zr, Ni, and Be) over a larger bandwidth and the number of photons. The radiation detected was in the spectral range expected for nonlinear Thomson scattering. A total of $10^{10}\ \text{photons}$ per shot were measured (integrated over the spectrum and the spatial distribution).

\subsubsection{Perspectives}
\label{chap6secAsec5}

Nonlinear Thomson scattering from the laser-plasma interaction remains a complex research topic and only a few experiments have been performed. Important numerical and experimental work has to be done to understand the mechanism. This research will be motivated by the fact that this source of radiation can become a powerful x-ray source. Indeed, the development of petawatt-class lasers will open novel perspectives for nonlinear Thomson scattering. Assuming that $a_0 = 20$ can be reached, the produced radiation is expected to have the following features. The source is extended from the XUV to the x-ray part of the spectrum with photon energies up to $\sim$ 10 keV. In addition, the number of photons will be significantly increased: $\sim 1$ photon per electron, per period, and per shot, and the radiation will be collimated within a half angle of $\sim150$ mrad. Such a source has the main advantage to be very simple to realize. Further developments would rely on the control of the preacceleration of the electron inside the plasma in order reduce the divergence of the x-ray beam, enhancing the brightness. Flattop or annular transverse laser profile should be also interesting to maintain electrons in the high-intensity regions and increase the number of x-ray photons. Finally, several groups reported on the possibility to produce attosecond pulses via nonlinear Thomson scattering \cite{PRE2003Lee, PoP2006Lan}. In fact, the nonlinear Thomson scattering from a single electron is naturally a train of attosecond pulses \cite{PRE2003Lee}. However, when summing the radiation from all electrons, the property disappears unless electrons radiate coherently (instead of incoherent radiation, as presented throughout this section). If this latter condition is satisfied, it becomes possible to generate single intense attosecond pulse \cite{PoP2006Lan}. To fulfill the coherence condition, the use of ultrathin solid targets has been suggested \cite{PoP2005Lee}. A scheme named ``lasetron" for generating coherent zeptosecond pulses has also been proposed by \textcite{PRL2002Kaplan}, and is based on the wiggling of electrons in a subwavelength-size solid particle or thin wire submitted to two counterpropagating circularly polarized petawatt laser pulses.\\

\subsection{Thomson backscattering}

\begin{figure}
\includegraphics[width=8.5cm]{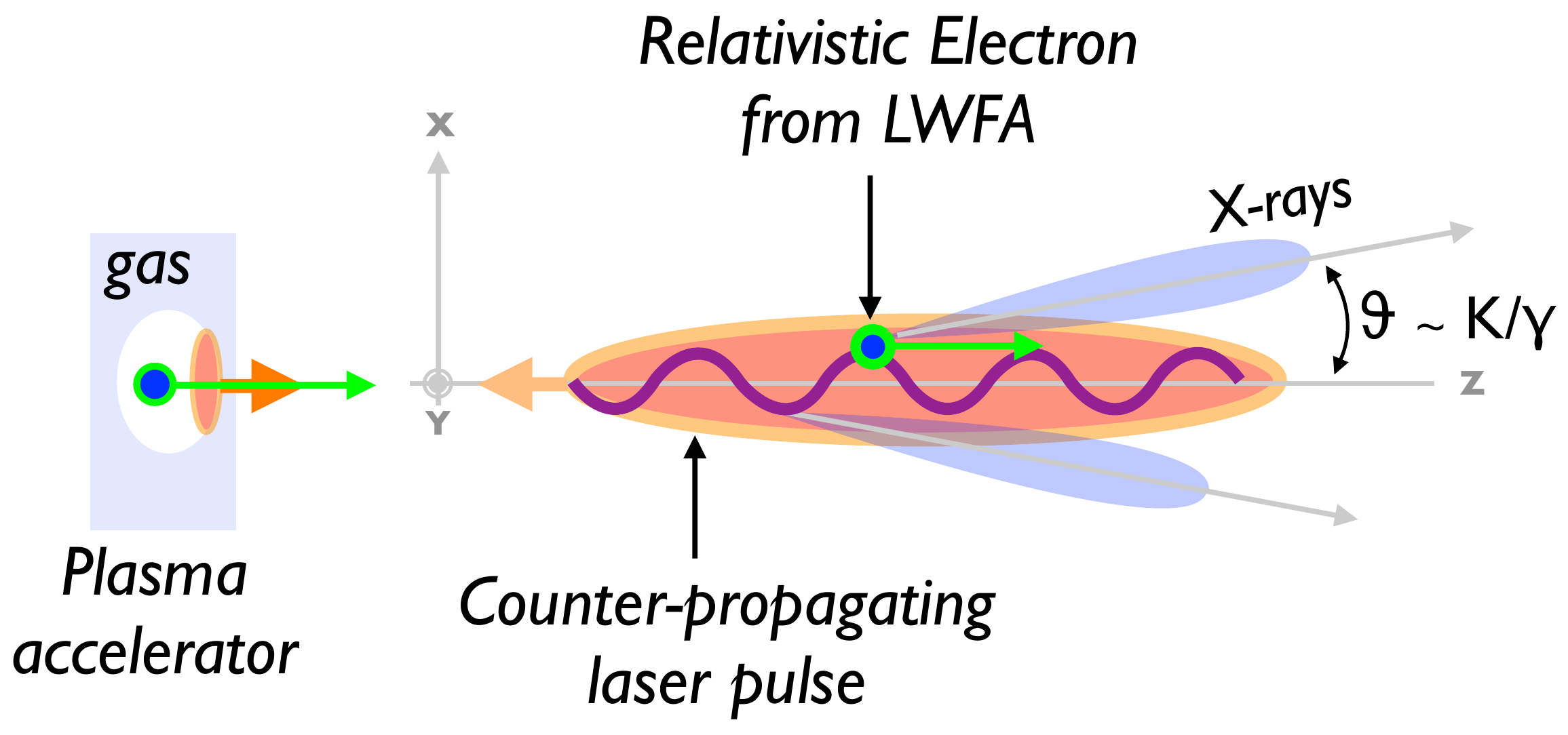}
\caption{Schematic of the all-optically driven Thomson backscattering source. A laser-plasma accelerated electron beam is injected in a counterpropagating laser pulse. Electrons oscillate transversally and emit x rays or $\gamma$ rays due to this motion.}
\label{chap6fig1}
\end{figure}

X-ray or $\gamma$-ray radiation can be produced by scattering an electromagnetic wave off a counterpropagating relativistic electron bunch. This scheme was first proposed in 1963 \cite{PL1963Arutyunian, PRL1963Milburn} and reinvestigated in the 1990s thanks to the advance in laser technology and high-brightness electron accelerators, suggesting the possibility to realize a source of high-brightness pulsed x rays or $\gamma$ rays \cite{JAP1992Sprangle, PRE1993Esarey, NIMA1993Esarey, NIMA1994Kim, PRE1995Ride, NIMA1996Esarey, PRL1996Hartemann, PRE1996Hartemann, NIMA1999Yang, PRE2001Hartemann, PRSTAB2002Li, Hartemann, PRL2004Gao, PRE2005Krafft}. More recently, when the research on electron acceleration from laser-plasma interaction became mature, it was proposed to use these laser-plasma accelerated electron bunches to develop an all-optically driven scheme \cite{MST2001Catravas, IEEE2003Hafz, IEEE2005Leemans, APB2005Tomassini, PRSTAB2007Hartemann}. The principle of this scheme is represented in Fig. \ref{chap6fig1}. Two laser pulses are required: the first drives the plasma accelerator, as discussed in Sec. \ref{chap3}, and the second scatters off the accelerated electrons. In order to derive the typical properties and analytical features of the Thomson backscattered radiation, the electron dynamics in a counterpropagating laser pulse in the simplest model of plane electromagnetic wave is presented. Then, numerical and experimental results are discussed. Foreseen perspectives of this method are described in the conclusion.

\subsubsection{Electron orbit in a counterpropagating laser pulse}
\label{chap6secBsec1}

We consider a laser-plasma accelerator producing a monoenergetic electron bunch of energy $\mathcal{E}=\gamma_imc^2$ with a velocity directed along $\vec{e}_z$ (toward the positive values). The counterpropagating laser pulse is modeled by a linearly polarized plane electromagnetic wave of wave vector $\vec{k}_i=-2\pi/\lambda_L\:\vec{e}_z$ and frequency $\omega_i=2\pi c/\lambda_L$, and with a normalized vector potential given by $\vec{a}=a_0\cos(\omega_i t+k_iz)\vec{e}_x$. The equation of motion of the test electron and the Hamiltonian $\hat{\mathcal{H}}$ describing the test electron dynamics in the electromagnetic wave are given by Eqs. (\ref{chap6eq1}) and (\ref{chap6eq2}). We use the same normalized units, with the choice $m=c=e=\omega_i=1$. As in Sec. \ref{chap6secAsec1}, the system is integrable and action-angle variables can be found to solve the electron motion. Here we analyze the symmetries of the system to directly yield the conserved quantities according to Noether theorem.
The Hamiltonian $\hat{\mathcal{H}}$ depends on the canonical momentum $\hat{\vec{P}}$ and on the potential vector $\vec{a}(\varphi)$ through the variable $\varphi=\hat{t}+\hat{z}$. Hence $\hat{\mathcal{H}}$ is independent of $\hat{x}$ and $\hat{y}$, which implies that the transverse canonical momentum is a constant of motion. In our problem, it is assumed that the electron has a velocity directed along $\vec{e}_z$ exclusively before entering the laser pulse. Therefore the first constant of motion reads
\begin{equation}
\hat{\vec{P}}_{\bot}=\hat{\vec{p}}_{\bot}-\vec{a}=\vec{0}.
\end{equation}
In addition, $\hat{\mathcal{H}}$ depends on $\hat{t}$ and $\hat{z}$ only through $\varphi=\hat{t}+\hat{z}$. Thus $\partial\hat{\mathcal{H}}/\partial \hat{t}=\partial\hat{\mathcal{H}}/\partial \hat{z}$, which leads to the second constant of motion $\text{C}$,
\begin{equation}
\label{chap6eq3}
\gamma+\hat{p}_z=\text{C}.
\end{equation}
The constant $\text{C}$ is obtained using the initial conditions $\gamma=\gamma_i$, $\hat{p}_z=\sqrt{\gamma_i^2-1}$.
The integration of the constants of motion gives the electron orbit:
\begin{align}
\hat{x}(\varphi)&= \frac{a_0}{\text{C}}\sin(\varphi),\\
\hat{y}(\varphi)&= 0,\\
\hat{z}(\varphi)&= \Bigg\{\frac{1}{2}-\frac{1+a_0^2/2}{2\text{C}^2}\Bigg\}\varphi-\frac{a_0^2}{8\text{C}^2}\sin(2\varphi),\\
\gamma(\varphi)&= \frac{\text{C}}{2}+\frac{1+a_0^2\cos^2(\varphi)}{2\text{C}},
\end{align}
where $\text{C}=\gamma_i+\sqrt{\gamma_i^2-1}=2\gamma_i-1/(2\gamma_i)+o(1/\gamma_i^2)$.

The trajectory obtained cannot exactly take the form $x(z)=K/(\gamma k_u)\sin(k_uz)$ corresponding to the standard sinusoidal trajectory studied in Sec. \ref{chap2}. Here $\gamma$ is not constant along the orbit and Eq. (\ref{chap6eq3}) implies that $\gamma$ and $p_z$ vary oppositely. The oscillation motion $z$ takes, however, a form similar to the sinusoidal trajectory, leading to the same characteristic figure-eight motion in the average rest frame of the electron. The description of Sec. \ref{chap2} can be recovered with the approximation $\varphi\simeq2\:\hat{z}\simeq2\:\hat{t}$. The counterpropagating laser pulse can be seen as an undulator with a spatial period given by
\begin{equation}
\lambda_u=\lambda_L/2,
\end{equation} 
and a strength parameter $K$ given by
\begin{equation}
K=a_0 =0.855 \sqrt{I [10^{18}\:\text{W/cm}^2] \lambda_L^2[\mu\text{m}]}.
\end{equation}

\subsubsection{Radiation properties}
\label{chap6secBsec2}

The qualitative features of the radiation remain the same as in Sec. \ref{chap2} and can be parametrized by $K$, $\lambda_u$, and $\gamma\simeq\gamma_i$ (for $a_0\ll\gamma_i$). The spectrum of the emitted radiation depends on the amplitude of the parameter $K$. For $K\ll1$, the electromagnetic wave acts as an undulator. In the average rest frame of the electron, Thomson scattering occurs in the linear regime. In the laboratory frame, the radiation is emitted at the Doppler shifted  fundamental frequency corresponding to a photon energy $\hbar\omega$ in the forward direction ($\theta=0$). As $K\rightarrow1$, harmonics of the fundamental start to appear in the spectrum. For $K\gg1$, the electromagnetic wave acts as a wiggler, and the spectrum contains many harmonics closely spaced and extends up to a critical energy $\hbar\omega_c$ (a nonlinear Thomson scattering occurs in the average rest frame of the electron). These energies are given by
\begin{align}
\nonumber
\hbar\omega &= (4\gamma^2 hc/ \lambda_L) / (1+K^2/2) \;\;\;\; \text{for} \: K <1,\\
\hbar\omega_c &= 3K \gamma^2 hc/ \lambda_L \;\;\;\; \text{for} \: K\gg1.
\end{align}
This gives in practical units
\begin{align}
\nonumber
\hbar\omega[\text{eV}] &=  4.96 \:\gamma^2/ \lambda_L[\mu\text{m}] \;\;\;\; \text{for} \: K \ll1,\\
\hbar\omega_c[\text{eV}] &= 3.18 \:\gamma^2 \sqrt{I[10^{18}\:\text{W/cm}^2]} \;\;\;\; \text{for} \: K\gg1.
\end{align}

The radiation is collimated within a cone of typical opening angle $\theta_{r}=1/\gamma$ in the undulator case. For a wiggler, the radiation is collimated within a typical opening angle $K/\gamma$ in the electron motion plane $(\vec{e}_x,\vec{e}_z)$ and $1/\gamma$ in the orthogonal plane $(\vec{e}_y,\vec{e}_z)$.

As described in Sec. \ref{chap2}, the duration of the emitted radiation is approximately equal to the duration of the electron bunch. The duration of the electron bunch being femtoseconds at the position of interaction~\cite{NatPhys2011Lundh}, the duration of the x-ray pulse is femtoseconds as well.

The number of emitted photons follows the expressions of Sec. \ref{chap2}.  In practical units, the number of photons emitted per period and per electron (at the mean photon energy $E=0.3\hbar \omega_c$ for the wiggler limit) is given by
\begin{align}
\nonumber
N_\gamma  &= 1.53 \times 10^{-2} K^2 \;\;\;\; \text{for} \: K<1,\\
N_\gamma  &= 3.31 \times 10^{-2}K \;\;\;\; \text{for} \: K\gg1.
\end{align}

From these expressions the features of the emitted radiation can be estimated. For a 100 MeV electron bunch ($\gamma\simeq 200$) containing a charge of 100 pC and a counterpropagating laser pulse with $a_0=0.5$ and $\lambda_L =800$ nm, the electron oscillation period is $\lambda_u=400$ nm and the emitted radiation has an energy $\hbar\omega \sim 200 \:\text{keV}$.  The radiation is collimated within a cone of typical solid angle of $5\:\text{mrad}\times5\:\text{mrad}$ and $2\times 10^6$ photons are emitted per oscillation period. Considering a laser pulse containing around ten optical cycles, the number of emitted photons is $\sim 10^7$.

\subsubsection{Numerical results}
\label{chap6secBsec3}

\begin{figure}
\includegraphics[width=8.5cm]{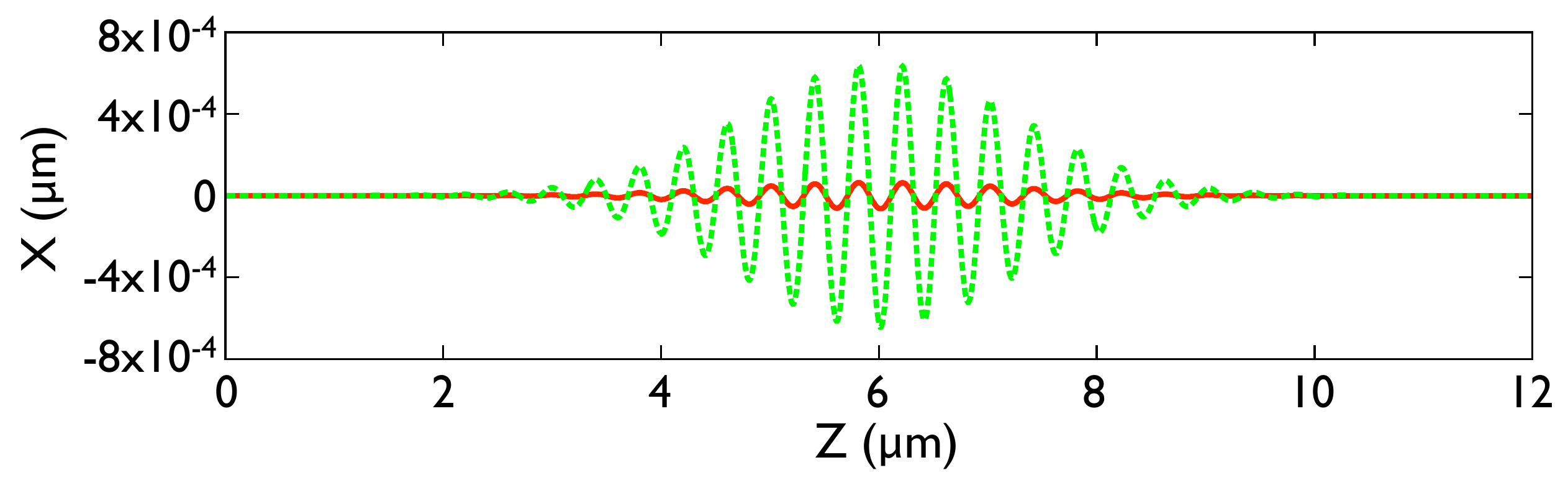}
\caption{Electron trajectories. The orbit is plotted as the red solid line for the undulator case ($a_0=0.2$) and as the green dashed line for the wiggler case ($a_0=2$). Before the interaction, the electron has $\gamma=200$ and $\vec{p}=\sqrt{\gamma^2-1}\vec{e}_z\:$ in both situations. The counterpropagating laser pulse has a Gaussian temporal profile with FWHM duration of 20 fs and a central wavelength of 800 nm.}
\label{chap6fig2}
\end{figure}

The electron equation of motion can be numerically integrated. The case of an electron initially propagating in the $\vec{e}_z$ direction (toward the positive values) with $\gamma = 200$ is investigated. The counterpropagating laser pulse is modeled by a linearly polarized plane wave with a Gaussian temporal envelope with $\tau_L = 20$ fs at full width at half maximum (FWHM), at the wavelength 0.8 $\mu$m. Two laser strength parameters are considered: $a_0 = 0.2$ for the undulator case and $a_0 = 2$ for the wiggler case.

Figure \ref{chap6fig2} shows the orbits of a test electron traveling in the counterpropagating laser pulse for each situation. The transverse motion consists of an oscillation at a period $\lambda_u = 0.4$ $\mu$m and its amplitude follows the envelope of the laser pulse and increases as $a_0$ increases. The maximum amplitudes are $6.4\times10^{-5}$ and $6.4\times10^{-4}$ $\mu$m for $a_0 = 0.2$ and $2$, respectively. 

\begin{figure}
\includegraphics[width=8.5cm]{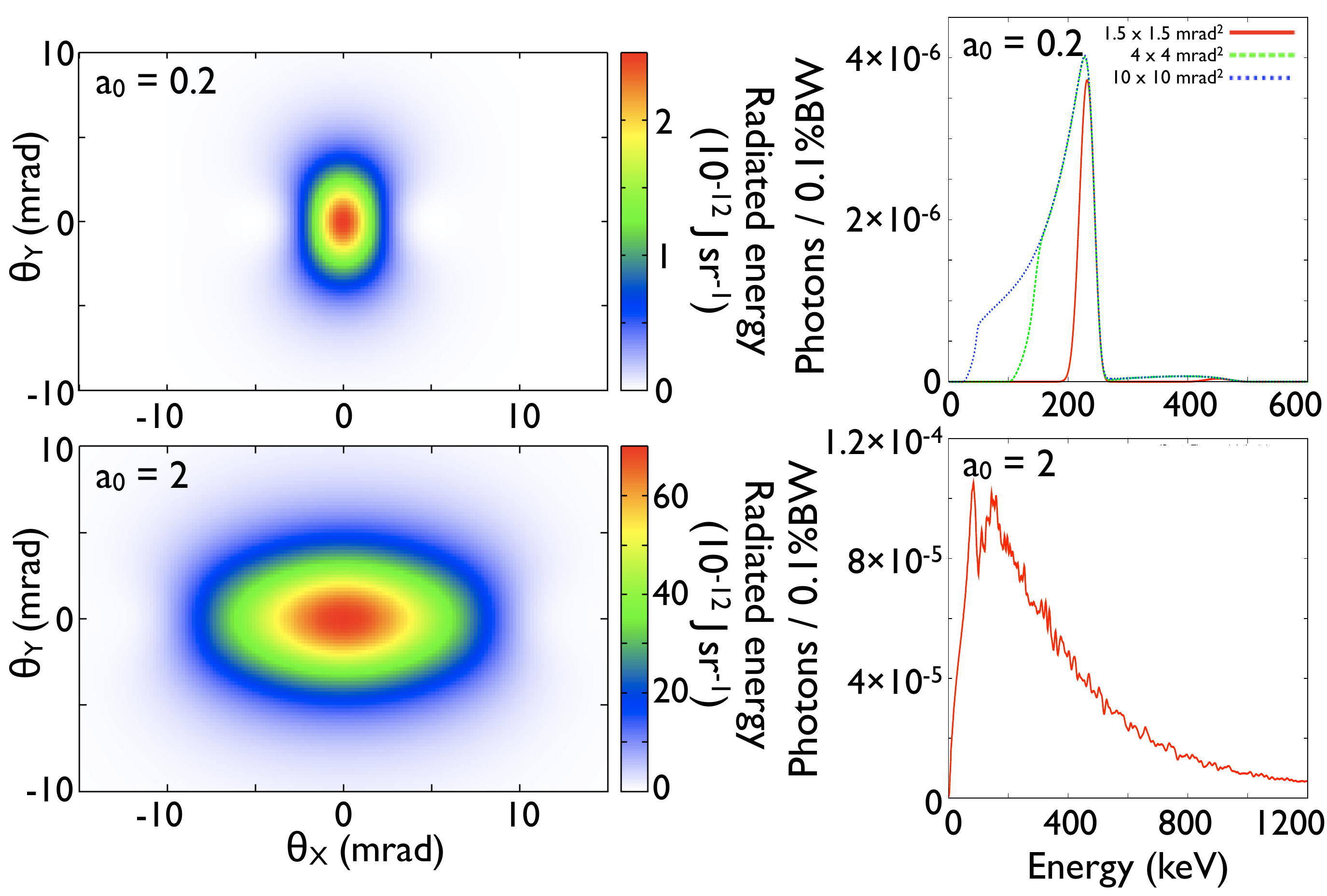}
\caption{On the left are represented the angular distributions of the radiation (radiated energy per unit solid angle) and the corresponding spectra are shown on the right (in number of photons per 0.1\% bandwidth, per electron, and per shot). The undulator case, $a_0=0.2$, is on the top whereas the wiggler case, $a_0=2$, is on the bottom. In the first situation, three different solid angles of integration have been considered to highlight the broadening of the spectrum. In the wiggler situation, the integration is performed over the total angular distribution.}
\label{chap6fig3}
\end{figure}

The spatial distributions of the radiation produced by the electron along these trajectories are presented in Fig. \ref{chap6fig3} (left). For $a_0=0.2$ (top), the radiation has a divergence of typical opening angle $1/\gamma\sim5$ mrad. For $a_0=2$ (bottom), the radiation has a divergence of typical opening angle $a_0/\gamma\sim10$ mrad in the plane of the electron motion and $1/\gamma\sim5$ mrad in the orthogonal direction. Figure \ref{chap6fig3} (right) represents the spectra of the emitted radiation, integrated over three selected solid angles for the undulator case $a_0=0.2$ ($1.5\times1.5$, $4\times4$, and $10\times10$ mrad$^2$), and over the overall angular distribution for the wiggler case $a_0=2$. For $a_0=0.2$, the spectrum is nearly monochromatic at the energy $\hbar\omega \sim 200$ keV in the forward direction, with a width of the emission line given by $\Delta \lambda = \lambda / N$ (where $N\sim7-8$ is the number of oscillation periods in the FWHM laser pulse duration of 20 fs), and it is broadened after integration over the emission angles due to the angular dependence of the radiated wavelength $\lambda \simeq (\lambda_u/2\gamma^2)(1+K^2/2+\gamma^2\theta^2)$ [see Sec. \ref{chap2}, Eq. (\ref{chap2eq3})]. For $a_0=2$, the spectrum becomes broadband with a critical energy of about 300 keV. An estimate of the total photon number can be obtained on the basis of the number of electrons that are currently accelerated in a laser-plasma accelerator. Assuming $1 \times 10^8$ electrons gives $\sim 5 \times 10^{4}\ \text{photons}$ within a spectral bandwidth of $\sim$ 10-15 \% in the undulator case (selection of the $1.5\times1.5$ mrad$^2$ solid angle). In the wiggler case, $\sim 2 \times 10^{3}\ \text{photons}/0.1\%\text{BW}$ at 600 keV and a total photon number in the range $10^{7-8}$ are expected.

\subsubsection{Experimental results}
\label{chap6secBsec4}

Many experiments of Thomson backscattering have been performed using electrons from conventional accelerators. The first experiment to actually use laser-backscattered photons as a beam in a physics measurement was conducted at SLAC in 1969 \cite{PRL1969Ballam}. Many other experiments followed this approach to produce x rays and $\gamma$ rays \cite{IEEE1983Sandorfi, JAP1995Ting, PRL1996Bula, PRL1996Glotin, PRL1997Burke, Science1996Schoenlein, IEEE1997Leemans, PRL1997Litvinenko, NIMA2000Kotaki, PRSTAB2000Pogorelsky, JJAP2001Yorozu, APB2003Yorozu, PRSTAB2003Sakai, APB2004Anderson, PRL2006Babzien, PRSTAB2010Albert}, and nowadays several $\gamma$-ray facilities are based on this method.

A significant milestone has been achieved with the first demonstration of femtosecond x-ray generation by Thomson backscattering (using electrons from conventional accelerators) in 1996, by Schoenlein and coworkers \cite{Science1996Schoenlein, IEEE1997Leemans}. They reported on the production of femtosecond x-ray pulses by crossing, at a right angle, a 50 MeV electron beam with a 100 fs duration, 0.8 $\mu$m wavelength, terawatt laser. The laser pulse contained 60 mJ and was focused on a 90-$\mu$m-diameter electron bunch (1.3 nC in a 20 ps FWHM bunch). In this experiment, $\sim 5\times 10^{4}$ photons were produced with a maximum energy of 30 keV. However, due to the geometry and the mismatch of the electron bunch and laser pulse durations, the number of electrons participating in the production of radiation was limited to the portion of the bunch selected by the laser pulse. The x-ray pulse duration was theoretically predicted to be $\sim 300$ fs due to the extension of the overlapping region and the radiation was found to be collimated within $\sim 35$ mrad.

\begin{figure}
\includegraphics[width=8.5cm]{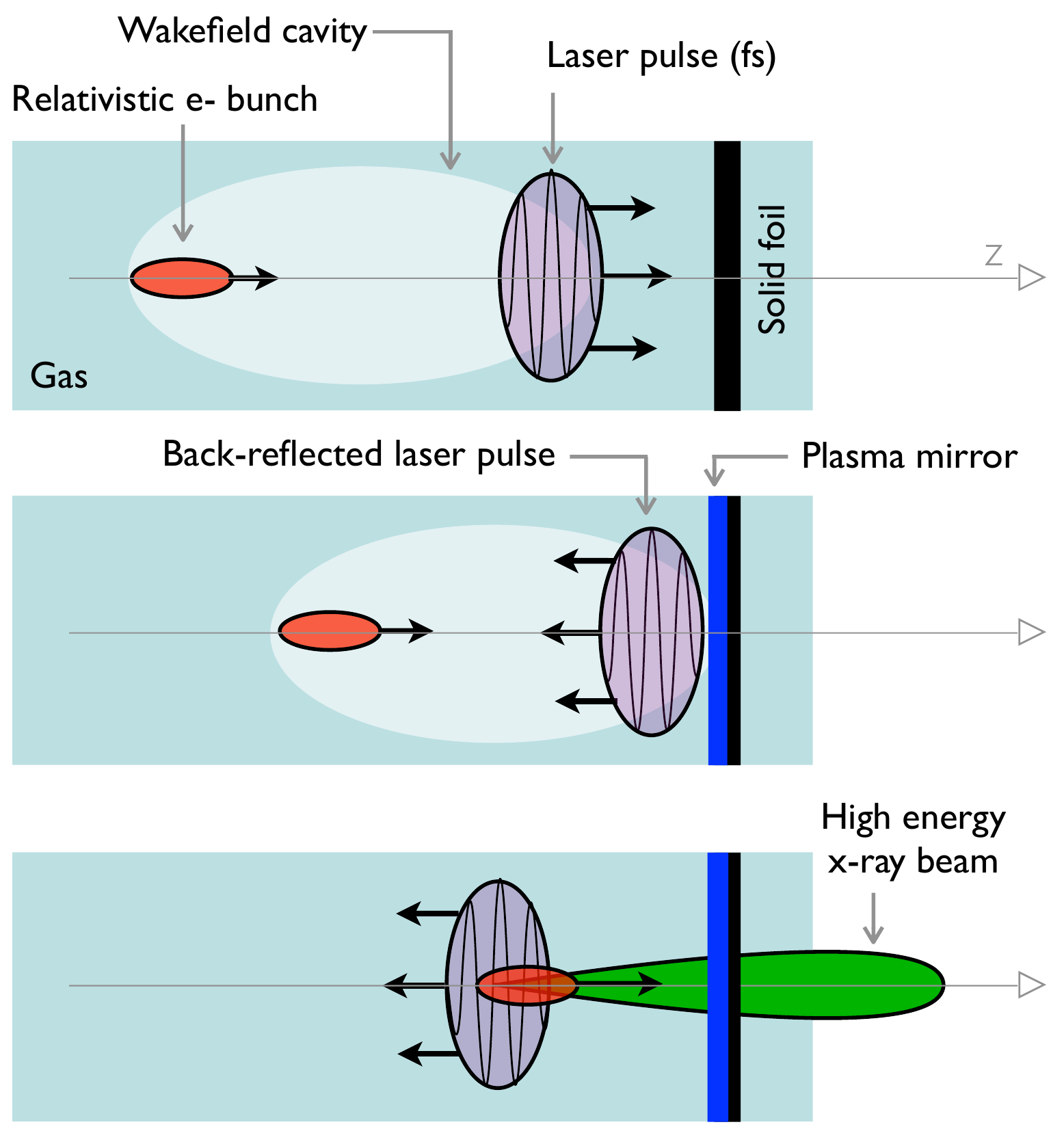}
\caption{Principle of the all-optical Thomson source based on the combination of a laser-plasma accelerator and a plasma mirror.}
\label{corde_fig31}
\end{figure}

Whereas many experiments have been performed using conventional accelerators, only two experiments produced Thomson backscattering radiation using an all-optically driven scheme \cite{PRL2006Schwoerer, NatPhot2012TaPhuoc}.
The first, performed in 2006, was based on the use of two laser pulses in a counterpropagating geometry \cite{PRL2006Schwoerer}. The first pulse (85 fs, 800 nm, and $a_0=3$) was used to produce an electron bunch with a Maxwellian spectrum whose temperature equals 6 MeV and with a total charge in the picocoulomb range. The second laser pulse (85 fs, 800 nm, and $a_0=0.8$) was counterpropagating and focused on the electron bunch. In this experiment,  about $3 \times 10^{4}$ x-ray photons in the 400 eV - 2 keV range were observed within a 80 $\mu$sr detection cone at 60 mrad with respect to the laser axis.  However, this type of experiment remains challenging because of the difficulty to overlap in time and space the electron bunch and the counterpropagating laser pulse.

Recently, a much more simple and efficient method, based on the use of one laser pulse only, has been demonstrated \cite{NatPhot2012TaPhuoc}. This method has, for the first time, allowed the generation of all-optically driven high-energy femtosecond x-ray beams. The scheme relies on the marriage of a laser-plasma accelerator and a plasma mirror. The principle is illustrated in Fig. \ref{corde_fig31}. An intense femtosecond laser pulse, focused into a millimeter-scale gas jet, drives a wakefield cavity in which electrons are trapped and accelerated. Then, the laser strikes a foil placed at nearly normal incidence with respect to the laser and electron beam axis. Ionized by the rising edge of the laser pulse, the foil turns into a plasma mirror which efficiently reflects the laser pulse (the reflectivity is up to 70$\%$). Naturally overlapped in time and space with the backreflected laser pulse, the relativistic electrons oscillate in the laser field and emit a bright femtosecond x-ray beam by Thomson backscattering. 

\begin{figure}
\includegraphics[width=8.5cm]{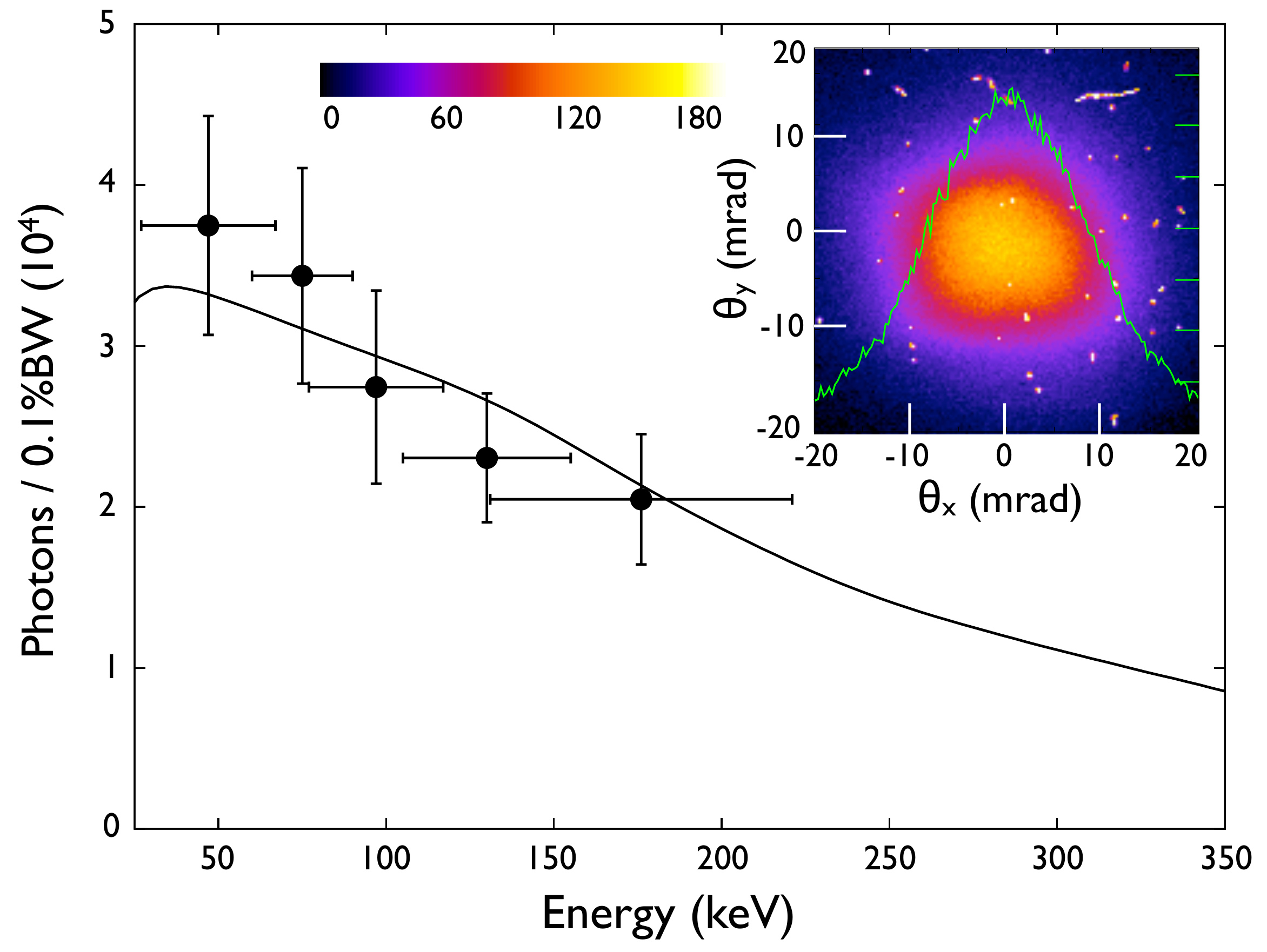}
\caption{Spectrum of the Thomson backscattering radiation measured using a set of copper and aluminum filters. The inset is the x-ray beam angular profile measured using a phosphor screen.}
\label{corde_fig32}
\end{figure}

The experiment was performed at the Laboratoire d'Optique Appliqu\'ee. A laser pulse (1 J, 35 fs, 810 nm, and $a_0\simeq1.5$) was focused onto the front edge of a supersonic helium gas jet (2.1 mm plateau and 600 $\mu$m gradients on both sides) to generate the electron bunch. The electrons were accelerated in the forced laser wakefield regime \cite{Science2002Malka}. The electron spectrum was broadband with a maximum energy of 100 MeV and a total charge of about 120 pC. A foil (either glass 1 mm or CH 300 $\mu$m) was placed in the region of the exit gradient of the gas jet. At the interaction of the backreflected laser pulse with the electron bunch, a bright high-energy femtosecond x-ray beam is generated via Thomson backscattering. The x-ray spectrum, obtained by measuring the x-ray signal transmitted by filters of different thicknesses and materials, is presented in Fig. \ref{corde_fig32}. With the electron spectrum broadband, the produced radiation spectrum was broadband as well. The spectrum extends, as expected, up to a few hundreds of keV and the total number of photons is about $10^8$. The x-ray spectrum was reproduced numerically considering the measured electron spectrum and charge, and a 15 fs (FWHM) duration and $a_0=1.2$ backreflected laser pulse. Figure \ref{corde_fig32} (inset) presents the x-ray beam profile. The radiation was found to be collimated within a cone of divergence 18 mrad FWHM. Finally, the source size, measured using the knife-edge technique, was estimated to be less than 3 $\mu$m (FWHM).  The peak brightness is estimated to be of the order of $10^{21}\ \text{photons}/(\text{s}\ \text{mm}^2\: \text{mrad}^2\: 0.1\%\text{BW})$ at 100 keV, which is more than 10000 brighter than Thomson $\gamma$-ray sources from conventional accelerators \cite{Science1996Schoenlein, PRSTAB2010Albert}. This high brightness is obtained thanks to the micron source size and the femtosecond duration reached in this all-optical scheme.

\subsubsection{Perspectives}
\label{chap6secBsec5}

Thomson backscattering is a remarkable source for the production of energetic radiation, since it allows one to convert electrons into x rays or $\gamma$ rays in an efficient manner.  Among the foreseen development, a promising route would be to take advantage of the latest progress made in laser-plasma acceleration. Monoenergetic electron bunches with tunable energy have recently been produced with an optical injection scheme \cite{Nature2006Faure}. The all-optically driven Thomson backscattering source, based on these electron bunches, would allow one to produce tunable and nearly monochromatic radiation. This would represent significant progress since to date none of the existing laser-based femtosecond x-ray sources can produce monochromatic x-ray beams. Thomson backscattering offers as well the possibility to generate $\gamma$-ray radiation. For example, considering an electron energy of 1 GeV and a counterpropagating laser pulse of wavelength $\lambda_L=800$ nm, $\gamma$ rays with an energy as high as 24 MeV can be produced. In this photon energy range, 100 keV and beyond, radiography applications, of interest for industrial or medical applications, will be possible. Taking advantage of the very small source size, high-resolution images can be obtained. Such high-energy radiation sources would also have many applications in nuclear physics. An additional foreseen source development relies on the production of a high repetition rate x-ray source, delivering bright photon beams in the tens of keV range. In that case, electron bunches in the few tens of MeV will be produced using compact 100 Hz-kHz laser systems. Generating such a compact femtosecond x rays source at high repetition rate, accessible to a university-scale laboratory, will certainly have unprecedented impact for applications in time-resolved x-ray science, high-resolution radiography, and phase contrast imaging.

Finally, a fully optical free-electron laser scheme based on Thomson backscattering has also been proposed \cite{PRSTAB2008Petrillo} and will be discussed in Sec. \ref{chap7}. This scheme remains a feasibility study since it requires an electron bunch quality which has not been achieved yet. However, such a source, if it becomes realizable, would revolutionize work in the community of x-ray radiation.

\section{Coherent Radiation: Toward a Compact X-Ray Free-Electron Laser}
\label{chap7}

In this section, the topic of free-electron laser (FEL) is introduced, especially in the XUV to x-ray range, and we discuss the perspectives to get FEL light from laser-accelerated electron bunches. The free-electron laser concept was first proposed by \textcite{JAP1971Madey} who, having experience with insertion devices (undulators and wigglers) for light sources, realized that a laserlike amplifier could be constructed by combining a high-quality electron beam with an undulator or wiggler and an input radiation beam. A few years later, the principle was demonstrated experimentally by Madey and co-workers \cite{PRL1977Deacon}.

The electron beam inside the undulator acts as an active medium and provides the amplification of the radiation beam. The undulator is the coupling component, which permits energy exchange between the radiation and electrons. In practice, injecting an electron beam in an undulator will naturally lead to incoherent synchrotron emission, as discussed in the previous sections. This incoherent emission provides the input radiation beam and can be amplified by the FEL process when certain conditions are fulfilled \cite{PA1980Kondratenko, OC1984Bonifacio, JOSAB1985Murphy, PRL1994Bonifacio, OC1998Saldin}. The mechanism can be explained as follows. A single electron experiences the radiation from other electrons in the bunch. This interaction between the bunch and its own radiation can lead to a microbunching of the electron distribution within the bunch at the fundamental wavelength of the radiation and its harmonics. This microbunching in return leads to coherent emission from the overall electron bunch, whose power is orders of magnitude higher than the usual incoherent synchrotron radiation. However, this free-electron laser mechanism requires restrictive conditions on the electron beam quality and an important number $N$ of oscillations to take place. This situation, where the incoherent synchrotron or spontaneous emission is amplified in a single pass, is commonly named the self-amplified spontaneous emission (SASE) configuration, and brings a fluctuating radiation output (shot to shot) with poor temporal coherence. In the opposite way, injecting in the undulator an input coherent radiation beam, \textit{i.e.}, an external seeding beam, gives a stable radiation output with good temporal coherence. Nevertheless, the latter case requires a coherent source at a given radiation wavelength with a power higher than the synchrotron one \cite{EPL2009Lambert}, and such coherent sources are not yet available in the x-ray domain. Hence, the SASE configuration seems to be the simplest route toward the production of hard x rays.

Many projects with a compact free-electron laser in the x-ray range of the spectrum (X-FEL) and operating FEL facilities are based on the SASE approach. The first facility which has provided SASE radiation in the soft x-ray domain is the Free-Electron Laser at Hamburg (FLASH) which is currently working at wavelengths down to $\lambda=6$ nm \cite{NJP2009Tiedtke}. 
The Linac Coherent Light Source \cite{NatPhot2009McNeil, NatPhot2010Emma, LCLS} and the Spring-8 Angstrom Compact Free-Electron Laser \cite{NatPhot2011Pile, NatPhot2012Ishikawa} have demonstrated FEL saturation at angstrom wavelengths, while the European X-ray Free-Electron Laser \textcite{XFEL} is under preparation. On the other hand, seeded free-electron lasers are considered to improve the beam quality  [for example, seeded FLASH (sFLASH) and FERMI@Elettra \cite{FERMI}], but they generally aim to less energetic photons (soft x rays).

For our purposes, the electron beam is generated using a laser-plasma accelerator and the undulator can be a conventional meter-scale undulator (see Sec. \ref{chap5}), an electromagnetic wave undulator (see Sec. \ref{chap6}), or the plasma undulator of the blowout regime (see Sec. \ref{chap4}). This latter differs significantly from the others because the strength parameter of the undulator depends on the amplitude of the betatron oscillation and on the electron energy. The amplification process has been studied for electrons in a ion channel, and has been named ion channel laser (ICL) for this case. In addition, the spectral ranges of interest are from XUV to x rays.

In the following, we first present a short description of the underlying physics of the FEL process and summarize the conditions needed to be applied on the electron bunch for setting the FEL amplification. The two different modes of operation are discussed: the amplification of a seeding radiation beam and the amplification of the shot noise, \textit{i.e.}, of the spontaneous synchrotron radiation (SASE regime). Finally, the required conditions for the FEL amplification are used to investigate its feasibility from laser-accelerated electron bunches for three specific cases of undulator: the conventional one, the electromagnetic wave one, and the ICL one. We focus especially on the conventional undulator, since it represents the most promising route for near-future developments toward a compact x-ray free-electron laser.

\subsection{The FEL amplifier}
\label{chap7secA}

\subsubsection{Principle of the free-electron laser process}
\label{chap7secAsec1}

To understand the physical mechanism of FEL amplification, it is instructive to begin by the interaction of a single electron with the radiation beam and the undulator \cite{Brau, Saldin, PRSTAB2007Huang}. In the following, a conventional undulator is considered, \textit{i.e.}, a periodic structure of magnets generating a periodic static magnetic field (as shown in Sec. \ref{chap5}), in order to present the FEL amplification process and the required conditions on the electron beam parameters.

The undulator forces the electron to follow a sinusoidal trajectory whose period corresponds to the undulator period $\lambda_u$ and whose amplitude is equal to $K\lambda_u/(2\pi\gamma)$. The term $K=e\lambda_uB_0/(2\pi mc)$ is the undulator strength parameter, $B_0$ is the peak magnetic field generated by the undulator, and $\gamma=\mathcal{E}/mc^2$ is the relativistic factor of the electron with $\mathcal{E}$ the electron energy. The electron propagates in the $z$ direction and the magnetic field of the undulator is oriented in the $y$ direction, such that the motion of the electron is confined in the $Oxz$ plane. In terms of velocity, the motion of the electron inside the undulator is written as
\begin{equation}
\label{chap7eq1}
\beta_{x}(z)=\frac{K}{\gamma}\cos(k_uz),
\end{equation}
where $k_u=2\pi/\lambda_u$, $\beta_x=\vec{\beta}.\vec{e}_x$, and $\vec{\beta}=\vec{v}/c$ is the electron velocity normalized to the speed of light $c$. We then consider the effect of the radiation beam. This latter is modeled by a linearly polarized plane electromagnetic wave whose electric field is $\vec{E}=E_0\sin(kz-\omega t)\vec{e}_x$ (one-dimensional analysis), where $k=2\pi/\lambda$ is the norm of the wave vector, $\omega=2\pi c/\lambda$ is the frequency, $\lambda$ is the radiation wavelength, and $E_0$ is the electric field amplitude. Assuming that the radiation field is small compared to the undulator field ($K\gg a_0=eE_0/mc\omega$), the expression of the electron velocity (\ref{chap7eq1}) can be used in order to obtain the rate of electron energy change using the kinetic energy theorem,
\begin{align}
\label{chap7eq2}
\nonumber
\frac{d\mathcal{E}}{dz}\simeq\frac{d\mathcal{E}}{cdt}&= -e\vec{\beta}.\vec{E}=-\frac{eKE_0}{\gamma}\cos(k_uz)\sin(kz-\omega t)\\
&= -\frac{eKE_0}{2\gamma}[\sin\psi+\sin(kz-\omega t-k_uz)],
\end{align}
where $\psi=kz-\omega t+k_uz$ is the ponderomotive phase. An energy exchange can occur between the electron and the radiation if the sine terms are nonzero after averaging over $z$.  The phase $\psi$ can remain constant if the increase of the undulator phase $k_uz$ compensates the decrease of the radiation phase $kz-\omega t$, so that the first sine term is manifestly nonzero after averaging. This latter condition is written, for the phase $\bar{\psi}=kz-\omega \bar{t}+k_uz$ defined with the averaged arrival time $\bar{t}=\int dz/\bar{v}_z$, as
\begin{align}
\label{chap7eq3}
\frac{d\bar{\psi}}{dz}=0=k_u+k-\frac{\omega}{\bar{v}_z}.
\end{align}
The above resonance condition states that when the electron advances one undulator period, the radiation field has slipped by one radiation wavelength $\lambda$. It can be put in the following form:
\begin{equation}
\label{chap7eq4}
\lambda=\frac{\lambda_u}{2\gamma^2}(1+K^2/2).
\end{equation}
We emphasize here that this equation is nothing else but the fundamental wavelength of the electromagnetic field radiated by the electron moving in the undulator [see Eq. (\ref{chap2eq3}) in Sec. \ref{chap2}]. This reinforces the physical intuition that energy exchange can occur between the electron and the radiation when the condition (\ref{chap7eq4}) is fulfilled. A more careful analysis of the sine terms of Eq. (\ref{chap7eq2}) [see \textcite{Brau2, Wiedemann2}] shows that energy exchange can also occur if $\lambda_n=(\lambda_u/2n\gamma^2)(1+K^2/2)$, \textit{i.e.}, for harmonics of the electromagnetic field radiated by the electron. This exchange at harmonics of the fundamental wavelength is possible because of the $z$ oscillatory motion of the electron in the undulator. In addition, the analysis shows that Bessel functions appear in the multiplicative factor, so that near the resonance at the fundamental wavelength Eq. (\ref{chap7eq2}) becomes
\begin{equation}
\label{chap7eq2b}
\frac{d\mathcal{E}}{dz}=-\frac{eA_uE_0}{\sqrt{2}\gamma}\sin{\psi},
\end{equation}
where $A_u=K[J_0(x)-J_1(x)]/\sqrt{2}$ with $x=K^2/[4(1+K^2/2)]$, $J_0$ and $J_1$ are Bessel functions (for $K\ll1$, $A_u\simeq K/\sqrt{2}$ and for $K\gg1$, $A_u\simeq K/2$), and the bar over $\psi$ is now implicitly assumed.
In the following, we will not consider harmonics and concentrate on the resonance at the fundamental wavelength.

The ponderomotive phase $\psi$ plays a major role in the FEL dynamics and it is relevant to locate the electron using the variable $\psi$ instead of $z$, and to use $z$ as the new time variable. If $\mathcal{E}_0$ is the energy at resonance satisfying (\ref{chap7eq4}) and $\Delta\mathcal{E}=\mathcal{E}-\mathcal{E}_0$ is the energy of the electron relative to the resonance energy, then for $\Delta\mathcal{E}\ll\mathcal{E}_0$ Eqs. (\ref{chap7eq3}) and (\ref{chap7eq2b}) become
\begin{align}
\label{chap7eq5a}
\frac{d\psi}{dz}&= \frac{k}{\gamma_z^2\mathcal{E}_0}\Delta\mathcal{E},\\
\label{chap7eq5b}
\frac{d\Delta\mathcal{E}}{dz}&= -\frac{eA_uE_0}{\sqrt{2}\gamma}\sin\psi,
\end{align}
where $\gamma_z=1/\sqrt{1-\overline{\beta_z}^2}$. By differentiating Eq. (\ref{chap7eq5a}), the pendulum equation is obtained,
\begin{equation}
\label{chap7eq6}
\frac{d^2\psi}{dz^2}+\frac{ekA_uE_0}{\sqrt{2}\gamma\gamma_z^2\mathcal{E}_0}\sin\psi=0.
\end{equation}
Hence, the electric field of the radiation $E_0$ (all other parameters being constant) can be interpreted as the amplitude of the effective or ponderomotive potential of the pendulum. There are two classes of trajectory for the pendulum equation: trapped and untrapped. The curve in phase space ($\psi$, $\mathcal{E}$) at the limit of both regimes is the separatrix. The higher the electric field is, the higher the phase space volume of trapped conditions (volume inside the separatrix) and the induced acceleration or deceleration for the electron.

\bigskip
This analysis of the interaction of a single electron with the radiation beam and the undulator can be used to describe qualitatively the collective effect of FEL amplification occurring with a bunch of electrons. The initial condition is the electron distribution in the phase space ($\psi$, $\mathcal{E}$). As discussed in Sec. \ref{chap2secF}, the electron bunch coming from a laser-plasma accelerator has a random distribution. For simplicity, we consider the case of a uniform distribution in phase $\psi$ (such that there is no synchrotron radiation) and electrons at exact resonance $\mathcal{E}=\mathcal{E}_0$.

Imposing that the electric field of the radiation $E_0$ is constant (the so-called small signal gain regime), there are as many electrons which lose energy as electrons which gain energy. As a result, there is no energy gain for the radiation. In this regime, increasing the radiation energy requires $\mathcal{E}>\mathcal{E}_0$ and decreasing the radiation energy requires $\mathcal{E}<\mathcal{E}_0$.

However, the FEL amplification takes place in the high-gain regime and the process significantly differs. When the electrons of the initially uniform beam start to move inside the ponderomotive potential, the electron distribution is modulated with a period $2\pi$ in $\psi$, which corresponds at a fixed time $t$ to a modulation with period $\lambda$ in $z$. This small modulation implies that the electron beam radiates partially coherently, such that the electric field of the radiation $E_0$ increases and the mean energy of the electron beam decreases  [in Eq. (\ref{chap7eq5b}), the energy variation averaged over all electrons is nonzero because $\langle \sin\psi \rangle\neq 0$]. As a consequence, on the one hand, the amplitude of the ponderomotive potential increases and, on the other hand, electrons lose more energy on average than predicted by the single electron model. Because the ponderomotive amplitude increases, the volume of trapped conditions also increases and electrons which initially were on untrapped trajectories can be finally trapped in the ponderomotive potential well. Therefore, after a stage of modulation of the electron distribution, electrons tend to be trapped in the wells of the ponderomotive potential. The amplification process can be summarized as the following:
\begin{itemize}
\item The electromagnetic radiation induces a modulation of the electron beam. At the beginning, the modulation is sinusoidal and the regime is linear. The higher the electric field amplitude $E_0$ is, the faster the amplitude of the sinusoidal modulation grows.
\item The electric field amplitude growth is proportional to the beam modulation. Hence, it results in an exponential amplification, $E_0\propto \exp(z/2l_g)$, where $l_g$ is the gain length in power.
\item The modulation of the electron density becomes on the order of the mean electron density, \textit{i.e.}, becomes nonlinear. This happens when electrons become trapped in the wells of the ponderomotive potential. The electron beam is microbunched at the radiation wavelength $\lambda$ and radiates coherently. The largest fraction of output radiation is generated during this last stage. This leads to an important loss of energy for the electron beam in such a way that it becomes off-resonance. This corresponds to the saturation: the radiation energy cannot be increased further in normal operation.
\item In addition, letting the process go on, electrons off-resonance fall in accelerating phases and extract energy from the electromagnetic radiation. When they recover sufficient energy to satisfy the resonance condition, they can amplify the radiation again. Hence, the radiation amplitude starts to oscillate after the saturation point.
\end{itemize}
\begin{figure}
\includegraphics[width=8.5cm]{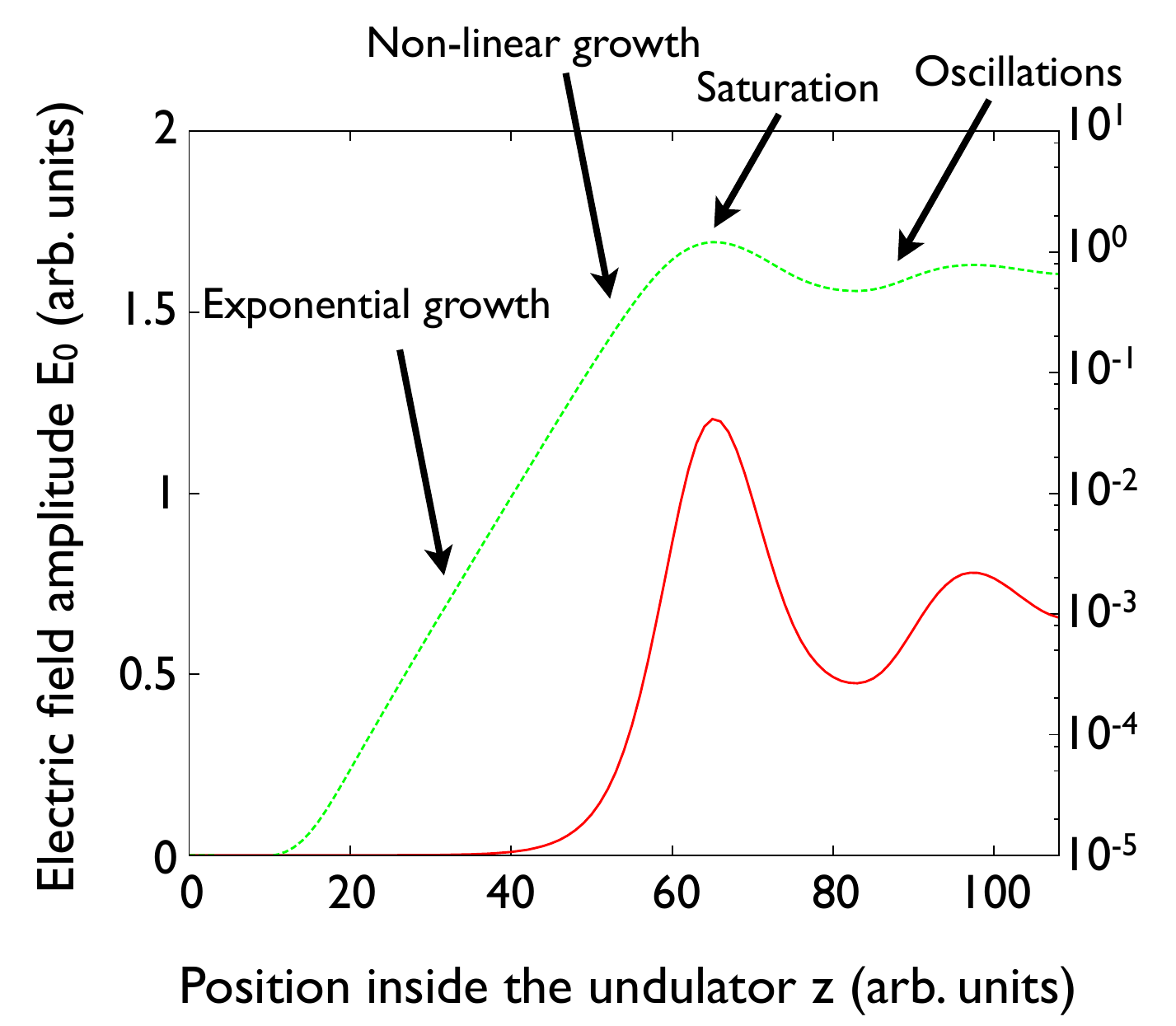}
\caption{Schematic of the FEL amplification process. A typical profile for the electric field amplitude $E_0$ along the undulator axis in linear scale (red solid line, left y axis) and logarithmic scale (green dashed line, right y axis).}
\label{chap7fig1}
\end{figure}
This FEL amplification process is illustrated in Fig. \ref{chap7fig1}, where the radiation amplitude is plotted as a function of $z$. This emphasizes the different stages of the process: linear mode of amplification, nonlinear mode, saturation, and then oscillations.

\bigskip
The efficiency of the amplification process, defined as the ratio of the output total radiation energy to the total energy of the electron bunch, is limited at saturation by the fact that electrons lose energy and become off-resonance. As presented next, in the ideal 1D case the efficiency at saturation equals the Pierce parameter $\rho$ which is small compared to unity in all cases. Recalling that the resonance condition reads $\lambda=(\lambda_u/2\gamma^2)(1+K^2/2)$, the loss of electron energy (decreasing of $\gamma$) can be compensated by varying undulator parameters (for example, the undulator gap and so $K$), which allows the process to remain in resonance after the saturation point and to continue to radiate at the wavelength $\lambda$ in a fully coherent manner. This approach is called \textit{tapering} and it permits the FEL efficiency $\eta$ to be increased well above the untapered one $\eta\simeq\rho$ and even to approach efficiency of unity.

\subsubsection{Required conditions on the electron beam parameters}
\label{chap7secAsec2}

In our simple illustration of the amplification process, any practical effects such as energy spread in the electron beam have not been considered and tolerances of the FEL process to deviations from the ideal case have not been discussed. Here the conditions requested from the electron beam parameters in order to allow amplification of an electromagnetic field are summarized.
The Pierce parameter $\rho$ (also named efficiency parameter and FEL universal parameter) is defined as
\begin{equation}
\label{chap7eq7bis}
\rho=\frac{1}{\gamma} \left (\frac{A_u\lambda_u}{4\lambda_p} \right )^{2/3}=\frac{1}{2\gamma} \left (\frac{I}{I_A} \right )^{1/3} \left (\frac{A_u\lambda_u}{2\pi\sigma} \right )^{2/3},
\end{equation}
where $\lambda_p=2\pi c/\sqrt{n_ee^2/m\epsilon_0}$ is the plasma wavelength (without relativistic correction), $I$ is the electron beam current, $I_A=4\pi\epsilon_0mc^3/e\simeq17.0\:$kA is the Alfv\'en current, $\sigma$ is the rms radius of the transverse distribution of the electron beam, and $A_u=K[J_0(x)-J_1(x)]/\sqrt{2}$ is defined in Sec. \ref{chap7secAsec1}. In practical units, the Pierce parameter reads
\begin{align}
\rho&= 1.78 \times 10^{-5}  \frac{A_u^{2/3}}{\gamma} \lambda_u^{2/3}[\text{cm}] n_e^{1/3}[\text{cm}^{-3}]\\
&= 2.65 \times 10^{1}  \frac{A_u^{2/3}}{\gamma} \lambda_u^{2/3}[\text{cm}]  \frac{I^{1/3}[\text{kA}]}{\sigma^{2/3}[\mu\text{m}]}.
\end{align}

\vspace{0.5cm}
\noindent
\textit{a. Steady-state 1D FEL theory}
%\paragraph{Steady-state 1D FEL theory\\}

In the steady-state 1D FEL theory, the Pierce parameter provides the following conditions applied on the electron beam  and the output values of the main parameters:
\begin{itemize}
\item the rms relative energy spread of the electron beam has to be smaller than $\rho$,
\begin{equation}
\label{chap7eq7}
\frac{\Delta\mathcal{E}}{\mathcal{E}}\lesssim \rho,
\end{equation}
\item the relative detuning of the mean energy $\mathcal{E}_m$ of the electron beam from the resonance ($\mathcal{E}_0$ is the energy at resonance corresponding to a given radiation wavelength $\lambda$) has to be smaller than $\rho$,
\begin{equation}
\frac{\mathcal{E}_m-\mathcal{E}_0}{\mathcal{E}_0} \lesssim \rho,
\end{equation}
\item the power gain length $l_g$ [defined by $P\propto\exp(z/l_g)$, where $P$ is the radiation power] is given by
\begin{equation}
l_g=\frac{\lambda_u}{4\pi\sqrt{3}\rho},
\end{equation}
\item the efficiency of energy conversion from the electron beam to the radiation $\eta=I{_{\text{radiation}}}/(N_e\gamma mc^2)$ is finally given at saturation by
\begin{equation}
\eta\approx\rho.
\end{equation}
\end{itemize}
The output radiation does not depend on the energy of the input radiation if the length of the undulator is chosen in such a way that saturation is attained at its exit. This is because, whatever the input radiation is, the process ends when electrons have lost sufficient energy to be off-resonance, and the amount of energy that they lose to become off-resonance is roughly given by $\rho\mathcal{E}$. The total radiation energy $I{_{\text{radiation}}}$ scales as $N_e^{4/3}$ since the Pierce parameter $\rho$ depends on the electron current as $I^{1/3}$.\\
The undulator length providing the saturation depends on the energy of the input radiation, and there is no simple expression to estimate it. Nevertheless, its typical value is on the order of 20 gain lengths, $l_{\text{sat}}\sim 20\: l_g$.\\
According to the above formulas, the Pierce parameter $\rho$ has to be as high as possible. Indeed, this leads to less stringent conditions on the energy spread and detuning, smaller gain length, and so smaller undulator length and higher efficiency at saturation. The parameter $\rho$ increases with the current or density as $I^{1/3}$ or $n_e^{1/3}$, with the strength parameter as $A_u^{2/3}$ and decreases with $\gamma$ as $1/\gamma$ for constant undulator period $\lambda_u$. However, for constant radiation wavelength $\lambda$, $\rho$ increases with $\gamma$ as $\gamma^{1/3}$ (the dependence on current or density is the same, while the dependence on $K$ becomes more complex).

\bigskip
Our next step will be to determine the limits of validity of the steady-state 1D FEL theory. Working inside the limits and the conditions of the 1D theory constitutes the best configuration for a high FEL performance. For out-of-limit configurations, one can use the Ming Xie formulas \cite{NIMA2000Xie} (given below) which coarsely account for 3D effects, or simulations with codes such as Genesis \cite{NIMA1999Reiche} for example, which give precise and quantitative answers.

\vspace{0.5cm}
\noindent
\textit{b. Three-dimensional effects}
%\paragraph{Three dimensional effects\\}

Three-dimensional effects and the associated limits of the 1D theory arise when considering the transverse dimensions. First, do the electron beam and the radiation correctly overlap? Indeed, in three dimensions diffraction occurs and radiation can escape the electron beam and become lost for the amplification process. To avoid this, the Rayleigh length of the radiation has to be higher than the gain length (to ensure gain guiding). Second, the electron beam having a finite emittance, it has an angular spread: electrons have slightly different directions, resulting in different values for $\gamma_z=1/\sqrt{1-\overline{\beta_z}^2}$ and in an effective energy spread. Imposing that this effective energy spread is smaller than $\rho$ provides a condition on the \textit{unnormalized} rms transverse beam emittance $\epsilon$ (see Sec. \ref{chap2secG}). The above conditions or limits of the 1D theory can be written as
\begin{align}
z_r\sim4\pi\frac{\sigma^2}{\lambda}&\gtrsim l_g,\\
\gamma_z^2\frac{\epsilon^2}{\sigma^2}\lesssim\rho  \;\;\; &\longrightarrow \;\;\; \epsilon^2\lesssim \frac{\lambda\sigma^2}{2\pi\sqrt{3}l_g}.
\end{align}
The Ming Xie formulas allow these effects on the FEL process to be estimated, by giving the modified gain length and the modified efficiency at saturation \cite{NIMA2000Xie},
\begin{align}
l_g^{\text{1D}}&\longrightarrow l_g^{\text{Ming\:Xie}}=l_g^{\text{1D}}(1+\Lambda),\\
\Lambda&=0.45\Lambda_d^{0.57}+0.55\Lambda_{\epsilon}^{1.6}+3\Lambda_{\gamma}^2,\\
\eta^{\text{1D}}&\longrightarrow \eta^{\text{Ming\:Xie}}\sim\eta^{\text{1D}}\Bigg(\frac{l_g^{\text{1D}}}{l_g^{\text{Ming\:Xie}}}\Bigg)^2,
\end{align}
where $\Lambda_d$, $\Lambda_\epsilon$, and $\Lambda_\gamma$ are parameters taking into account, respectively, diffraction, emittance, and energy spread effects, and are given by
\begin{align}
\Lambda_d&= \frac{l_g^{\text{1D}}\lambda}{4\pi\sigma^2},\\ 
\Lambda_{\epsilon}&= 4\pi\frac{\epsilon^2l_g^{\text{1D}}}{\lambda\sigma^2},\\
\Lambda_{\gamma}&= \frac{1}{\sqrt{3}\rho}\frac{\Delta\mathcal{E}}{\mathcal{E}}.
\end{align}
These formulas show how 3D effects degrade the FEL performance, increasing the gain length and decreasing the efficiency. However, they do not take into account time-dependent effects and space-charge effects.

\vspace{0.5cm}
\noindent
\textit{c. Time-dependent effects}
%\paragraph{Time dependent effects\\}

Time-dependent effects describe how the short duration of electron bunches and variation of electron beam parameters inside the bunch influence the FEL amplification process. The previous formulas assume a steady-state electron beam, \textit{i.e.}, infinitely long with constant density, for which slippage between radiation and electrons can be neglected. Realistic bunches have finite duration and the electron density varies along the bunch. The cooperation length is defined as the distance slipped by the radiation with respect to electrons in one gain length,
\begin{align}
l_c&= l_g(c-v_z)/c=\frac{\lambda_u}{4\pi\sqrt{3}\rho}\frac{1}{2\gamma_z^2}\\
&= \frac{\lambda}{4\pi\sqrt{3}\rho}.
\end{align}
Electrons interact with each other via the radiation field. However, this radiation field slowly slips on the bunch. The cooperation length indicates the size inside the bunch where interaction between electrons can occur during a gain length. As a direct consequence of this notion, one could consider that the bunch is made of independent sub-bunches of length $l_c$.\\
The steady-state model is valid if the electron beam parameters are approximately constant in a cooperation length inside the bunch, and if the electron bunch length is much longer that the cooperation length. This latter condition reads
\begin{equation}
c\tau\gg\frac{\lambda}{4\pi\sqrt{3}\rho},
\end{equation}
where $\tau$ is the duration of the electron bunch.\\
For bunch durations smaller than the above limit, the FEL process is highly degraded because radiation escapes the bunch in less than a gain length (edge effects). This regime is called weak superradiance \cite{NIMA1985Bonifacio, OC1988Bonifacio, PRA1989Bonifacio, NIMA1990Bonifacio}. Superradiance, as defined by \textcite{PR1954Dicke}, is a spontaneous emission from a coherently prepared system, whose energy scales as the square of the number of electrons. FEL superradiance is slightly different, because electrons begin to bunch when interacting with the spontaneous emission and evolve toward a coherently prepared system. The radiation energy of the weak superradiance regime then scales as $N_e^2$ instead of the $N_e^{4/3}$ steady-state dependence, but is much lower than the steady-state radiation energy. In contrast, for long electron bunches, strong superradiance can occur if the weak superradiance radiation emitted by the trailing edge of the bunch propagates and slips through the overall bunch and is subsequently amplified. In that case, the peak power is much greater than the steady-state saturation value.

The notion of cooperation length also indicates that the condition on the energy spread does not have to be applied on the overall electron bunch. It is sufficient that each sub-bunch of length $l_c$ has an energy spread smaller than $\rho$. The energy spread calculated within a cooperation length is called the slice energy spread. The condition (\ref{chap7eq7}) is relaxed and replaced by
\begin{equation}
\Big(\frac{\Delta\mathcal{E}}{\mathcal{E}}\Big)_{\text{slice}}\lesssim \rho.
\end{equation}
Requiring transverse emittance can also be replaced with the same conditions applying slice transverse emittance, defined in the same way as slice energy spread.

\vspace{0.5cm}
\noindent
\textit{d. Space-charge effects}
%\paragraph{Space charge effects\\}

Another effect which influences the FEL performance is space charge. When the FEL amplification process tends to microbunch the electron beam at the radiation wavelength, longitudinal space-charge forces, on the contrary, tend to debunch the beam. The inverse plasma frequency $\omega_p^{-1}$ (time scale of plasma oscillations) can indicate over which distance debunching occurs, while the gain length characterizes the distance over which bunching forces are significant. In the electron rest frame, the longitudinal distance is $\gamma$ times higher than in the laboratory frame, so the density is divided by $\gamma$ and the plasma frequency by $\gamma^{1/2}$. Hence plasma oscillations take place on a time scale $\gamma^{1/2}/\omega_p$ in the rest frame, and because of time dilatation, they take place on a time scale $\gamma^{3/2}/\omega_p$ in the laboratory frame, corresponding to a covered distance of $\gamma^{3/2}c/\omega_p$. Longitudinal space-charge forces are then negligible when
\begin{equation}
\gamma^{3/2}c/\omega_p \gtrsim l_g,
\end{equation}
where $\omega_p=\sqrt{n_ee^2/m\epsilon_0}$ is the plasma frequency without any relativistic correction.\\
Transverse space-charge forces can also influence the FEL performance. In the envelope equation for the electron beam, the focusing field, emittance term, and the space-charge term appeared. To neglect transverse space-charge forces, the corresponding term in the envelope equation has to be small compared to the other terms, \textit{i.e.}, the focusing one and the emittance one \cite{Humphries}.

\vspace{0.5cm}
\noindent
\textit{e. Undulator errors and wakefields}
%\paragraph{Undulator errors and wakefields\\}

The practical realization of a free-electron laser consists in many undulator sections and transport devices. The magnetic field of undulators has small errors and sections can be slightly misaligned, resulting in a degradation of the FEL performance. In addition, a high-current electron beam propagating between the two resistive walls of small-gap undulators induces a wakefield which can modify the beam properties along the propagation, altering the FEL process. See \textcite{PRSTAB2007Huang} for a description of these effects.

\vspace{0.5cm}
\noindent
\textit{f. Quantum FEL}
%\paragraph{Quantum FEL\\}

The FEL process described above is based on a classical treatment, so it is relevant to ensure that quantum phenomena are unimportant by determining the limit for which the dynamics is effectively described by classical mechanics. The FEL dynamics becomes quantum if the momentum recoils $\Delta\vec{p}=-\hbar\vec{k}$ of an electron emitting a photon of wave vector $\vec{k}$ is greater than the FEL bandwidth $\rho|\vec{p}|=\rho\mathcal{E}/c$, \textit{i.e}, if the electron becomes off-resonance after emitting a photon. Hence, in a quantum FEL, each electron radiates only one photon. In all realizable cases, $\hbar k\ll\rho p$ and quantum effects are completely negligible so that the classical FEL theory correctly describes the process.

\bigskip
To conclude, we emphasize that 1D formulas are useful for simple estimations, and Ming Xie formulas \cite{NIMA2000Xie} for corrections due to energy spread, diffraction, and emittance. For quantitative study of a given configuration and for time-dependent effects and space-charge force effects, self-consistent, 3D, and time-dependent codes such as Genesis can be used \cite{NIMA1999Reiche}.

\subsubsection{Seeding or self-amplified spontaneous emission configurations}
\label{chap7secAsec3}

In the above description of the FEL amplification process we did not discuss where the input radiation field came from. The FEL radiation can be started either by an initial seeding field or by a random modulation of the electron beam (SASE regime). 

The above description of the FEL amplification process directly applies to the seeding case. This case provides high-quality output radiation, but requires that a relatively intense coherent radiation field at the given wavelength $\lambda$ can already be generated, which is not the case for high-energy photons as x rays. In principle, high-order harmonics (HHG) generated in gas allow one to seed FEL up to the extreme ultraviolet part of the spectrum at the present state of the art. A recent experiment has demonstrated the FEL amplification of a seeding beam at 160 nm by 3 orders of magnitude \cite{NatPhys2008Lambert}. These schemes, and, in particular, HHG on solid targets \cite{PRA2000Tarasevitch, NatPhys2006Dromey, RMP2009Teubner}, could be extended to x-ray photons in the near future and could provide a seed for X-FEL. However, the seeding power has to be higher than the power of the incoherent synchrotron or spontaneous emission \cite{EPL2009Lambert}, and this latter is proportional to the desired photon energy. This means that it will be more and more difficult to have a suitable seeding beam for smaller radiation wavelength and higher photon energy because the required power will be higher.

The case of an initially randomly modulated electron beam (SASE configuration) is different and the theoretical description is more complex. No radiation input enters with the electron beam inside the undulator. As discussed in Sec. \ref{chap2secF}, electrons inside bunches delivered by conventional or laser-plasma accelerators are randomly distributed in space. Whereas if the electron distribution in space is uniform no radiation is emitted, for a random one, incoherent radiation is emitted (synchrotron radiation, whose power is proportional to the number of electrons) and this incoherent radiation provides a seed for the FEL amplification process. Moreover, the incoherent radiation is emitted at the wavelength given by Eq. (\ref{chap7eq4}), and therefore the resonance condition is automatically satisfied. The main advantage of this method is straightforward: it allows one to extend FEL radiation to high-energy photons because no external coherent radiation is needed. A major experiment demonstrating self-amplified spontaneous emission gain, exponential growth, and saturation was achieved by \textcite{Science2001Milton} and highlighted the possibility to develop an operational X-FEL. Nowadays, most of the X-FEL projects ($\lambda<1$ nm) are based on this approach. However, because the input is a fluctuating quantity shot to shot, the quality of the output radiation is very low compared to the seeding case, in which the output inherits the quality of the external input. For the SASE case, the output power has very large fluctuations and very poor temporal coherence. The temporal profile and the spectrum are constituted of many spikes; the spikes of the temporal profile have typical durations of $l_c/c$ while the spikes of the spectrum have typical relative width of $\lambda_u/l_b$, where $l_b$ is the bunch length (the FEL process smooths the parameters of the electron beam and the radiation over the scale of the cooperation length).

\subsection{Free-electron laser from a laser-plasma accelerator}
\label{chap7secB}

From the different required conditions on the electron beam parameters (see Sec. \ref{chap7secAsec2}), the feasibility of a FEL from a laser-plasma accelerator can be investigated for each specific undulator or wiggler type. The above results which are strictly valid for the conventional undulator give only good estimations for the other ones [for example, see \textcite{NIMA2008Bacci} for the comparison between electromagnetic wave and conventional undulators].

\subsubsection{With a conventional undulator}
\label{chap7secBsec1}

Conventional undulators have typically centimeter period and strength parameter of unity. Hence, according to the resonance condition (\ref{chap7eq4}), using electron bunches at the 100 MeV level, the FEL radiation wavelength is in the UV to XUV range, whereas at the 1 GeV level it reaches the soft x-ray and hard x-ray ranges. In the following, we consider two different scenarios, X-FEL 1 and 2, which are speculative and can be seen as future perspectives in the short term and long term, respectively, of laser-plasma accelerators. All input parameters of these scenarios and the corresponding output FEL parameters are given in Table \ref{chap7tab1}. Table \ref{chap7tab1} contains the results obtained by 3D time-dependent Genesis simulations in the SASE configuration.

The choice of the electron beam parameters is motivated by extension of the 100 MeV range state-of-the-art laser-plasma accelerator performance to higher electron energies. The electron beam peak current is maintained constant when going to higher electron energies because the requirement of small energy spread imposes an optimized beam loading which flattens the accelerating electric field along the bunch \cite{PRL2009Rechatin2}. The optimized peak current scales as the normalized vector potential $a_0$ of the laser pulse and is independent of the electron density of the plasma $n_e$. Because the electron accelerator efficiency scales as $1/a_0$, $a_0$, and therefore the electron beam peak current, should not be increased when designing a higher energy laser-plasma accelerator. The absolute energy spread and the normalized emittance can remain constant in the acceleration process if beam loading is optimized and if quasistationary forces are applied to the electron beam. Therefore, in principle, the low absolute energy spread (a few MeV) \cite{PRL2009Rechatin1} and normalized emittance $\epsilon_N=1\:\pi.\text{mm}.\text{mrad}$ \cite{PRL2010Brunetti} achievable for electrons at the 100 MeV range could also be obtained at 1 and 5 GeV.

Note that the bunch rms transverse size $\sigma$ is experimentally chosen by the transport configuration. Of course, this choice determines the rms angular spread $\theta=\epsilon/\sigma$. In order to have the Pierce parameter as high as possible, $\sigma$ has to be small, but, on the other hand, emittance, diffraction, and space-charge effects increase while $\sigma$ decreases. In addition, practical considerations on the transport realization along the undulator put a limit on the smallest transverse size achievable.

\begin{table*}
\caption{Parameters of different scenarios for realizing free-electron lasers from laser-driven plasma-based accelerators in the XUV and x-ray domains with the corresponding FEL quantities obtained from Genesis simulations in SASE configuration. For all scenarios, the undulator period $\lambda_u$ is equal to 1 cm and the strength parameter $K$ is equal to 1.}
\smallskip
\begin{tabular*}{\textwidth}{ l*{3}{ @{\extracolsep{0pt plus12pt}} l} }
\hline
\hline
& X-FEL 1 & X-FEL 2\\ \hline
Electron beam input parameters \\ \hline
Electron energy & 1 GeV &  5 GeV \\ 
Peak current & 10 kA & 10 kA\\ 
Bunch rms duration &  4 fs & 4 fs\\
Bunch rms transverse size & 30 $\mu$m & 30 $\mu$m\\
rms energy spread & 0.2\% & 0.04\%\\
Normalized rms transverse emittance & 1 $\pi.\text{mm}.\text{mrad}$ & 1 $\pi.\text{mm}.\text{mrad}$ \\ \hline
FEL output parameters\\ \hline
Radiation wavelength $\lambda$ & 1.96 nm & 0.0783 nm \\
Pierce parameter $\rho$ & 0.00225 & 0.000450 \\
Undulator length & 12.5 m & 100 m \\
Peak power & 25 GW & 15 GW \\
Pulse energy & 50 $\mu$J & 25 $\mu$J\\
rms pulse duration & 2 fs & 2 fs \\
\hline
\hline
\end{tabular*}
\label{chap7tab1}
\end{table*}

\bigskip
Since the topic of short wavelength free-electron laser using conventional undulators has been extensively studied by means of conventional accelerators, we discuss here the differences which arise when using electron bunches produced by laser-plasma accelerators. The first difference is straightforward: because of the short duration of the electron bunch, currents can be higher  which result in higher $\rho$, smaller gain lengths, smaller undulator lengths, higher efficiency at saturation, less stringent conditions on the energy spread and detuning, but also in higher space-charge effects.\\
Second, the short duration of the electron bunch can become comparable or smaller than the cooperation length divided by $c$, leading to important edge effects which degrade the FEL performance.\\
Third, the electron bunch transport from the plasma source to the undulator is specific: even if the emittance is comparable to that of conventional accelerators, at the exit of the plasma, the transverse size is very small (a few microns) when the divergence is higher (a few milliradians). The transverse size grows fast: 10 cm after the plasma it reaches a few hundreds of microns, and 1 m after it reaches a few millimeters. For such size, chromatic aberrations of quadrupoles (which are proportional to the electron beam energy spread) are large, leading to an enormous growth of emittance, and because electrons cover different distances, the longitudinal bunch length also grows. It is therefore necessary to design an ultracompact transport system which acts near the plasma source and does not let the beam transverse size grow. Space-charge forces, higher in the case of laser-plasma accelerators, also strongly affect the beam transport, increasing emittance, longitudinal bunch length, and total energy spread [they induce a linear energy chirp: the head of the bunch moves faster than the tail \cite{APB2007Gruner, PRSTAB2009Gruner}].

\begin{figure}
\includegraphics[width=8.5cm]{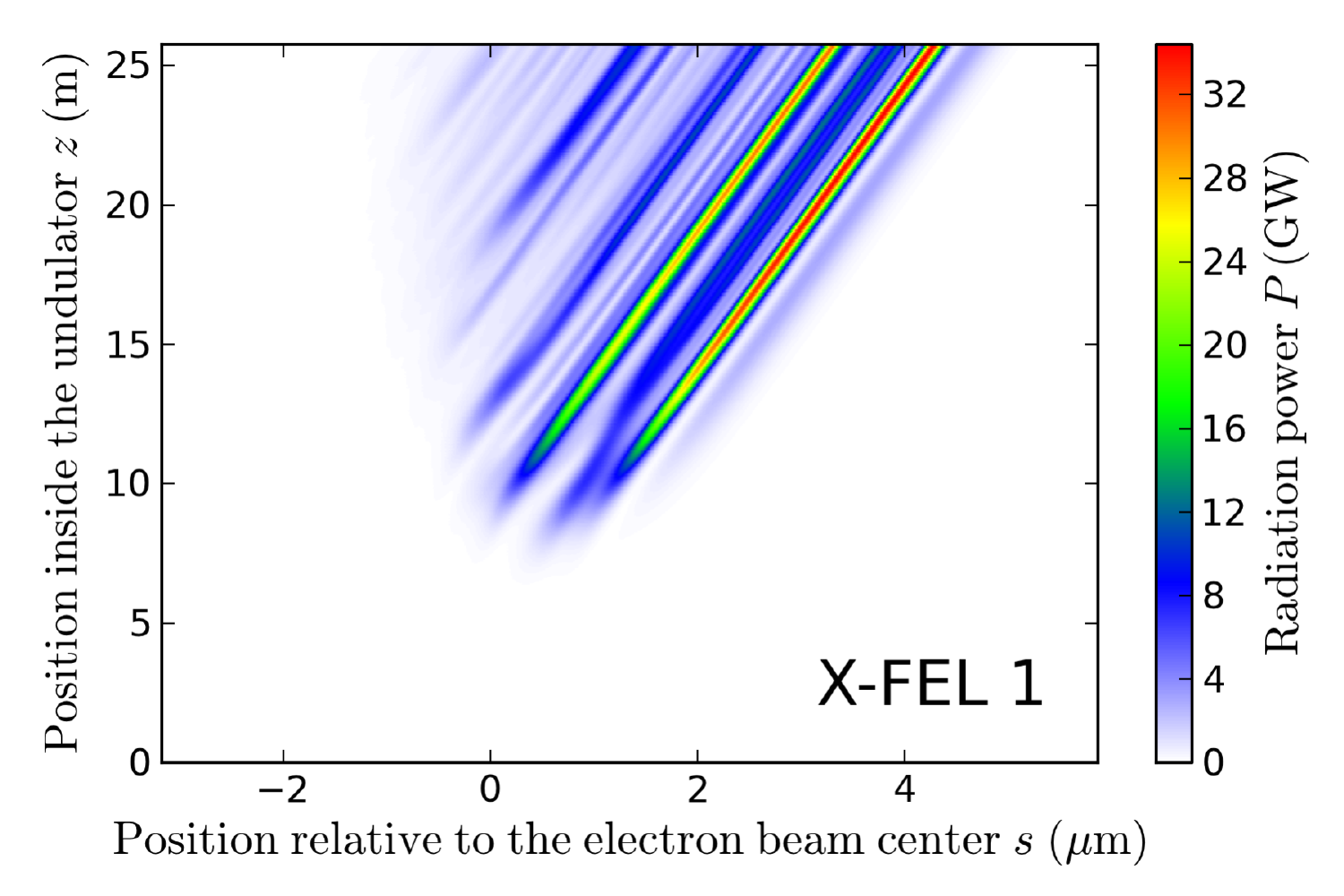}
\includegraphics[width=8.5cm]{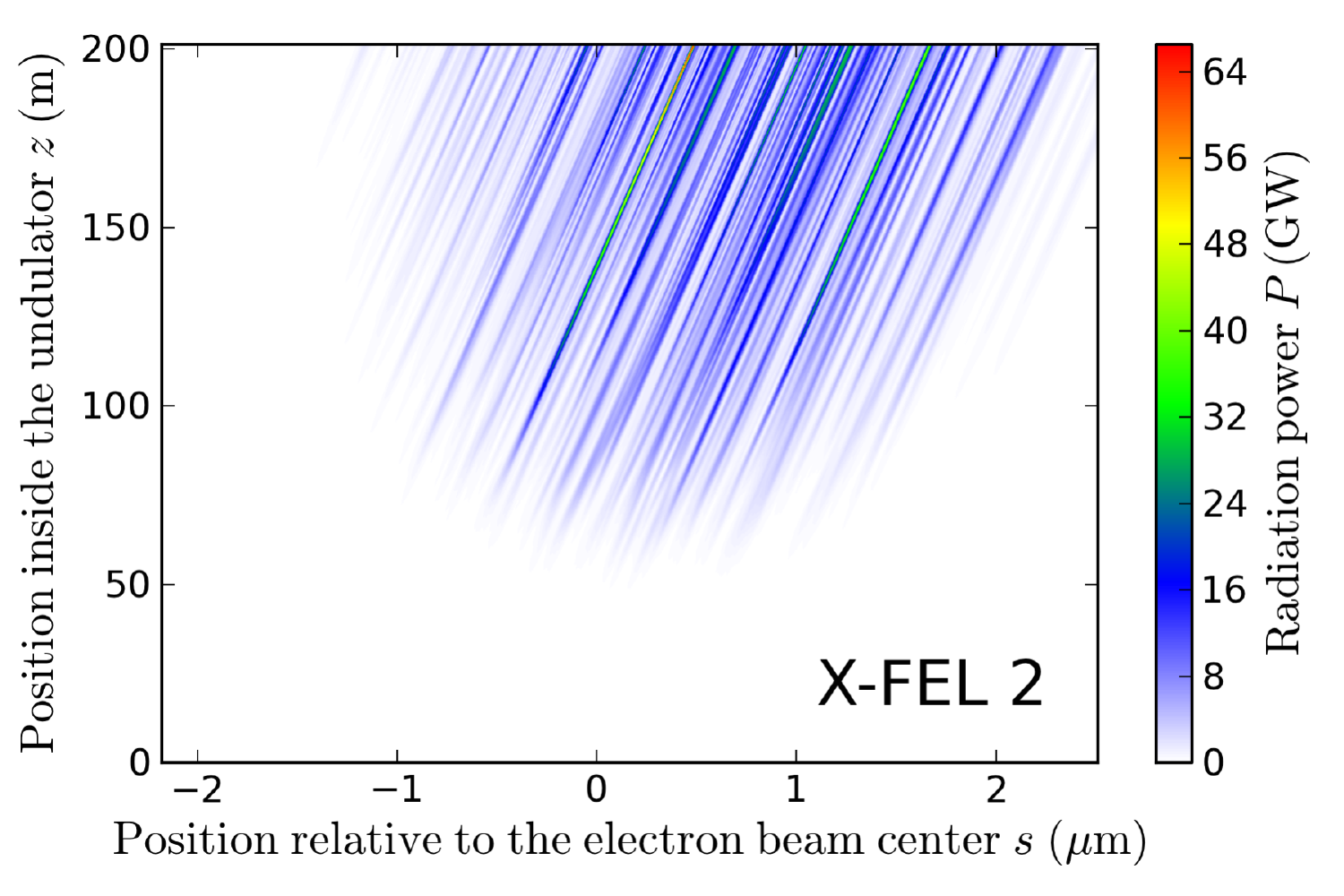}
\caption{Longitudinal profile of radiation power as a function of the position inside the undulator obtained from Genesis simulations. The corresponding electron beam and undulator parameters are indicated in Table \ref{chap7tab1}. Top: X-FEL 1. Bottom: X-FEL 2. The position relative to the electron beam center is defined as $s=ct_e-ct$, where $t_e$ and $t$ are, respectively, the arrival time of the electron beam center and of the radiation, at a given position $z$ in the undulator.}
\label{chap7fig2}
\end{figure}

\bigskip
We comment now on the results obtained for the scenarios considered in Table \ref{chap7tab1}. They are illustrated in Fig. \ref{chap7fig2}, which shows the longitudinal profile of radiation power as a function of the position inside the undulator for each scenario. In both configurations, X-FEL 1 and 2, the small values of the Pierce parameter impose stringent conditions on the electron beam energy spread. As can be seen in Table \ref{chap7tab1}, the energy spread is close to $\rho$ in both cases, and using a higher electron beam energy spread strongly reduces the FEL performance. In X-FEL 1, the output FEL radiation consists of a few SASE spikes, meaning that the bunch length is of the order of a few cooperation lengths and that edge effects do not degrade the FEL performance. Moreover, diffraction, emittance, and longitudinal space-charge effects are found to have negligible impact on FEL output, while the main degrading effect comes from the energy spread. Strong FEL radiation, with 25 GW peak power and 50 $\mu$J pulse energy, is produced at a wavelength of 1.96 nm with an undulator length of 12.5 m.\\
In X-FEL 2, the radiation wavelength reaches the angstrom range, a spectral region of great interest for diffraction experiments, more particularly on biological samples. In these conditions, the bunch length equals many cooperation lengths and the notion of slice energy spread is helpful: it is sufficient that only the slice energy spread is below 0.04\%, not the total energy spread. Here, because of the higher $\gamma$ and the smaller $\rho$, the requirements on the relative slice energy spread and on the undulator length are more stringent. The output FEL radiation has a 15 GW peak power and 25 $\mu$J pulse energy at a wavelength of 0.0783 nm, for an undulator length of 100 m. In this situation, diffraction, longitudinal space-charge and edge effects are found to be negligible while both emittance and energy spread have an impact on the FEL performance, which may explain the slightly lower performance than in the previous scenario.

\bigskip
In summary, laser-plasma accelerators have promising perspectives in the realization of compact free-electron lasers, by combining them with conventional undulators. The scenarios X-FEL 1 and 2 highlight the possibility to produce strong FEL radiation from a laser-plasma accelerator. The most challenging condition is the relative energy spread of the electron beam, which has to be decreased to attain the x-ray part of the spectrum. In addition, high-quality electron beams at the GeV level are required [1 GeV electron beams have already been generated \cite{NatPhys2006Leemans} but not with the same quality, \textit{i.e.}, energy spread and stability, as at the 100 MeV level]. The notion of slice energy spread and slice emittance can facilitate reaching the FEL requirements for the highest-energy case.

\subsubsection{With an electromagnetic wave undulator}
\label{chap7secBsec2}

In Sec. \ref{chap6}, the incoherent radiation of an electron beam submitted to a counterpropagating electromagnetic wave was presented. This wave acts as an undulator or wiggler whose strength parameter $K$ equals the dimensionless vector potential amplitude $a_0$ of the laser pulse and whose period equals $\lambda_L/2$, where $\lambda_L$ is the wavelength of the counterpropagating laser pulse. Currently, only two laser systems can deliver such a pulse with $a_0$ on the order of unity: Ti:Sa and CO$_2$, with wavelengths of 0.8 and 10 $\mu$m, respectively. From the resonance condition (\ref{chap7eq4}), electron beams at the 10 MeV level radiate in the keV (nanometer) range for CO$_2$ lasers and in the 10 keV (angstrom) range for Ti:Sa lasers. At the 100 MeV level, the radiation is in the 100 keV range and 1 MeV range for, respectively, CO$_2$ and Ti:Sa lasers, and at the GeV level the radiation is in the 10 and 100 MeV range for, respectively, CO$_2$ and Ti:Sa lasers. The use of an electromagnetic wave undulator in the FEL process was proposed in the 1980s \cite{IEEE1987Danly, IEEE1987Gea, IEEE1988Gallardo, NIMA1988Mima} and reinvestigated recently \cite{NIMA2008Bacci, PRSTAB2006Bacci, PROC2007Maroli, PRSTAB2008Petrillo}. It was also suggested to couple an electron beam coming from a laser-plasma accelerator and a relativistic laser pulse, in the conditions where FEL collective amplification takes place, and this system has been called an all-optical free-electron laser (AOFEL) \cite{PRSTAB2008Petrillo}.

According to Pierce parameter scalings (\ref{chap7eq7bis}), for a given radiation wavelength $\lambda$, using a smaller undulator period $\lambda_u$ and smaller $\gamma$ is unfavorable: the Pierce parameter $\rho$ decreases and the requirements on the electron beam parameters are more stringent for the electromagnetic undulator than for the conventional undulator (remember that for constant $\lambda$, $\rho$ scales as $\gamma^{1/3}$). The most realistic scenarios for the electromagnetic undulator use high values of $\lambda$, \textit{i.e.} CO$_2$ laser and electron beams at the 10MeV level.

\textcite{PRSTAB2008Petrillo} simulated the FEL collective amplification taking place in the Thomson backscattering of a relativistic laser pulse by an electron beam, this latter generated by a laser-plasma accelerator. First, they use a 2.5D particle in cell simulation (3D in the fields and 2D in the coordinates) to obtain electron beam parameters. They considered the case of an injection realized by a down ramp in the plasma density profile and showed that, after selecting a particular slice of the electron beam of longitudinal size 0.5$\:\mu$m, it is possible to obtain a slice energy spread of 0.4\%, a slice transverse emittance of $0.3\:\pi.\text{mm}.\text{mrad}$, and a current of $20\:$kA at electron energies of about 30 MeV. Then, inserting the exact shape of the slice in the FEL code Genesis 1.3 [using the equivalence between conventional and electromagnetic undulators developed by \textcite{NIMA2008Bacci}], they obtained for the SASE case the radiation output as a function of the undulator length: for a CO$_2$ laser pulse with normalized vector potential $a_0=0.8$, coherent radiation at wavelength $\lambda=1.35\:\text{nm}$ with subfemtosecond duration and $200\:$MW peak power is generated for a saturation length of a few millimeters. These results open perspectives for an x-ray AOFEL using millimeter-scale electromagnetic wave undulators. However, their simulation does not take into account the short propagation from the plasma exit to the electromagnetic wave undulator, which could degrade the beam parameters at such low energies and high currents.

\bigskip
To conclude, the AOFEL has the main advantage of being compact and not needing a transport system. Because of the short undulator period, high photon energies can be achieved with small electron energies. Although it allows one to theoretically generate hard x rays and $\gamma$ rays, the FEL amplification for such wavelength demands extremely stringent conditions on energy spread and emittance (the Pierce parameter is smaller for higher photon energies). Hence, it is more reasonable to use electron energies at the 10 MeV level and CO$_2$ laser pulses in order to succeed in operating a free-electron laser. Experimentally, good quality electron beams (with percent energy spread) have been demonstrated at the 100 MeV level \cite{PRL2009Rechatin2}, but not yet at the 10 MeV level. It is generally thought that, because acceleration conserves \textit{absolute} energy spread, smaller \textit{relative} energy spread is easier to obtain at higher electron energies. Hence, achieving very small relative energy spread at the low-energy 10 MeV level seems to be difficult. From the perspective of an AOFEL, it is therefore necessary to experimentally investigate the production of high-quality electron beams at rather small electron energies, when currently the tendency is toward higher and higher electron energies.

\subsubsection{With a plasma undulator}
\label{chap7secBsec3}

As shown in Sec. \ref{chap4}, the ion cavity, generated behind the laser pulse when the bubble or blowout regime is attained, acts as an undulator or wiggler whose period is $\lambda_u=\sqrt{2\gamma}\lambda_p$ and whose strength parameter is $K=r_\beta k_p\sqrt{\gamma/2}$. Assuming that we can prepare experimental conditions in which the electron beam keeps a constant energy inside the ion cavity ($\gamma$ is constant and $\lambda_u$ is too), then FEL amplification can in principle take place. For the case of a ion channel, the process has been studied \cite{PRL1990Wittum, PFB1992Whittum} and is called an ion channel laser (ICL). Additional conditions have to be fulfilled: all electrons have to oscillate in the same plane (for the radiation to be polarized identically for all electrons) and with the same strength parameter (to radiate at the same wavelength) \cite{PRE2002Esarey, PoP2003Kostyukov}.

The main difference between ICL and FEL is that, in ICL, the strength parameter $K$ depends on both the energy and the transverse amplitude of motion $r_{\beta}$, when FEL it is a constant defined uniquely by the undulator parameters. In ICL, different values of $K$ lead to an effective energy spread in $\gamma_z=\gamma/\sqrt{1+K^2/2}$. Imposing that this energy spread is smaller than the Pierce parameter $\rho$ leads to the condition $\Delta r_{\beta}/r_{\beta}\lesssim \rho (2+K^2)/K^2$ for the admissible relative spread in the amplitude of motion $r_{\beta}$. Hence, the realization of ICL needs a configuration in which there is no acceleration to satisfy the condition of constant $\gamma$, where there is an off-axis injection with amplitude $r_{\beta}$ and corresponding relative amplitude spread roughly below $\rho$. To our knowledge, a method to implement such a configuration is currently not known. In addition, conditions analogous to those presented in Sec. \ref{chap7secAsec2} regarding the natural energy spread, emittance, and undulator length, also have to be fulfilled.

\section{Conclusion}
\label{chap9}

In this paper, mechanisms that can produce short-pulse x-ray or gamma-ray radiation using laser-accelerated electrons have been described. We have seen that all these sources are based on the radiation from a relativistically moving charge, and that even if the details of the electron orbits are different for each source, the main features can be obtained using only five parameters: the relativistic factor $\gamma$ of electrons, the number of electrons $N_e$, the strength parameter $K$, the period $\lambda_u$, and the number of periods $N$ of the undulator or wiggler. In addition, the incoherent radiation provided by these sources can eventually be amplified to coherent radiation via the FEL mechanism for sufficiently high-quality electron beams; the use of conventional undulators represents the most promising route toward such goal. Here a summary of the features of these sources is given.

Table \ref{chap9tab1} gives typical parameters and features as well as scaling laws for the incoherent sources based on the use of a plasma undulator (betatron radiation), a conventional undulator or an electromagnetic wave undulator (Thomson backscattering). For the produced sources, interaction parameters accessible with currently available tens of TW-class lasers have been considered. The photon spectra obtained with each mechanism show that an energy range from eV to MeV can be covered. The main difference between the different sources reviewed resides in their photon energy. Depending on the spectral range required for a desired application, Table \ref{chap9tab1} helps to select the most suitable type of source.

\begin{table*}
\caption{The first subtable gives typical features for the different sources achievable with a 50 TW-class laser. The values represent the orders of magnitude of the parameters. The second subtable provides the scalings of the radiation features with the relevant parameters $K$, $\lambda_u$, and $\gamma$ derived in Sec. \ref{chap2} for sinusoidal trajectories. The last subtable presents the scalings of the relevant parameters with the practical parameters for each schemes derived from simple ideal models which have been described in each corresponding section. The practical parameters of each source are indicated in brackets. We refer to each section for the definition of the quantities used in this table.}
\bigskip
\begin{tabular*}{\textwidth}{ l*{8}{ @{\extracolsep{0pt plus12pt}} l} }
Typical features & $\gamma$ & $\lambda_u$ & $K$ & $N$ & $N_e$ & $\hbar\omega/\hbar\omega_c$ & $\theta_r$ & $N_\gamma $\\ 
\hline
Betatron & 200 & 150 $\mu$m & 10 & 3 & $10^9$ &  5 keV & 50 mrad & $\sim 10^9$\\ 
Conventional undulator & 400 & 1 cm& 1 & 100 & $10^8$ &  25 eV & 2.5 mrad &$\sim 10^8$  \\ 
Thomson backscattering & 400 & 0.4 $\mu$m & 1 & 10 & $10^8$ & 650 keV & 2.5 mrad & $\sim 10^7$\\ 
\smallskip
\end{tabular*}
\begin{tabular*}{\textwidth}{ l*{1}{ @{\extracolsep{0pt plus12pt}} l} }
Scalings of the radiation features\\
\hline
Fundamental radiation energy $\hbar\omega$ for undulators ($K<1$) & $(2\gamma^2 hc/\lambda_u)/(1+K^2/2)$\\
Critical radiation energy $\hbar\omega_c$ for wigglers ($K\gg1$) & $\frac{3}{2}K\gamma^2 hc/\lambda_u$\\
Typical opening angle of the radiation $\theta_r$ for undulators ($K<1$) & $1/\gamma$\\
Typical opening angle of the radiation $\theta_r$ for wigglers ($K>1$) & $K/\gamma$\\
Number of emitted photons per electron and per period $N{_\gamma}$,\\
for undulators ($K<1$) & $1.53\times10^{-2}K^2$\\
for wigglers ($K\gg1$) & $3.31\times10^{-2}K$\\
\smallskip
\end{tabular*}
\begin{tabular*}{\textwidth}{ l*{2}{ @{\extracolsep{0pt plus12pt}} l} }
Scalings of the relevant parameters & $K$ & $\lambda_u$\\
\hline
Betatron ($n_e$, $r_{\beta}$, and $\gamma$) & $\sqrt{\gamma/2}k_pr_{\beta}$ & $\sqrt{2\gamma}\lambda_p$  \\
Conventional undulator ($B_0$, $\lambda_u$, and $\gamma$) & $eB_0\lambda_u/(2\pi mc)$ & $\lambda_u$ \\
Thomson backscattering ($a_0$, $\lambda_L$, and $\gamma$) & $a_0$ & $\lambda_L/2$\\
\end{tabular*}
\label{chap9tab1}
\end{table*}

\bigskip
In the XUV domain, the conventional undulator coupled with a laser-plasma accelerator has been discussed. The emitted radiation can be monochromatic if the electron bunch is monoenergetic. Up to $\sim 10^{9}$ photons collimated within a few milliradians could be produced in a single shot using $10^9$ electrons (160 pC) in the hundreds of MeV energy range and an undulator of strength parameter $K=1$ with 100 periods. The radiation wavelength can be tuned by varying the electron energy. Because the acceleration and the wiggling are dissociated, a stable and/or tunable regime of acceleration can be used, resulting in a stable and/or tunable radiation source. From the perspective of GeV electron beams, this scheme will allow one to reach the keV range. However, this source is more complex to realize because the electron bunch must be transported into the undulator. The transport can degrade the electron bunch duration (and so the x-ray pulse duration) and the transverse emittance, leading to a broader bandwidth in the radiation spectrum and a smaller brightness. Finally, this scheme deserves to be developed as it is on the path toward a free-electron laser based on laser-plasma accelerators.

In the soft x-ray range, we discussed nonlinear Thomson scattering. This radiation is produced by electrons oscillating in an intense laser field. While this source has a spectrum extending in the few hundreds of eV range using tens of TW-class lasers, it may become an interesting mechanism to generate keV x rays with PW-class lasers.

In the range of a few keV, the betatron mechanism has been demonstrated to be an efficient and simple method to produce short x-ray pulses. This source can produce up to $10^9\ \text{photons}$ per shot, collimated within less than 50 mrad (FWHM). On the one hand, controlling the electron orbits within the cavity in order to ensure a better energy transfer from the electron to the radiation could allow one to extend the spectral range to the tens of keV range, while still working with 50 TW-class lasers. On the other hand, petawatt-class lasers and/or capillaries would be suitable to create larger ion cavities with lower plasma density and to accelerate electrons to the GeV range. This should allow one to produce x-ray beams whose divergence is in the milliradian range and photon energies in the tens and hundreds of keV range, with a higher number of photons.

Above 10 keV and up to a few MeV, the mechanism used is Thomson backscattering. Here high-energy radiation is produced thanks to the short period of the electron motion. This scheme is similar to the case of a laser-plasma accelerator coupled with a conventional undulator, except that it benefits from the short period of the electron motion resulting in emitted photons of high energies and it does not need transport. The photon energy can be chosen by tuning the electron energy. This scheme has been demonstrated recently and it is the most promising for generating tunable incoherent bright hard x-ray and $\gamma$-ray radiation.

\bigskip
The different schemes reviewed here produce incoherent radiation whose energy is proportional to the number of electrons contributing to the emission. These sources can be compared, in some ways, to conventional synchrotron radiation. The main advantage provided by the laser-plasma approach is, of course, the reduced size (from hundreds of meters or kilometers to the laboratory room) and the reduced cost (from multihundred million dollars or multibillion dollars to multimillion dollars). Moreover, the pulse duration of laser-based sources is femtosecond when it is commonly picosecond for conventional synchrotrons even if, with the latest technology, it can be reduced down to femtosecond (and this has a certain cost). Another advantage is the natural synchronization of these sources with exciting laser pulses needed in pump-probe experiments, since both the pump and the probe come from the same laser system. Again, external femtosecond synchronization in synchrotron facilities is feasible with recent methods, but it remains difficult. The femtosecond x-ray sources based on the relativistic laser-plasma interaction presented in this review are attractive mainly because of their low cost, and this is why perspectives cannot be extended indefinitely to the use of higher intensity, since the cost will become comparable with conventional accelerators and synchrotron facilities. However, compared to synchrotrons, laser- and plasma-based sources present significant shot-to-shot fluctuations in terms of photon energy, number of photons, and divergence. They can therefore not cover a range of applications as large as synchrotrons. These sources remain at the research stage with limited applications  for the moment. Important challenging developments are required to produced stable and robust x-ray beams appropriate for applications. In addition, the repetition rate is problematic: while synchrotron facilities work at MHz repetition rate allowing one to considerably increase the signal-to-noise ratio, laser-based sources work at 10 Hz on paper, but in practice they work in the single-shot regime in most 50 TW-class laser facilities to prevent rapid damage on optics such as  in compressors. This means that theses sources are efficient for a single-shot experiment or experiments requiring not too much data accumulation.

\begin{figure}[b!]
\includegraphics[width=8.5cm]{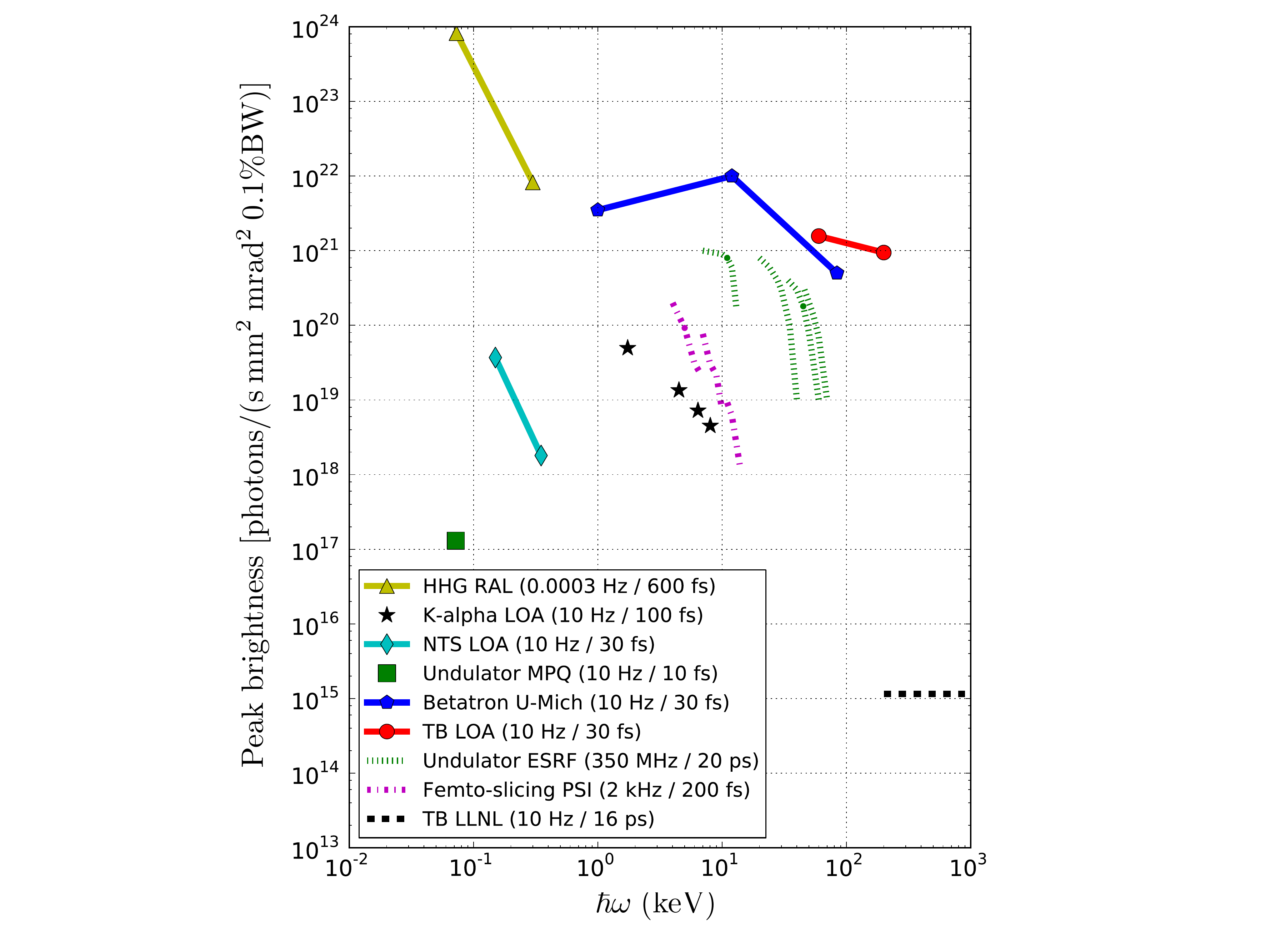}
\caption{Peak brightness for different types of x-ray sources: high harmonic generation from relativistic laser and overdense plasma interaction (triangle and yellow line) [experiment conducted with the Vulcan Petawatt laser system at Rutherford Appleton Laboratory (RAL) by \textcite{NatPhys2006Dromey}], $K_\alpha$ radiation from laser-produced plasmas (black star) [experiment conducted at the Laboratoire d'Optique Appliqu\'ee (LOA) by \textcite{PRE1994Rousse}], nonlinear Thomson scattering (NTS) from relativistic laser and underdense plasma interaction (diamond and cyan line) [experiment conducted at LOA by \textcite{PRL2003TaPhuoc}], conventional undulator radiation from laser-plasma accelerators (green square) [experiment conducted with the ATLAS laser system at the Max-Planck-Institut f\"ur Quantenoptik (MPQ) by \textcite{NatPhys2009Fuchs}], betatron radiation from laser-plasma accelerators (pentagon and blue line) [experiment conducted with the Hercules laser system at the University of Michigan (U-Mich) by \textcite{NatPhys2010Kneip}], Thomson backscattering (TB) radiation from laser-plasma accelerators (circle and red line) [experiment conducted at LOA by \textcite{NatPhot2012TaPhuoc}], undulator radiation from conventional accelerators (green dotted line) [data from the European Synchrotron Radiation Facility (ESRF); see the \href{http://www.esrf.eu/Accelerators/Performance/Brilliance}{ESRF website}], femto-slicing-based undulator radiation from conventional accelerators (magenta dash-dotted line) [data from the femto-laser station at the Paul Scherrer Institut (PSI); see the \href{http://www.psi.ch/sls/microxas/femto}{PSI website} and \textcite{PRL2007Beaud}] and Thomson backscattering (TB) from conventional accelerators (black dashed line) [experiment conducted at the Lawrence Livermore National Laboratory (LLNL) by \textcite{PRSTAB2010Albert}]. The repetition rate and the pulse duration used in the peak brightness estimation are given in the legend, and permit one to convert peak brightness into average brightness.}
\label{fig_brightness}
\end{figure}

In Fig. \ref{fig_brightness}, the x-ray sources are compared to other existing x-ray sources in terms of their peak brightness. This includes x-ray sources based on the laser-plasma interaction such as high harmonic generation from relativistic laser and overdense plasma interaction \cite{PRA2000Tarasevitch, NatPhys2006Dromey, PRL2007Dromey, NatPhys2007Thaury, RMP2009Teubner}, and $K_\alpha$ radiation from laser-produced plasmas \cite{Science1991Murnane, PhysFluids1993Kieffer, PRE1994Rousse}, as well as x-ray sources based on conventional accelerators: the undulator radiation source from the European Synchrotron Radiation Facility (ESRF), the femto-slicing-based undulator radiation source from the femto-laser station at the Paul Scherrer Institut (PSI), and the Thomson $\gamma$-ray source from the Lawrence Livermore National Laboratory (LLNL) \cite{PRSTAB2010Albert}. Figure \ref{fig_brightness} shows that laser-plasma-based x-ray sources have competitive peak brightness, in particular, the betatron radiation and the all-optical Thomson source, because of their femtosecond duration and their micron source size.

As free-electron lasers are the natural evolution of synchrotrons, FEL based on laser-plasma interaction could be the next step in the development of laser- and plasma-based x-ray sources. In the free-electron laser process, the emitted radiation is amplified and becomes coherent with the number of photons increased by orders of magnitude. The most promising scenario uses a conventional undulator coupled to a laser-plasma accelerator. Because stringent conditions apply to the electron beam parameters such as energy spread and transverse emittance, challenging developments are required in the laser-plasma accelerator domain, including ultracompact transport system. Thanks to the high brightness of FEL radiation, many experiments could be performed in single shot or with small data accumulation such that the low repetition rate of laser facilities becomes a more benign disadvantage than for incoherent sources.

\bigskip
In conclusion, this article gives a comprehensive review of the different mechanisms proposed for the production of femtosecond x-ray radiation from laser-plasma accelerators. These sources have been described within the same formalism, considering different configurations of accelerator and undulator or wiggler and extracting the relevant parameters such as $K$, $\lambda_u$, and $\gamma$. Finally, the possibility to implement a free-electron laser using these configurations has been discussed and the electron beam quality required for such realization has been presented.

\begin{acknowledgments}
We are grateful to C. Rechatin, O. Lundh, J. Faure, M. Ribiere  and S. Sebban for fruitful discussions.
We acknowledge the Agence Nationale pour la Recherche, through the COKER Project No. ANR-06-BLAN-0123-01, the European Research Council through the PARIS ERC project (under Contract No. 226424), and LASERLAB-EUROPE/LAPTECH, EC FP7 Contract No. 228334 for their financial support.
\end{acknowledgments}

\end{document}